\documentclass[aps,prb,longbibliography,preprint]{revtex4-1}

\usepackage{amsmath,amssymb,bm,graphicx}
\usepackage{xcolor}
\usepackage{upgreek}
\usepackage[colorlinks=true,urlcolor=blue,linkcolor=blue,citecolor=blue,bookmarks=false]{hyperref}

\begin{document}


\title{Generic aspects of skyrmion lattices in chiral magnets}

\author{Andreas Bauer}
\email[]{andreas.bauer@frm2.tum.de}
\affiliation{Physik-Department, Technische Universit\"{a}t M\"{u}nchen, James-Franck-Stra{\ss}e, D-85748 Garching, Germany}

\author{Christian Pfleiderer}
\affiliation{Physik-Department, Technische Universit\"{a}t M\"{u}nchen, James-Franck-Stra{\ss}e, D-85748 Garching, Germany}

\begin{abstract}
Magnetic skyrmions are topologically non-trivial spin whirls that may not be transformed continuously into topologically trivial states such as ferromagnetic spin alignment. In recent years lattice structures composed of skyrmions have been discovered in certain bulk chiral magnets with non-centrosymmetric crystal structures. The magnetic phase diagrams of these materials share remarkable similarities despite great variations of the characteristic temperature, field, and length scales and regardless whether the underlying electronic state is that of a metal, semiconductor, or insulator.
\end{abstract}

\vskip2pc

\maketitle

\section{Introduction and outline}

In 1961 British nuclear physicist Tony Skyrme proposed a theoretical model in which neutrons and protons arise as topological solitons of pion fields, i.e., fermions are derived from bosonic fields~\cite{1961:Skyrme:ProcRSocLondA, 1961:Skyrme:ProcRSocLondA2, 1962:Skyrme:NuclPhys}. Representing the, perhaps, first example of what is now broadly referred to as fractionalization, the implications of Skyrme's model only began to be fully appreciated two decades later, when Witten and Adkins demonstrated its relevance for real experiments~\cite{1983:Adkins:NuclPhysB}. Since the days of this early work many different variants of Skyrme's original notion have been worked out in entirely different fields of physics. These states and excitations are now rather generously called \textit{skyrmions}. Examples include areas as diverse as particle physics~\cite{1983:Adkins:NuclPhysB, 1986:Zahed:PhysRep, 1997:Diakonov:ZPhysA, 1997:Chabanat:NuclPhysA, 1998:Chabanat:NuclPhysA}, the quantum Hall state at half-filling~\cite{1993:Sondhi:PhysRevB, 1995:Schmeller:PhysRevLett, 2006:Yang:PhysRevB}, Bose-Einstein condensates~\cite{1998:Ho:PhysRevLett, 2001:Khawaja:Nature, 2009:Leslie:PhysRevLett}, and liquid crystals~\cite{2011:Fukuda:NatCommun}. However, in recent years skyrmions are probably most actively investigated in the area of solid state magnetism, where certain spin textures are referred to as skyrmions. These magnetic textures display a non-trivial real-space topology, i.e., it is not possible to continuously transform them into conventional (topologically trivial) forms of spin order such as ferromagnetism or antiferromagnetism. 

While skyrmions were theoretically predicted to exist in non-centrosymmetric magnetic materials with uniaxial anisotropy as early as 1989~\cite{1989:Bogdanov:SovPhysJETP, 1994:Bogdanov:JMagnMagnMater}, it was despite concerted efforts rather unexpected, when skyrmions in magnetic materials were identified experimentally for the first time in the cubic transition metal compounds MnSi~\cite{2009:Muhlbauer:Science} and Fe$_{1-x}$Co$_{x}$Si~\cite{2010:Munzer:PhysRevB} in the form of a lattice structure. Since then similar topologically non-trivial spin textures have been reported to exist for a rapidly growing number of rather different bulk and thin film systems. The interest driving this search for further materials stabilizing skyrmions is quite diverse, ranging from fundamental questions on the possible break down of Fermi liquid theory~\cite{2001:Pfleiderer:Nature,2004:Pfleiderer:Nature,2013:Ritz:Nature} all the way to new forms of spintronics applications~\cite{2013:Nagaosa:NatureNano}. From a practical point of view the most important implication of the non-trivial topology is their emergent electrodynamics leading to an exceptionally efficient coupling between the spin textures and spin currents~\cite{2010:Jonietz:Science, 2013:Fert:NatureNano}. Further, the very detailed understanding of the spin excitations achieved to date suggests strongly that tailored microwave devices may be designed through the combination of different materials~\cite{2012:Seki:Science, 2012:Mochizuki:PhysRevLett, 2013:Okamura:NatCommun, 2015:Schwarze:NatureMater, 2015:Kezsmarki:arXiv}.

A precondition for further advances is a detailed understanding of the magnetic phase diagrams of these compounds. In turn, this chapter provides a review of the most extensively studied class of skyrmion materials to date, namely cubic chiral magnets crystallizing in the space group $P2{_1}3$. We begin in Sec.~\ref{ChiralMagnets} with a brief introduction to the basic properties of this class of compounds focusing on the salient properties of the skyrmion lattice state. This is followed by an introduction to the Ginzburg-Landau model of these materials in Sec.~\ref{Theory}. The main part of this chapter in Sec.~\ref{PhaseDiagrams} is dedicated to an account of the determination of magnetic phase diagrams based on measurements of thermodynamic bulk properties. Despite great variations of the characteristic temperature, field, and length scales between the different materials of interest, the magnetic phase diagrams observed are remarkably similar. This brings us to a summary of the main consequences that arise from the non-trivial topological winding of skyrmions in Sec.~\ref{EmergentED}, in particular their emergent electrodynamics. The chapter closes in Sec.~\ref{Conclusions} with a brief account of topologically non-trivial spin structures as recently discovered in other materials.


\section{Skyrmion lattice in cubic chiral magnets}
\label{ChiralMagnets}

\begin{figure}
\includegraphics[width=1.0\linewidth]{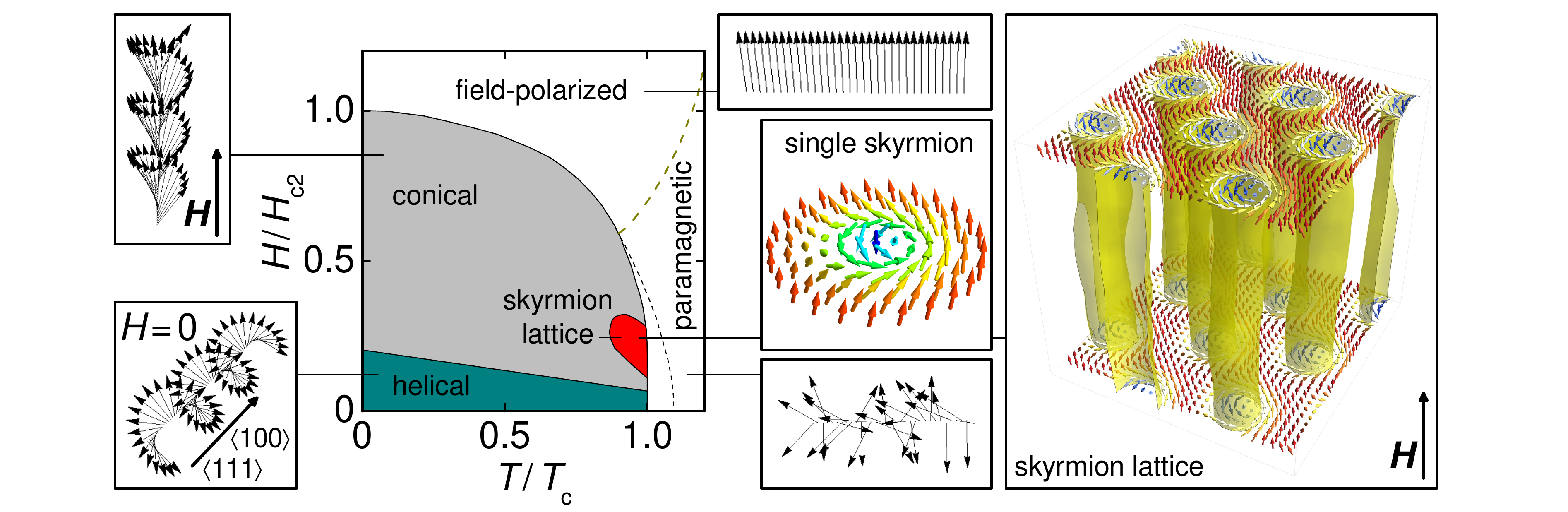}
\caption{Spin structures the cubic chiral magnets. Typical magnetic phase diagram (center) and schematic spin structures of the helical, the conical, the paramagnetic, and the field-polarized state. In a phase pocket (red) in finite fields just below the helimagnetic ordering temperature, $T_{c}$, a regular arrangement of topologically non-trivial spin whirls is observed, a so-called skyrmion lattice. Schematic depictions by Markus Garst and from Ref.~\cite{2013:Milde:Science}.}
\label{figure1}
\end{figure}

The helimagnetism of the materials reviewed in this chapter is homochiral with a modulation wavelength that is large as compared to typical lattice constants. The latter represents an important precondition for the description of the magnetic properties in a continuum model and the characterization of the topological properties. Well-known representatives are the (pseudo-)binary $B20$ transition metal monosilicides and monogermanides MnSi, Mn$_{1-x}$Fe$_{x}$Si, Mn$_{1-x}$Co$_{x}$Si, Fe$_{1-x}$Co$_{x}$Si, FeGe, MnGe, and mixtures thereof, as well as the insulator Cu$_{2}$OSeO$_{3}$. All of these compounds crystallize in the space group $P2_{1}3$, which lacks inversion symmetry such that two crystalline enantiomers stabilize.

The long-wavelength helimagnetic order observed in these compounds originates in a well-understood set of hierarchical energy scales, as already pointed out in Landau-Lifshitz, Vol.\ VIII, Sec.~52,~\cite{1980:Landau:Book}. On the strongest scale exchange interactions favor parallel spin alignment. On intermediate scales isotropic Dzyaloshinskii-Moriya spin-orbit interactions arise due to the lack of inversion symmetry of the crystal structure favoring perpendicular spin alignment~\cite{1957:Dzyaloshinsky:SovPhysJETP, 1960:Moriya:PhysRev, 1964:Dzyaloshinsky:SovPhysJETP}. In competition with the stronger exchange a helical modulation is stabilized~\cite{1976:Ishikawa:SolidStateCommun, 1976:Motoya:SolidStateCommun}. The chirality of the Dzyaloshinskii-Moriya interaction and thus of the helical modulation is fixed by the enantiomer of the crystal structure~\cite{1983:Shirane:PhysRevB, 1985:Ishida:JPhysSocJpn}. Finally, on the weakest energy scale higher-order spin-orbit coupling terms, also referred to as crystal electric field effects or cubic anisotropies, determine the propagation direction of the helical modulations~\cite{1980:Bak:JPhysCSolidState}.

The hierarchy of energy scales is directly reflected in a rather universal magnetic phase diagram, as schematically depicted in Fig.~\ref{figure1}. As summarized below, the same phase diagram is observed regardless whether the materials are metals, semiconductors, and insulators (MnGe is perhaps the only exception as discussed in Sec.~\ref{Conclusions}). In particular, the phase diagram appears to be insensitive to the quantitative values of the transition temperatures, transition fields, and helix wavelengths, which vary by roughly two orders of magnitude between different compounds. 

At sufficiently high temperatures the magnetic properties are characteristic of exchange-enhanced paramagnetism with large fluctuating moments~\cite{1985:Ishikawa:PhysRevB}. At low temperatures and zero magnetic field multi-domain helical order is observed with equal domain populations, where the helical propagation vector is determined by weak cubic magnetic anisotropies, fourth-order in spin-orbit coupling. Under small applied magnetic fields the domain population changes, until the helical state undergoes a spin-flop transition at a transition field $H_{c1}$~\cite{1993:Lebech:Book}. The spin-flop phase is broadly referred to as conical state, with a single-domain state of spin spirals propagating along the magnetic field direction. The expression conical phase alludes to the notion, that the spins tilt towards the field direction while twisting helically along to the field direction. When increasing the magnetic field further this conical angle closes and a transition takes place to a field-polarized state above $H_{c2}$~\cite{1975:Bloch:PhysLettA}. We will return to a more detailed discussion of the transitions at $H_{c1}$ and $H_{c2}$ below.

In recent years the perhaps largest scientific interest has been attracted by a small phase pocket at intermediate fields just below the helimagnetic transition temperature, $T_{c}$. Historically this phase pocket has been referred to as A-phase. The existence of the A-phase, first discovered in MnSi, had already been reported in the 1970s~\cite{1970:Fawcett:InternJMagnetism, 1976:Kusaka:SolidStateCommun}. However, the detailed microscopic spin structure was only identified in 2008 (publication in 2009), when small-angle neutron scattering established the first realization of a skyrmion lattice in a bulk solid state system~\cite{2009:Muhlbauer:Science}. 

The skyrmion lattice consists of a regular hexagonal arrangement of spin whirls, that may essentially be described by the phase-locked superposition of three helices under $120^{\circ}$ in a plane perpendicular to the applied magnetic field in combination with a ferromagnetic component along the field. Of particular interest is the non-trivial topology of this spin texture, meaning, it cannot be continuously transformed into a topologically trivial state such as a paramagnet, ferromagnet, or helimagnet. The associated winding number of the structure, $\mathit{\Phi}$, is an integer and the integrated value of the skyrmion density, $\phi_{i}$, per magnetic unit cell, given by
\begin{equation}
\phi_{i} = \frac{1}{8\uppi}\epsilon_{ijk}\hat{\psi}\cdot\partial_{j}\hat{\psi}\times\partial_{k}\hat{\psi}
\end{equation}
where, $\epsilon_{ijk}$ is the antisymmetric unit tensor and $\hat{\psi} = \bm{M}(\bm{r})/M(\bm{r})$ is the orientation of the local magnetization. Along the field direction the quasi two-dimensional spin structure repeats itself, forming skyrmion lines as depicted in the right panel of Fig.~\ref{figure1}. Perhaps most intriguing, the interaction of each skyrmion with an electron spin corresponds to one quantum of emergent flux and an emergent electrodynamics presented in Sec.~\ref{EmergentED}.

\begin{figure}
\includegraphics[width=1.0\linewidth]{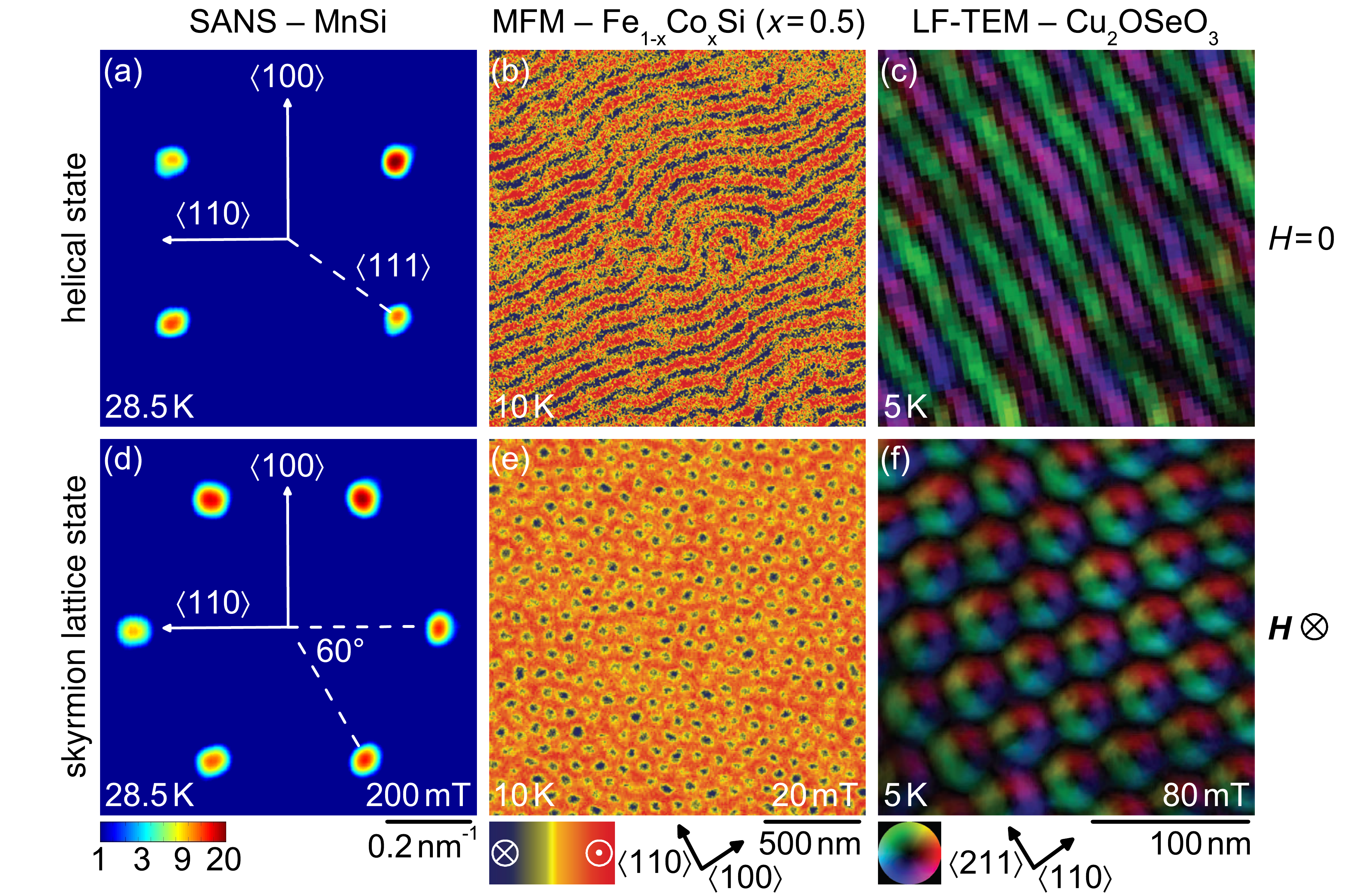}
\caption{Helical and skyrmion lattice state as observed in reciprocal and real space. \mbox{(a)--(c)}~Helical state in zero magnetic field. \mbox{(d)--(f)}~Skyrmion lattice state in finite field. Data from small angle neutron scattering (SANS)~\cite{2009:Muhlbauer:Science,2011:Adams:PhysRevLett}, magnetic force microscopy (MFM)~\cite{2013:Milde:Science}, and Lorentz force transmission electron microscopy (LF-TEM)~\cite{2010:Yu:Nature,2012:Seki:Science} are shown. The color-coded in-plane orientation in the LF-TEM data was obtained by a transport-of-intensity (TIE) analysis.}
\label{figure2}
\end{figure}

Experimentally, the existence of skyrmions was first recognized in the form of the skyrmion lattice as observed in reciprocal space using small-angle neutron scattering (SANS) in bulk samples~\cite{2009:Muhlbauer:Science, 2010:Munzer:PhysRevB, 2010:Pfleiderer:JPhysCondensMatter, 2012:Adams:PhysRevLett, 2013:Moskvin:PhysRevLett}. Further detailed SANS studies on MnSi revealed the presence of weak higher-order scattering, indicating a weak particle-like character of the skyrmions. The evolution of this higher-order scattering as a function of temperature and field proved the long-range crystalline nature of the skyrmion lattice and, in particular, the phase-locked multi-$Q$ nature of the modulation at heart of the non-trivial topological winding~\cite{2011:Adams:PhysRevLett}. These measurements were soon followed-up by real-space imaging studies using Lorentz force transmission electron microscopy (LF-TEM). This method is sensitive to in-plane components of the magnetic moments. However, it may only be used to study thinned bulk samples~\cite{2010:Yu:Nature, 2011:Yu:NatureMater, 2012:Seki:Science, 2012:Tonomura:NanoLett}, whereas magnetic force microscopy (MFM) allowed the detection of the stray magnetic field above the surface of bulk samples~\cite{2013:Milde:Science}. As the most recent achievement of real-space imaging, the spin arrangement in the skyrmion lattice could even be reconstructed in three dimensions by means of electron holography~\cite{2014:Park:NatureNano}.

Typical data from SANS, MFM, and LF-TEM recorded on different chiral magnets are shown for the helical and the skyrmion lattice state in Fig.~\ref{figure2}. In the helical state at zero magnetic field SANS experiments show intensity maxima along the easy axes of the helical propagation vector $\bm{q}$, typically either $\langle100\rangle$ or $\langle111\rangle$~\cite{1976:Ishikawa:SolidStateCommun, 1980:Bak:JPhysCSolidState}. Real-space images reveal stripy patterns with $\bm{q}$ perpendicular to the stripes~\cite{2006:Uchida:Science}. The skyrmion lattice state in finite fields in SANS experiments, see Fig.~\ref{figure2}(d), is characterized by a sixfold scattering pattern in a plane perpendicular to the applied magnetic field that is only fully revealed if the magnetic field is applied parallel to the neutron beam. In earlier experiments the magnetic field and the neutron beam had been applied perpendicular to each other leading to erroneous interpretations~\cite{1984:Ishikawa:JPhysSocJpn, 1993:Lebech:Book, 1995:Lebech:JMagnMagnMater, 2006:Grigoriev:PhysRevB}. Note that the wave vector in the skyrmion lattice has the same absolute value as in the helical state, $q = 2\uppi / \lambda_{\mathrm{h}}$. Thus, due to the hexagonal packing of the skyrmions in real space, the distance between neighboring skyrmion cores is a factor of $2/\sqrt{3} \approx 1.15$ larger than the helix wavelength. In real-space images, see Figs.~\ref{figure2}(e) and \ref{figure2}(f), a trigonal lattice of objects is observed. The magnetic moments in their cores are aligned antiparallel to the applied field, cf.\ blue color in Fig.~\ref{figure2}(e), i.e., the spin structure in the cubic chiral magnets in fact consists of anti-skyrmions. 

Interestingly, when the size of bulk samples along the field direction becomes comparable to the helical modulation length, the skyrmion lattice extents over increasingly larger parts of the magnetic phase diagram as demonstrated in LF-TEM studies~\cite{2011:Yu:NatureMater}. In contrast, the magnetic properties of epitaxially grown thin films of the same chiral magnets, forming equal crystalline domain populations with both chiralities in the same film, are still debated controversially~\cite{2012:Huang:PhysRevLett, 2013:Li:PhysRevLett, 2014:Sinha:PhysRevB, 2014:Wilson:PhysRevB}. Here, in addition to the effects resulting from the heterochirality and the reduced dimensionality, strain arising from the lattice mismatch with the substrate needs to be taken into account.


\section{Theoretical description}
\label{Theory}

The thermodynamic properties of the cubic chiral magnets may be described extremely well in the framework of a Ginzburg-Landau model of the free energy density, see also chapter by Markus Garst. It is convenient to distinguish two contributions, $f = f_{0} + f_{\mathrm{cub}}$, where the first term accounts for isotropic contributions and the second term accounts for the effects of magnetic anisotropies. More specifically, $f_{0}$ includes ferromagnetic exchange, the Dzyaloshinskii-Moriya interaction as the highest-order (isotropic) spin-orbit coupling term, and the Zeeman term as the response on an external magnetic field. It may be written as:
\begin{equation}
f_{0} = \frac{1}{2}\bm{\psi}(r-J\nabla^{2})\bm{\psi} + D\bm{\psi}(\nabla \times \bm{\psi}) + \frac{u}{4!}(\bm{\psi}^{2})^{2} - \mu_{0}\mu\bm{\psi}\bm{H}
\end{equation}
We choose the three component order parameter field, $\bm{\psi}$, with dimensionless units yielding a magnetization density $\bm{M} = \mu\bm{\psi}$ with $\mu = \mu_{\mathrm{B}}/\mathrm{f.u.}$, i.e., a single Bohr magneton per formula unit ($\mu_{\mathrm{B}} > 0$). The parameter $r$ tunes the distance to the phase transition, $J$ is the exchange stiffness and $u$ the lowest order mode-coupling parameter. The second term, $D\bm{\psi}(\nabla \times \bm{\psi})$, corresponds to the Dzyaloshinskii-Moriya interaction with the coupling constant $D$. This term is justified by the lack of inversion symmetry of the crystal structure. The last term describes the Zeeman coupling to an applied magnetic field $\bm{H}$. An ansatz for a single conical helix is:
\begin{equation}
\bm{\psi}(\bm{r}) = \psi_{0}\hat{\bm{\psi}}_{0} + \mathit{\Psi}_{\mathrm{hel}}\hat{\bm{e}}^{-}\mathrm{e}^{\mathrm{i}\bm{Q r}} + \mathit{\Psi}_{\mathrm{hel}}^{*}\hat{\bm{e}}^{+}\mathrm{e}^{-\mathrm{i}\bm{Q r}}
\end{equation}
Here, $\psi_{0}$ is the amplitude of the homogeneous magnetization and $\mathit{\Psi}_{\mathrm{hel}}$ is the complex amplitude of the helical order characterized by the pitch vector $\bm{Q}$. The vectors $\hat{\bm{e}}_{1} \times \hat{\bm{e}}_{2} = \hat{\bm{e}}_{3}$ form a normalized dreibein where $\hat{\bm{e}}^{\pm} = (\hat{\bm{e}}_{1} \pm \mathrm{i}\hat{\bm{e}}_{2})/\sqrt{2}$ and $\bm{Q} = Q\hat{\bm{e}}_{3}$.

This brings us to the second term of the free energy density, $f_{\mathrm{cub}}$, which contains spin-orbit coupling of second or higher order breaking the rotation symmetry of $f_{0}$ already in zero field. 
\begin{equation}
f_{\mathrm{cub}} = \frac{J_{\mathrm{cub}}}{2}\left[(\partial_{x}\psi_{x})^{2} + (\partial_{y}\psi_{y})^{2} + (\partial_{z}\psi_{z})^{2}\right] + ...
\end{equation}
This leading-order term of the cubic anisotropies, where $J_{\mathrm{cub}} \ll J$, implies that the easy axis of the helical propagation vector is either a $\langle100\rangle$ or a $\langle111\rangle$ direction as explored by Bak and Jensen~\cite{1980:Bak:JPhysCSolidState}. As the field is increased the Zeeman term gains importance and finally overcomes the cubic anisotropies, stabilizing the conical state with the propagation vector parallel to the magnetic field, in analogy to the spin-flop transition of a conventional antiferromagnet. In order to account for more subtle effects, further cubic anisotropies need to be considered consistent with the non-centrosymmetric space group $P2_{1}3$.

While the contributions in $f_{0}$ and $f_{\mathrm{cub}}$ are sufficient to describe the helical, the conical, the field-polarized, and the paramagnetic ground states, specific issues require consideration of the higher-order spin-orbit coupling terms mentioned above and other contributions. For instance, for an universal account of the collective spin excitations it is necessary to include dipolar interactions~\cite{2015:Schwarze:NatureMater}. Moreover, just above the paramagnetic-to-helimagnetic phase transition at $T_{c}$ non-analytic corrections to the free energy functional arise from strong interactions between isotropic chiral fluctuations. These interactions suppress the correlation length and the second-order mean-field transition resulting in a fluctuation-disordered regime just above $T_{c}$ and a fluctuation-induced first-order transition. The scenario relevant for cubic chiral magnets was originally predicted by Brazovskii~\cite{1975:Brazovskii:SovPhysJETP} and recently demonstrated in MnSi by a study combining neutron scattering, susceptibility, and specific heat measurements~\cite{2013:Janoschek:PhysRevB}. Depending on the strength of the interaction between the fluctuations, for other chiral magnets an extended Bak-Jensen or a Wilson-Fischer scenario may be relevant~\cite{2010:Grigoriev:PhysRevB2, 2011:Grigoriev:PhysRevB, 2014:Zivkovic:PhysRevB}.

\begin{figure}
\includegraphics[width=1.0\linewidth]{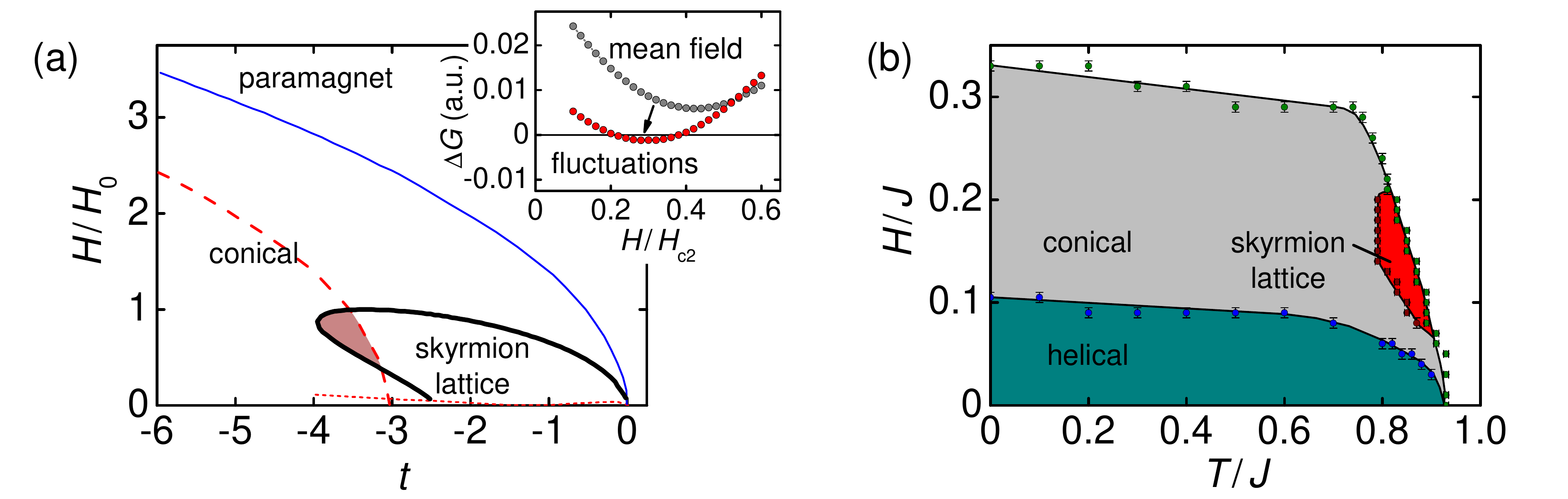}
\caption{Stabilization of the skyrmion lattice. (a)~Theoretical magnetic phase diagram as obtained from a Ginzburg-Landau ansatz. The inset shows that thermal fluctuations already in Gaussian order stabilize the skyrmion lattice at intermediate fields~\cite{2009:Muhlbauer:Science}. (b)~Magnetic phase diagram as obtained from Monte-Carlo simulations~\cite{2013:Buhrandt:PhysRevB}.}
\label{figure3}
\end{figure}

As a hidden agenda the fluctuation-induced first-order transition underscores that the skyrmion lattice state is stabilized by thermal fluctuations, as depicted in Fig.~\ref{figure3}(a). The leading-order correction arise from Gaussian fluctuations around the mean-field spin configurations of the conical and the skyrmion lattice state, respectively. Interestingly, both short-range and long-range fluctuations favor the skyrmion lattice for intermediate magnetic fields~\cite{2009:Muhlbauer:Science}. Consistently, the skyrmion lattice forms rather independently from the orientation of the underlying crystalline lattice, where the cubic anisotropies only lead to a slightly anisotropic temperature and field range of the skyrmion lattice phase pocket~\cite{2012:Bauer:PhysRevB,2012:Seki:PhysRevB2} and determine the precise orientation of the skyrmion lattice~\cite{2009:Muhlbauer:Science,2012:Adams:PhysRevLett}. 

Both the Brazovskii scenario and the stabilization of the skyrmion lattice by thermal fluctuations have recently been corroborated by classical Monte Carlo simulations~\cite{2013:Buhrandt:PhysRevB}. Here, a fully non-perturbative study of a three-dimensional lattice spin model, i.e., going beyond Gaussian order, reproduced the thermodynamic signatures associated with a Brazovskii-type fluctuation-induced first-order phase transition and, as shown in Fig.~\ref{figure3}(b), the experimental magnetic phase diagram.

All of these recent advances compare and contrast with the seminal studies of Bogdanov and coworkers, who anticipated the existence of skyrmions in non-centrosym\-metric materials with a uniaxial anisotropy and in the presence of a magnetic field~\cite{1989:Bogdanov:SovPhysJETP, 1994:Bogdanov:JMagnMagnMater}. In particular, based on mean-field calculations ignoring the importance of thermal fluctuations, they concluded for cubic compounds that the skyrmion lattice would be metastable. Moreover, recently they predicted more complex magnetic phase diagrams comprising, besides the phases discussed so far, of meron textures and skyrmion liquids~\cite{2006:Roessler:Nature, 2010:Butenko:PhysRevB}. Putative evidence for such complex phase diagrams has been reported in FeGe based on susceptibility~\cite{2011:Wilhelm:PhysRevLett, 2012:Wilhelm:JPhysCondensMatter}, specific heat~\cite{2013:Cevey:PhysStatusSolidiB}, and SANS data~\cite{2013:Moskvin:PhysRevLett}. However, as illustrated in Sec.~\ref{PhaseDiagrams}, all data reported to date for all cubic chiral magnets are qualitatively extremely similar. Thus, when consistently inferring the transition fields and temperatures by virtue of the very same conditions, the magnetic phase diagrams of all compounds including FeGe are highly reminiscent of each other supporting strongly a rather universal scenario as described in the following without evidence of these complexities..


\section{Magnetic phase diagrams}
\label{PhaseDiagrams}

In the following we focus on the determination of the magnetic phase diagrams of cubic chiral magnets based on magnetization, ac susceptibility, and specific heat data, where the conditions for determining the transition fields are confirmed by microscopic probes, notably extensive neutron scattering. In the first part of this section we present typical data, explain how transition fields or temperatures are defined, and illustrate that demagnetization effects may lead to significant corrections. This is followed in the second part by the presentation of magnetic phase diagrams of the most-extensively studied stoichiometric compounds MnSi, FeGe, and Cu$_{2}$OSeO$_{3}$ as well as the magnetic and compositional phase diagrams of the most extensively studied doped compounds, namely Mn$_{1-x}$Fe$_{x}$Si and Fe$_{1-x}$Co$_{x}$Si.

\subsection{Phase transitions in the susceptibility and specific heat}

\begin{figure}
\includegraphics[width=1.0\linewidth]{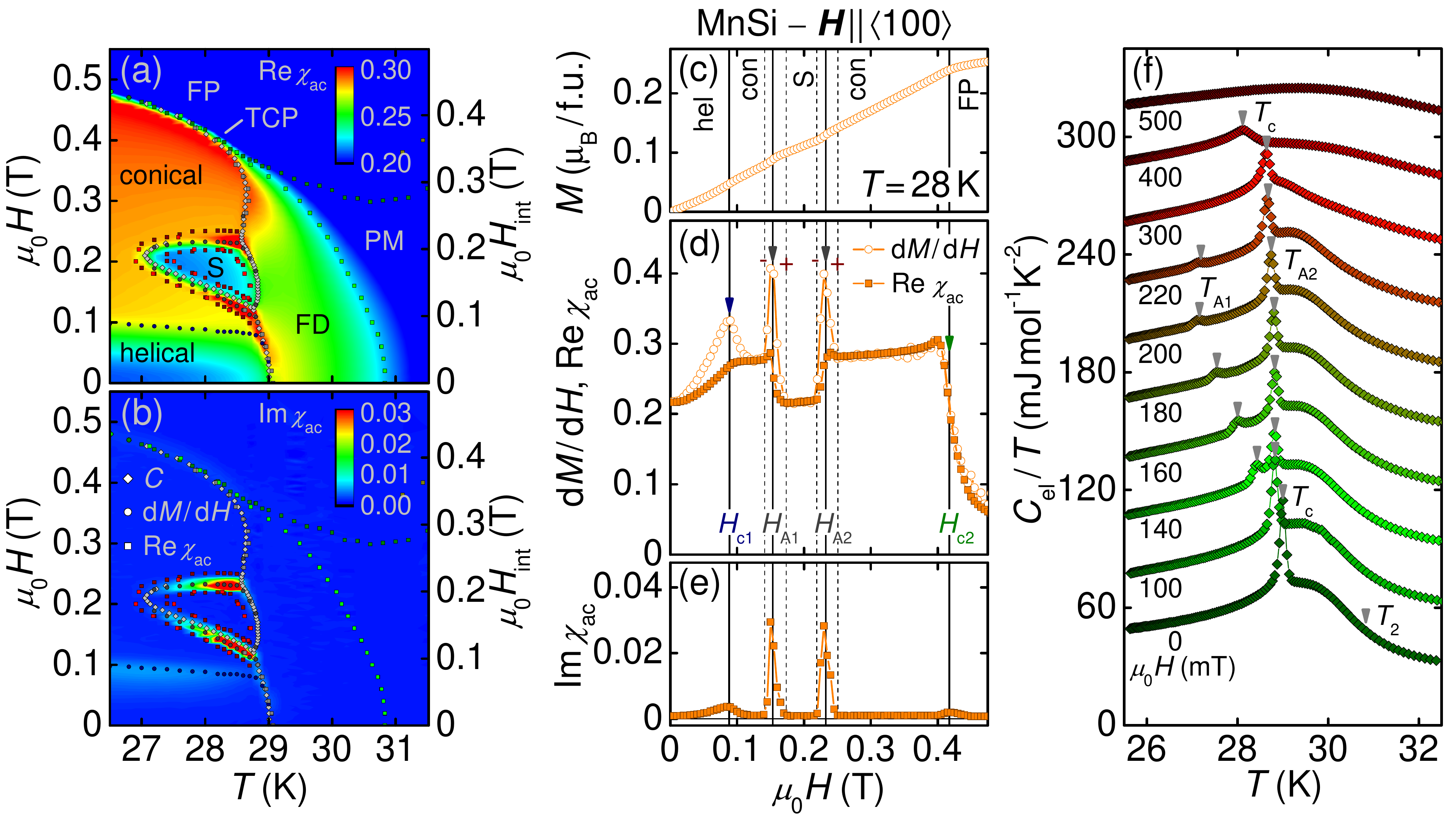}
\caption{Typical magnetization, ac susceptibility, and specific heat data of MnSi. (a)~Color map of the real part of the ac susceptibility. We distinguish the following regimes; helical, conical, skyrmion lattice~(S), fluctuation-disordered~(FD), paramagnetic~(PM), and field-polarized~(FP). A field-induced tricritical point (TCP) is located at the high-field boundary of the FD regime. (b)~Color map of the imaginary part revealing considerable dissipation only between the conical and the skyrmion lattice state. \mbox{(c)--(e)}~Typical data of the magnetization, the susceptibility calculated from the magnetization, $\mathrm{d}M/\mathrm{d}H$, as well as the real and imaginary part of the ac susceptibility as a function of field. Note the definitions of the various transition fields. (f)~Electronic contribution to the specific heat as a function of temperature for several applied magnetic fields. Data has been offset for clarity.}
\label{figure4}
\end{figure}

The different magnetic states in the cubic chiral magnets and the phase transitions between them give rise to distinct signatures in various physical properties. Experimentally, the magnetic ac susceptibility and specific heat are easily accessible for most compounds and allow the determination of a very detailed magnetic phase diagram, based on feature-tracking. This provides the starting point for further studies and motivated us to concentrate on these quantities in the following. 
As an overview, we start with colormaps of the real and imaginary part of the ac susceptibility, $\mathrm{Re}\,\chi_{\mathrm{ac}}$ and $\mathrm{Im}\,\chi_{\mathrm{ac}}$, in Figs.~\ref{figure4}(a) and \ref{figure4}(b), where blue shading corresponds to low and red shading to high values. As an example we show data for a cube-shaped single crystal of MnSi measured at an excitation frequency of 120\,Hz and an excitation amplitude of 0.5\,mT. The field was applied after zero-field cooling along an $\langle100\rangle$ axis, i.e., along the hard direction for the helical propagation vector.

In $\mathrm{Re}\,\chi_{\mathrm{ac}}$ the conical state is characterized by a plateau of high and rather constant susceptibility (orange to red shading). The reduced value at low fields is associated with the helical state. Just below the helimagnetic ordering temperature, $T_{c}$, a plateau of reduced susceptibility in finite fields is characteristic for a single pocket of skyrmion lattice state (light blue shading). Just above $T_{c}$ an area of relatively large susceptibility (green shading) is associated with the fluctuation-disordered (FD) regime that emerges as a consequence of the Brazovskii-type phase transition from paramagnetism to helimagnetism. At high temperatures or high fields, respectively, the system is in a paramagnetic (PM) or field-polarized (FP) state with low susceptibility (blue). A broad maximum observed in temperatures sweeps of $\mathrm{Re}\,\chi_{\mathrm{ac}}$ (not shown) marks the crossover between these two regimes~\cite{2010:Bauer:PhysRevB}. $\mathrm{Im}\,\chi_{\mathrm{ac}}$ only shows contributions at the phase transitions and, in particular, between the skyrmion lattice and conical state. Here, the finite dissipation suggests a regime of phase coexistence where the nucleation process of topologically non-trivial skyrmions within the conical phase and vice versa eventually triggers a first-order transition~\cite{2012:Bauer:PhysRevB, 2013:Milde:Science, 2013:Bauer:PhysRevLett}. In contrast, at the fluctuation-induced first-order transition between the skyrmion lattice and the fluctuation-disordered regime as a function of temperature no significant contribution to $\mathrm{Im}\,\chi_{\mathrm{ac}}$ is observed.

In order to define the different transition fields and temperatures, it is instructive to consider the typical field dependence of the magnetization, $M$, the susceptibility calculated from the measured magnetization, $\mathrm{d}M/\mathrm{d}H$, and the measured ac susceptibility for a temperature just below $T_{c}$ as shown in Figs.~\ref{figure4}(c) through \ref{figure4}(e). Starting at $H=0$, i.e., in the helical state, with increasing field the material undergoes transitions to the conical and the skyrmion lattice state before returning to the conical state and finally reaching the field-polarized state above $H_{c2}$. Below $H_{c2}$ the magnetization increases almost linearly as shown in Fig.~\ref{figure4}(c), where the changes of slope at the different phase transitions are best resolved in the derivative $\mathrm{d}M/\mathrm{d}H$ depicted as open symbols in Fig.~\ref{figure4}(d). Here, we compare the measured ac susceptibility, $\mathrm{Re}\,\chi_{\mathrm{ac}}$, with $\mathrm{d}M/\mathrm{d}H$ which may be viewed as zero-frequency limit of $\mathrm{Re}\,\chi_{\mathrm{ac}}$. 

At the transition between the helical and conical state and in the regimes between the conical and the skyrmion lattice state $\mathrm{d}M/\mathrm{d}H$ shows pronounced maxima that are not tracked by $\mathrm{Re}\,\chi_{\mathrm{ac}}$. In the former case this discrepancy may be attributed to the slow, complex, but well-understood reorientation of macroscopic helical domains. In the latter case the discrepancy is accompanied by strong dissipation, which may be inferred from $\mathrm{Im}\,\chi_{\mathrm{ac}}$ in Fig.~\ref{figure4}(e) and attributed to regimes of phase coexistence between the conical and the skyrmion lattice state as expected for first-order phase transitions. In these regimes both $\mathrm{Re}\,\chi_{\mathrm{ac}}$ and $\mathrm{Im}\,\chi_{\mathrm{ac}}$ show a pronounced dependence on the excitation frequency with a characteristic frequency that increases with temperature~\cite{2012:Bauer:PhysRevB, 2014:Levatic:PhysRevB}.

We define the helical-to-conical transition at $H_{c1}$ as the maximum of $\mathrm{d}M/\mathrm{d}H$ that typically coincides with a point of inflection in $\mathrm{Re}\,\chi_{\mathrm{ac}}$. The low-field and high-field boundary of the skyrmion lattice state, $H_{A1}$ and $H_{A2}$, may be fixed by maxima in $\mathrm{d}M/\mathrm{d}H$. The regimes of phase coexistence between the conical and the skyrmion lattice state are characterized by $\mathrm{d}M/\mathrm{d}H \neq \mathrm{Re}\,\chi_{\mathrm{ac}}$ and $\mathrm{Im}\,\chi_{\mathrm{ac}} \gg 0$, where the corresponding boarders are labeled $H_{A1}^{\pm}$ and $H_{A2}^{\pm}$, respectively. For $H < H_{A1}^{-}$ and $H > H_{A2}^{+}$ the constant susceptibility of the conical phase is observed, while in the skyrmion lattice state for $H_{A1}^{+} < H < H_{A2}^{-}$ the system displays a plateau of lower susceptibility. The second-order transition from the conical to the field-polarized state belonging to the $XY$ universality class is finally indicated by a point of inflection in both $\mathrm{d}M/\mathrm{d}H$ and $\mathrm{Re}\,\chi_{\mathrm{ac}}$. Similar criteria may be used to extract transition temperatures from data recorded as a function of temperature (not shown)~\cite{2012:Bauer:PhysRevB}.

Important related information on the nature of the phase transitions may be extracted from measurements of the specific heat. Using a quasi-adiabatic large heat pulse technique allows to determine transition temperatures with high precision~\cite{2012:Adams:PhysRevLett,2013:Bauer:PhysRevLett}. Fig.~\ref{figure4}(f) shows the electronic contribution to the specific heat, i.e., after subtraction of the phononic contribution, divided by temperature, $C_{\mathrm{el}}/T$, as a function of temperature for different applied field values. In zero field a sharp symmetric peak marks the onset of helimagnetic order at the fluctuation-induced first-order transition at $T_{c}$. The peak resides on top of a broad shoulder that displays for small fields a so-called Vollhardt invariance~\cite{1997:Vollhardt:PhysRevLett} at $T_{2}$, i.e., an invariant crossing point of the specific heat, $\left.\partial C/\partial H\right|_{T_{2}} = 0$, that coincides with a point of inflection in the magnetic susceptibility, $\left.T\partial^{2}M/\partial T^{2}\right|_{T_{2}} \approx \left.TH\partial^{2}\chi/\partial T^{2}\right|_{T_{2}} = 0$~\cite{2010:Bauer:PhysRevB}. At intermediate fields two symmetric peaks, labeled $T_{A1}$ and $T_{A2}$, track the phase boundaries of the skyrmion lattice state indicating two first-order transitions. In larger fields again one anomaly, labeled $T_{c}$, is observed. Increasing the field further causes a change of the shape of the anomaly from that of a slightly broadened symmetric delta peak to the asymmetric lambda anomaly of a second-order transition at a field-induced tricritical point (TCP). This field-induced change from first to second order is expected in the Brazovskii scenario, as the interactions between the chiral paramagnons become quenched under increasing magnetic fields. 

In the magnetic phase diagram, see Figs.~\ref{figure4}(a) and \ref{figure4}(b), the crossovers between the fluctuation-disordered and the paramagnetic regime as well as between the paramagnetic and the field-polarized regime as observed in temperature sweeps of the susceptibility emanate from this TCP. An analysis of the entropy released at the phase transitions (not shown) also corroborates the position of the TCP. It suggests that the skyrmion lattice state possesses an entropy that is larger than the surrounding conical state, consistent with a stabilization by thermal fluctuations~\cite{2013:Bauer:PhysRevLett}. The latter is supported by the detailed shape of the phase boundary between the fluctuation-disordered and the long-range ordered states, where the skyrmion lattice extents to higher temperatures as compared to the conical state.

\begin{figure}
\includegraphics[width=1.0\linewidth]{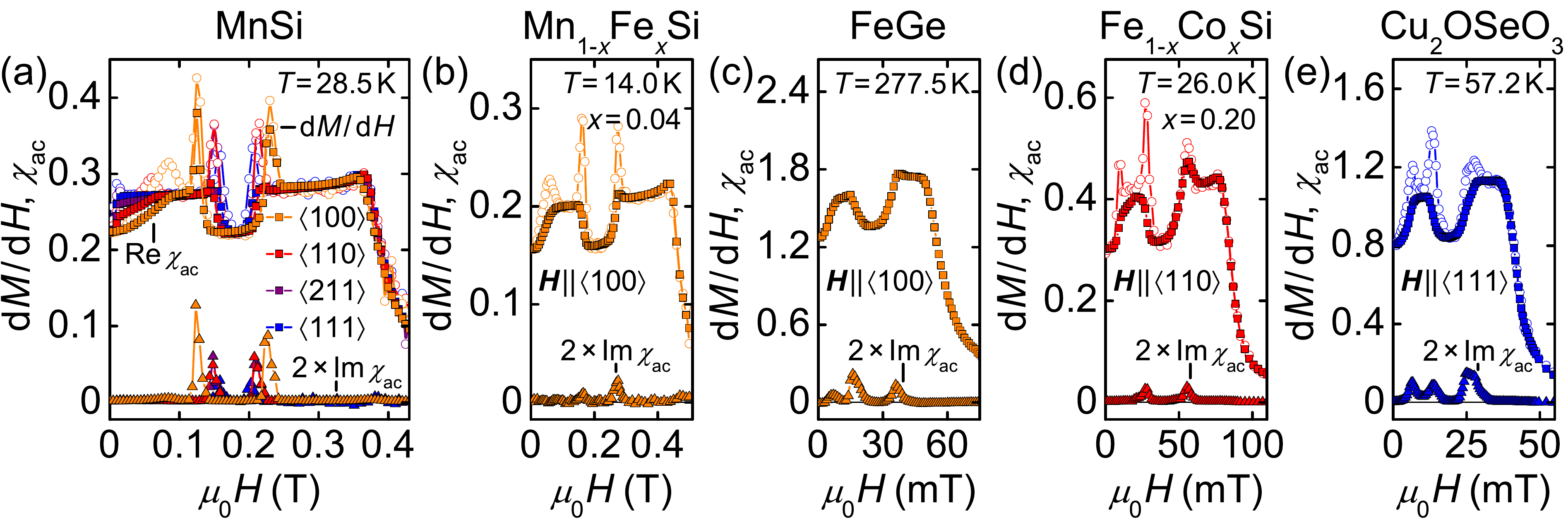}
\caption{Typical field dependence of the susceptibility for a temperature crossing the skyrmion lattice state. (a)~Real and imaginary part of the ac susceptibility as well as susceptibility calculated from the magnetization, $\mathrm{d}M/\mathrm{d}H$, for MnSi and fields along major crystallographic directions. Besides well-understood anisotropies of the helical-to-conical transition and the extent of the skyrmion lattice phase pocket, the magnetic properties of MnSi are essentially isotropic. \mbox{(b)--(e)}~Susceptibility for Mn$_{1-x}$Fe$_{x}$Si ($x = 0.04$), FeGe, Fe$_{1-x}$Co$_{x}$Si ($x = 0.20$), and Cu$_{2}$OSeO$_{3}$. Qualitatively very similar behavior is observed. Data in panel (c) taken from Ref.~\cite{2012:Wilhelm:JPhysCondensMatter}.}
\label{figure5}
\end{figure}

Following the detailed description of data recorded in MnSi with the magnetic field applied along $\langle100\rangle$, we now turn to Fig.~\ref{figure5} illustrating typical susceptibility data as a function of field for different field directions and materials. Fig.~\ref{figure5}(a) shows data of MnSi for field applied along the major crystallographic axes after zero-field cooling measured on two cubes, i.e., with unchanged demagnetization effects. In general, the magnetic behavior is very isotropic. Changing the field direction only influences the weakest energy scale in the system, the cubic anisotropies, and has two well-understood consequences for the magnetic phase diagram. First, the helical-to-conical transition field is smallest for the easy axis of the helical propagation vector $\langle111\rangle$ and largest for the hard axis $\langle100\rangle$. In addition, the transition is only second-order if it is symmetry-breaking and otherwise represents a crossover. Second, the extent of the skyrmion lattice in both temperature and field decreases as the conical state is favored by the cubic anisotropies, i.e., in MnSi it is largest for field along $\langle100\rangle$ and smallest for $\langle111\rangle$. It is important to note, that even for field along the easy axis of the helix the skyrmion lattice is observed for all chiral magnets questioning a stabilization of the skyrmion lattice by cubic anisotropies only. In fact, for doped compounds such as Fe$_{1-x}$Co$_{x}$Si or Mn$_{1-x}$Fe$_{x}$Si the anisotropies are usually less pronounced or even completely suppressed, presumably due to the large amount of chemical disorder present in the system~\cite{2010:Bauer:PhysRevB,2010:Munzer:PhysRevB}, and yet the skyrmion lattice state represents nonetheless a well-defined stable phase.

Figs.~\ref{figure5}(b) through \ref{figure5}(e) show typical susceptibility data for Mn$_{1-x}$Fe$_{x}$Si ($x = 0.04$), FeGe, Fe$_{1-x}$Co$_{x}$Si ($x = 0.20$), and Cu$_{2}$OSeO$_{3}$ highlighting the universal aspects of different cubic chiral magnets. Despite the different temperature, field, length, and moment scales the susceptibilities of the different materials are qualitatively highly reminiscent. Omitting quantitative information on temperature, field, and susceptibility, even an expert would struggle to distinguish data between the different materials.

It is finally essential to account for demagnetization effects, for instance when data recorded on samples with different sample shapes are combined in a single magnetic phase diagram. In general, the internal magnetic field, $\bm{H}_{\mathrm{int}}$, is calculated as $\bm{H}_{\mathrm{int}} = \bm{H}_{\mathrm{ext}} - \mathbf{N}\bm{M}(\bm{H}_{\mathrm{ext}})$ with the externally applied magnetic field $\bm{H}_{\mathrm{ext}}$ and the $3\times3$ demagnetization matrix $\mathbf{N}$ that obeys $\mathrm{tr}\left\{\mathbf{N}\right\} = 1$ in SI units. While a proper treatment of the dipolar interactions in the cubic chiral magnets requires to take several matrix entries into account~\cite{2015:Schwarze:NatureMater}, in most cases consideration of the scalar equation $H_{\mathrm{int}} = H_{\mathrm{ext}} - NM(H_{\mathrm{ext}})$ is sufficient, in which for field along the $z$-direction the matrix entry $N_{zz}$ is referred to as $N$. Note that for the measured ac susceptibility, $\chi_{\mathrm{ac}}^{\mathrm{ext}}$, not only the field scale but also the absolute value of the susceptibility depends on demagnetization effects via the applied excitation field $H_{\mathrm{ac}}^{\mathrm{ext}}$.

From a practical point of view many samples are essentially rectangular prisms for which effective demagnetization factors for fields applied along the edges may be calculated following Ref.~\cite{1998:Aharoni:JApplPhys}. In addition, in the cubic chiral magnets the susceptibility assumes essentially a constant value in the conical phase. Using the measured value, $\chi_{\mathrm{con}}^{\mathrm{ext}}$, as a first approximation for the entire helimagnetically ordered part of the magnetic phase diagram, i.e., for $T<T_{c}$ and $H<H_{c2}$, the magnetization may be expressed as $M(H_{\mathrm{ext}}) = \chi_{\mathrm{con}}^{\mathrm{ext}} H_{\mathrm{ext}} = \chi_{\mathrm{con}}^{\mathrm{int}} H_{\mathrm{int}}$. Hence, the internal and the externally applied magnetic fields are related by:
\begin{equation}
H_{\mathrm{int}} = H_{\mathrm{ext}}\left(1-N\chi_{\mathrm{con}}^{\mathrm{ext}}\right) = \frac{H_{\mathrm{ext}}}{1 + N\chi_{\mathrm{con}}^{\mathrm{int}}}
\end{equation}
We note that the internal value of the constant susceptibility of the conical state, $\chi_{\mathrm{con}}^{\mathrm{int}}$, is an important dimensionless measure for the effective strength of dipolar interactions in the chiral magnets~\cite{2015:Schwarze:NatureMater}. If the magnetic properties and a second quantity, e.g., electrical resistivity, are determined on samples with differing demagnetization factors, $N_{1}$ and $N_{2}$, the formula to calculate the internal field of the second sample may be written as:
\begin{equation}
H_{\mathrm{int,2}} = H_{\mathrm{ext,2}}\left(1-N_{2}\,\frac{\chi_{\mathrm{con},1}^{\mathrm{ext}}}{1-\chi_{\mathrm{con},1}^{\mathrm{ext}}(N_{1}-N_{2})}\right)
\end{equation}
In the field-polarized state above $H_{c2}$ one may, again in first approximation, assume the magnetization as saturated and thus $M(H_{\mathrm{ext}}) = \chi_{\mathrm{con}}^{\mathrm{ext}}H_{c2}^{\mathrm{ext}} = \chi_{\mathrm{con}}^{\mathrm{int}}H_{c2}^{\mathrm{int}}$ leading to a constant offset, $H_{\mathrm{int}} = H_{\mathrm{ext}} - N\chi_{\mathrm{con}}^{\mathrm{ext}}H_{c2}^{\mathrm{ext}}$.

Despite the crude nature of the approximation given above, this treatment proves to be sufficient to account for the most prominent effects of demagnetizing fields in the chiral magnets such as the shift of transition fields. Additionally, a smearing of phase transitions and very broad regimes of phase coexistence between the conical and the skyrmion lattice state may be observed in samples with large and, in particular, inhomogeneous demagnetization effects~\cite{2012:Bauer:PhysRevB}. Such unfavorable sample shapes are, for instance, thin platelets with their short edge along the field or irregular shapes in general. Materials with a large absolute value of the susceptibility intensify the issue.

\subsection{Magnetic phase diagrams for different materials}

\begin{figure}
\includegraphics[width=1.0\linewidth]{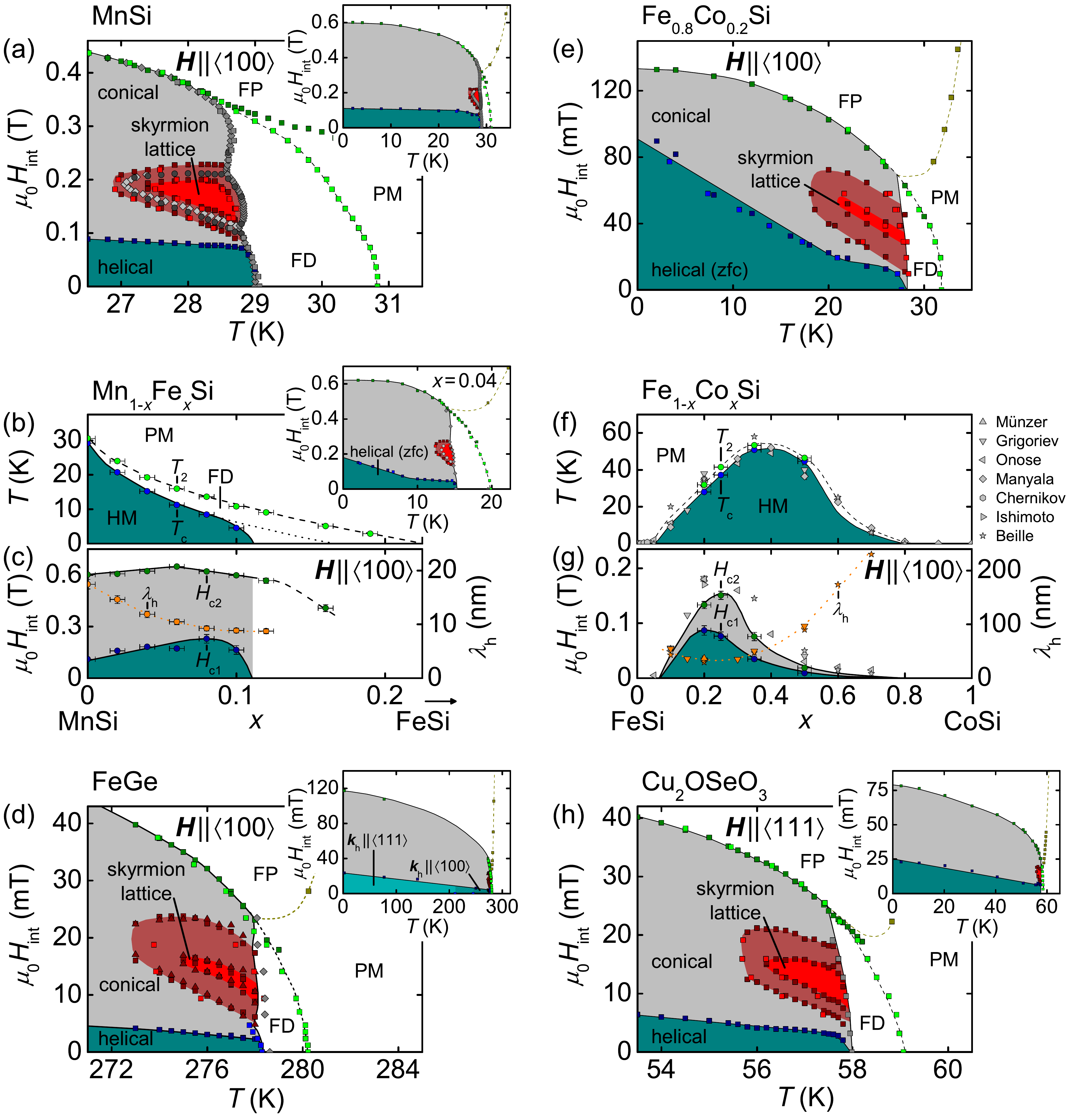}
\caption{Magnetic phase diagrams of selected cubic chiral magnets. (a)~MnSi. \mbox{(b),(c)}~Mn$_{1-x}$Fe$_{x}$Si. Substitutional doping of MnSi with Fe leads to a suppression of the ordering temperature and a decrease of the helix wavelength, $\lambda_{\mathrm{h}}$. The magnetic phase diagram, as shown in the inset for $x = 0.04$, stays qualitatively similar for $x \leq 0.10$. (d)~FeGe. Susceptibility data from Refs.~\cite{2011:Wilhelm:PhysRevLett, 2012:Wilhelm:JPhysCondensMatter}, specific heat data from Ref.~\cite{2013:Cevey:PhysStatusSolidiB}, and further information from Refs.~\cite{1970:Lundgren:PhysScr,1989:Lebech:JPhysCondensMatter,1993:Lebech:Book} were analyzed in the same manner as for all other compounds. (e)~Fe$_{0.8}$Co$_{0.2}$Si. \mbox{(f),(g)}~Fe$_{1-x}$Co$_{x}$Si. As a function of cobalt content $x$ the characteristic temperature, field, and length scales may be varied over a large range. Values are taken from Refs.~\cite{1983:Beille:SolidStateCommun, 1995:Ishimoto:PhysicaB, 1997:Chernikov:PhysRevB, 2000:Manyala:Nature, 2005:Onose:PhysRevB, 2007:Grigoriev:PhysRevB, 2010:Munzer:PhysRevB}. (h)~Cu$_{2}$OSeO$_{3}$. In contrast to the other materials, this local-moment insulator displays substantial magnetoelectric coupling. Still, the magnetic phase diagram is unchanged.}
\label{figure6}
\end{figure}

Using the definitions for the transition fields and temperature given in the previous subsection on susceptibility and specific heat data we have compiled magnetic and compositional phase diagrams of various cubic chiral magnets as shown in Fig.~\ref{figure6}. Data extracted from measurements of the derivative of the magnetization, the ac susceptibility, and the specific heat are shown as circles, squares, and diamonds, respectively. Light and dark colors represent data from temperature and field sweeps, respectively. Magnetic fields were applied after zero-field cooling. All field values are given on internal field scales, i.e., after correcting for demagnetization effects. In general the magnetic phase diagrams of the cubic chiral magnets are qualitatively extremely similar. We distinguish the following six regimes; helical, conical, skyrmion lattice~(S), fluctuation-disordered~(FD), paramagnetic~(PM), and field-polarized~(FP). In addition, we mark the regime of phase coexistence between the conical and the skyrmion lattice state by a faint red shading. Solid and dashed lines indicate phase transitions and crossovers, respectively, while dotted lines represent guides to the eye.

Fig.~\ref{figure6}(a) reproduces the magnetic phase diagram of MnSi for field along $\langle100\rangle$, i.e., the hard axis for the helical propagation vector, as discussed in the previous subsection. The inset shows the phase diagram across the entire parameter range of long-range helimagnetic order. We note that the helix wavelength, $\lambda_{\mathrm{h}}$, in MnSi increases from ${\sim}165\,\textrm{\AA}$ at $T_{c}$ to ${\sim}180\,\textrm{\AA}$ at lowest temperatures~\cite{1976:Ishikawa:SolidStateCommun, 2006:Grigoriev:PhysRevB, 2013:Janoschek:PhysRevB}.

Substitutional doping of iron at the manganese sites of MnSi results in a reduction of the helimagnetic ordering temperature while the critical field values in the zero-temperature limit change only weakly, cf.\ Figs.~\ref{figure6}(b) and \ref{figure6}(c). The magnetic phase diagram is qualitatively very similar to pure MnSi for $x \leq 0.10$ as shown in the inset of Fig.~\ref{figure6}(b) for Mn$_{1-x}$Fe$_{x}$Si with $x = 0.04$. The most notable difference concerns the helical state, which forms in Mn$_{1-x}$Fe$_{x}$Si only properly after zero-field cooling. In addition, $H_{c1}$ becomes essentially isotropic and increases with decreasing temperature. These effects, however, may be related to the increased amount of disorder present in the system. The helix wavelength and hence also the skyrmion lattice constant decreases by up to a factor of roughly 2 resulting in an increase of the skyrmion density by a factor of 4~\cite{2010:Pfleiderer:JPhysCondensMatter, 2014:Franz:PhysRevLett}. 

The complex quantum critical behavior that emerges at high iron concentrations, where static magnetic order is fully suppressed, is the topic of ongoing research~\cite{2010:Bauer:PhysRevB, 2011:Grigoriev:PhysRevB}. Doping with iron, cobalt, and nickel leads to an essentially identical modification of the magnetic behavior if scaled by the number valance electrons per formula unit~\cite{2010:Bauer:PhysRevB, 2010:Teyssier:PhysRevB}. Doping with chromium, i.e., reducing the number of valance electrons, leads to a suppression of $T_{c}$ comparable to iron doping~\cite{1998:Achu:JMagnMagnMater}. This behavior is consistent with the notion that the main effects of chemical doping are due to a rigid shift the Fermi level, as recently inferred from a combined study of ab initio calculations and the electric transport properties in Mn$_{1-x}$Fe$_{x}$Si~\ref{2014:Franz:PhysRevLett}.

We now turn to FeGe which is rather similar to MnSi, however, with a transition temperature near room temperature and $\lambda_{\mathrm{h}} = 700\,\textrm{\AA}$. Around 230\,K the easy direction of the helical pitch changes from $\langle100\rangle$ at high temperatures to $\langle111\rangle$ at low temperatures, where a large thermal hysteresis of ${\sim}35$\,K is observed~\cite{1989:Lebech:JPhysCondensMatter}. Recent publications~\cite{2011:Wilhelm:PhysRevLett, 2012:Wilhelm:JPhysCondensMatter, 2013:Cevey:PhysStatusSolidiB, 2013:Moskvin:PhysRevLett} claimed putative experimental evidence for the formation of a very complex magnetic phase diagram with multiple pockets and precursor phenomena around the skyrmion lattice state. The authors concluded that these findings prove that the skyrmion lattice is in fact not stabilized by thermal fluctuations but by a combination of uniaxial anisotropies and a softened modulus of the magnetization. 

In stark contrast, applying accurately the same definitions given in the previous subsection to the data published in Refs.~\cite{2011:Wilhelm:PhysRevLett, 2012:Wilhelm:JPhysCondensMatter, 2013:Cevey:PhysStatusSolidiB} provides the phase diagram shown in Fig.~\ref{figure6}(d). This phase diagram strongly resembles that of the other cubic chiral magnets. The broad regimes of phase coexistence may be attributed to large demagnetization effects as a consequence of the relatively large absolute value of the susceptibility in FeGe and the shape of the samples used in these studies; we extract $\chi_{\mathrm{con}}^{\mathrm{ext}} = 1.6$ and $N \approx 0.33$ from Ref.~\cite{2012:Wilhelm:JPhysCondensMatter} yielding $\chi_{\mathrm{con}}^{\mathrm{int}} = 3.4$. Most importantly, however, we observe no signatures of additional phase pockets or mesophases. We finally note that a temperature discrepancy of the maximum in the specific heat in Refs.~\cite{2012:Wilhelm:JPhysCondensMatter, 2013:Cevey:PhysStatusSolidiB} indicates that care has to be taken when combining data from different samples or measurement setups.

Figs.~\ref{figure6}(e) through \ref{figure6}(g) are dedicated to Fe$_{1-x}$Co$_{x}$Si, a pseudo-binary $B20$ system that displays helimagnetism in a large composition range, $0.05 \lesssim x \lesssim 0.8$~\cite{1983:Beille:SolidStateCommun, 1987:Motokawa:JMagnMagnMater, 2000:Manyala:Nature}, albeit the parent compounds FeSi and CoSi are a paramagnetic insulator~\cite{1967:Jaccarino:PhysRev} and a diamagnetic metal~\cite{1966:Shinoda:JPhysSocJpn}, respectively. Starting from the strongly correlated insulator FeSi~\cite{2008:Arita:PhysRevB}, an insulator-to-metal transition takes place around $x \approx 0.02$~\cite{1997:Chernikov:PhysRevB}. However, due to the comparatively high absolute value of the electrical resistivity and an upturn at low temperatures helimagnetic Fe$_{1-x}$Co$_{x}$Si is typically referred to as a strongly doped semiconductor~\cite{1983:Beille:SolidStateCommun, 2004:Manyala:NatureMater, 2005:Onose:PhysRevB}.

Compared to the stoichiometric helimagnets, Fe$_{1-x}$Co$_{x}$Si offers the opportunity to vary the characteristic parameters of the helimagnetism over a wide range by compositional tuning while the magnetic phase diagrams stays that of a typical cubic chiral magnet, cf.\ Fig.~\ref{figure6}(e). As summarized in Figs.~\ref{figure6}(f) and \ref{figure6}(g), the helimagnetic transition temperature reaches up to ${\sim}50$\,K, the critical fields assume values up to ${\sim}150$\,mT, and the helix wavelength ranges from about $300\,\textrm{\AA}$ to more than $2000\,\textrm{\AA}$. As for doped MnSi, a proper helical state is observed only after zero-field cooling. Fe$_{1-x}$Co$_{x}$Si displays easy $\langle100\rangle$ axes that, especially for larger cobalt contents, are less pronounced than for other cubic chiral helimagnets~\cite{2007:Grigoriev:PhysRevB}. For $x = 0.20$ a helical pitch along $\langle110\rangle$ was identified in Ref.~\cite{2010:Munzer:PhysRevB}. The latter study also revealed the existence of a skyrmion lattice in Fe$_{1-x}$Co$_{x}$Si that is sensitive to the field and temperature history. While the reversible pocket of skyrmion lattice state is comparable to other systems, field cooling may result in a metastable extension down to lowest temperatures allowing for conceptionally new types of experiments~\cite{2013:Milde:Science}. A similar behavior was later also discovered in low-quality MnSi samples under applied pressure~\cite{2013:Ritz:PhysRevB}. Moreover, depending on the field direction, two Skyrmion lattice domains with different in-plane orientations were observed leading to a twelvefold small-angle scattering pattern~\cite{2010:Adams:JPhysConfSer}.

Fig.~\ref{figure6}(h) finally shows the magnetic phase diagram of copper-oxo-selenite, Cu$_{2}$OSeO$_{3}$. The crystalline structure of this compound is more complex than that of the $B20$ transition metal systems, but also belongs to space group $P2_{1}3$~\cite{1976:Meunier:JApplCryst}. Magnetically, on the strongest scale Cu$_{2}$OSeO$_{3}$ shows local-moment ferrimagnetic order of the spin-$\frac{1}{2}$ Cu$^{2+}$ ions. Here, the ferromagnetically aligned moments on the Cu$^{\textrm{I}}$ sites couple antiferromagnetically to the ions on the Cu$^{\textrm{II}}$ sites leading to a 3:1 ratio~\cite{1977:Kohn:JPhysSocJpn} with exchange constants $J_{\mathrm{FM}} = -50$\,K and $J_{\mathrm{AFM}} = 65$\,K~\cite{2010:Belesi:PhysRevB}. No breaking of the ferrimagnetic coupling is observed up to 55\,T~\cite{2011:Huang:PhysRevB}. The ferrimagnetism is superimposed by a long-wavelength helical modulation based on the Dzyaloshinskii-Moriya interaction with $\lambda_{\mathrm{h}} = 620\,\textrm{\AA}$~\cite{2012:Seki:Science}. The resulting magnetic phase diagram is highly reminiscent of the helimagnetic $B20$ compounds with an easy $\langle100\rangle$ for the helical propagation vector and a delicate pinning within the skyrmion lattice state~\cite{2012:Adams:PhysRevLett, 2012:Seki:PhysRevB}. A study using resonant soft x-ray diffraction further suggested that the Cu$^{\textrm{I}}$ and Cu$^{\textrm{II}}$ sites may form individual but coupled skyrmion lattices that are rotated by a few degree with respect to each other giving rise to a moir\'{e} pattern~\cite{2014:Langner:PhysRevLett}. More recent work reveals, however, that this conjecture may be wrong.

Cu$_{2}$OSeO$_{3}$, albeit being a non-polar insulator, possesses a magnetically induced electrical polarization in finite fields and, in particular, within the skyrmion lattice state~\cite{2012:Seki:PhysRevB2}. The polarization resulting from this magnetoelectric coupling may be described in a $d$-$p$ hybridization model~\cite{2006:Jia:PhysRevB}, where the covalency between copper $d$ and oxygen $p$ orbitals is modulated according to the local magnetization direction via the spin-orbit interaction leading to a local electric dipole along the bond direction~\cite{2012:Seki:PhysRevB2}. Hence, though Cu$_{2}$OSeO$_{3}$ is actually a (heli-)ferrimagnetic magnetoelectric, it is often erroneously referred to as a multiferroic. The origin of this notion may be seen in the hitherto unique opportunity to manipulate a topologically non-trivial entity of magnetoelectric nature using various external control parameters, see for example Refs.~\cite{2012:White:JPhysCondensMatter, 2013:Mochizuki:PhysRevB, 2013:Okamura:NatCommun, 2014:Mochizuki:NatureMater, 2015:Okamura:PhysRevLett}.


\section{Emergent electrodynamics}
\label{EmergentED}

\begin{figure}
\includegraphics[width=1.0\linewidth]{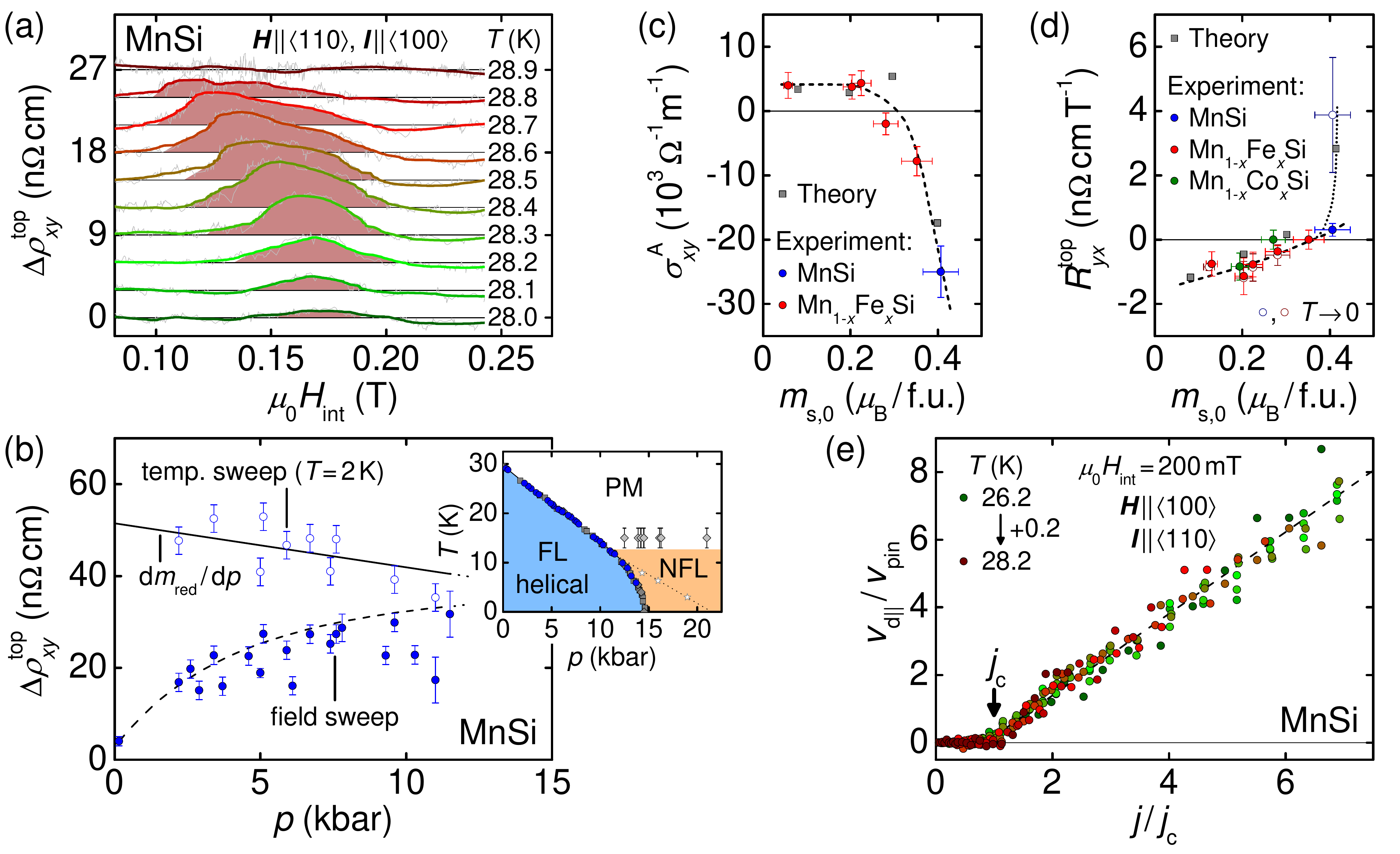}
\caption{Examples of the efficient coupling of spin currents to the skyrmion lattice. (a)~Topological Hall contribution, $\Delta\rho_{xy}^{\mathrm{top}}$, in MnSi as a function of field~\cite{2009:Neubauer:PhysRevLett}. (b)~Topological Hall contribution, $\Delta\rho_{xy}^{\mathrm{top}}$, as a function of hydrostatic pressure in MnSi~\cite{2013:Ritz:PhysRevB}. The intrinsic size (open symbols) may only be observed after field-cooling down to the lowest temperatures. The inset shows the pressure-temperature phase diagram of MnSi highlighting the extended regime of non-Fermi liquid (NFL) behavior~\cite{2007:Pfleiderer:Science,2013:Ritz:Nature}. \mbox{(c),(d)}~Anomalous Hall conductivity, $\sigma_{xy}^{\mathrm{A}}$, and topological Hall constant, $R_{yx}^{\mathrm{top}}$, as a function of the magnetic moment as varied, e.g., by iron or cobalt doping. First-principle calculations and experimental data are in excellent agreement~\cite{2014:Franz:PhysRevLett}. (e) Drift velocity of the skyrmion lattice, $v_{\mathrm{d}\parallel}$, as a function of current density, $j$. Ultra-low current densities in the order of $j_{\mathrm{c}} \sim 10^{6}\,\mathrm{A}/\mathrm{m}^{2}$ unpin the skyrmion lattice~\cite{2010:Jonietz:Science, 2012:Schulz:NaturePhys}.}
\label{figure7}
\end{figure}

A particularly exciting consequence of the non-trivial topology of the skyrmions concerns their coupling to spin currents. In the following we focus on the consequences in metallic compounds and we refer to the chapter by Markus Garst for a more detailed account. The spin structure of the skyrmion, as seen from the point of view of an electron traversing it, gives rise to real-space Berry phases which may be expressed as emergent magnetic and electric fields, $B^{e}_{i} = \frac{\hbar}{2}\epsilon_{ijk}\hat{\psi}\cdot\partial_{j}\hat{\psi}\times\partial_{k}\hat{\psi}$ and $E^{e}_{i} = \hbar\hat{\psi}\cdot\partial_{i}\hat{\psi}\times\partial_{t}\hat{\psi}$, respectively, with $\partial_{i} = \partial / \partial r_{i}$ and $\partial_{t} = \partial / \partial t$~\cite{2012:Schulz:NaturePhys}. As a consequence an additional topological contribution to the Hall effect may be observed in the skyrmion lattice state as illustrated in Fig.~\ref{figure7}(a)~\cite{2009:Neubauer:PhysRevLett}. 

Using the charge carrier spin polarization $P$ and assuming the absence of spin-flip scattering, while non-spin-flip scattering is captured by the normal Hall constant $R_{0}$, the topological Hall contribution may be estimated as $\Delta\rho_{xy}^{\mathrm{top}} = PR_{0}B_{\mathrm{eff}}$. The effective emergent field, $B_{\mathrm{eff}}$, is topologically quantized in the sense that it is given by the product of the emergent flux quantum that each skyrmion supports, $\phi_{0} = h/e$, and the skyrmion density $\phi$. Thus, the sign of the topological Hall contribution allows to distinguish, in principle, between skyrmions ($\mathit{\Phi} = +1$) and anti-skyrmions ($\mathit{\Phi} = -1$), such as in MnSi, provided the normal Hall constant $R_{0}$ is sufficient to express the details of the band structure~\cite{2009:Neubauer:PhysRevLett}.

In real materials the electronic structure at the Fermi surface may contribute in different ways and the spin polarization as well as the skyrmion lattice constant may change as a function of temperature or field. In addition, processes such as spin-flip scattering may cause a reduction compared to the intrinsic value of $\Delta\rho_{xy}^{\mathrm{top}}$. For instance, in MnSi the topological Hall contribution in the skyrmion lattice is of the order of $4\,\textrm{n}\Omega\,\textrm{cm}$ whereas an intrinsic topological Hall signal of the order of $50\,\textrm{n}\Omega\,\textrm{cm}$ is expected for its emergent field of $B_{\mathrm{eff}} = -13$\,T~\cite{2013:Ritz:PhysRevB}. Field-cooling the skyrmion lattice down to low temperatures allows to reduce the finite temperature effects, as it is for instance possible in high-pressure studies of MnSi. Here, as shown in Fig.~\ref{figure7}(b), the intrinsic value of $\Delta\rho_{xy}^{\mathrm{top}}$ could be inferred which in turn scales with the charge carrier spin polarization that follows the reduced magnetic moment $m_{\mathrm{red}} = m(p) / m(p = 0)$.

At higher pressures where static helimagnetic order in MnSi is fully suppressed at $p_{c} = 14.6$\,kbar more complex behavior has been observed, cf.\ inset of Fig.~\ref{figure7}(b)~\cite{1997:Pfleiderer:PhysRevB, 1997:Thessieu:JPhysCondensMatter, 2007:Pfleiderer:Science}. In particular, in a large pressure and field range the standard description of the metallic state, namely the Fermi liquid (FL) theory, breaks down~\cite{2001:Pfleiderer:Nature, 2003:Doiron-Leyraud:Nature}. In addition, neutron scattering reveals so-called partial magnetic order in a pocket above $p_{c}$~\cite{2004:Pfleiderer:Nature}. In combination with the lack of observable relaxation in muon data~\cite{2007:Uemura:NaturePhys}, it has been concluded that the spin correlations of the partial order are dynamic on a timescale between $10^{-10}$\,s and $10^{-11}$\,s. Finally, a clear connection between the topological Hall effect in the skyrmion lattice at ambient pressure and a large topological Hall signal that coincides with the non-Fermi liquid (NFL) regime above $p_{c}$ empirically suggests that spin correlations with non-trivial topological character drive the breakdown of Fermi liquid theory~\cite{2013:Ritz:Nature}.


Calculations based on density functional theory allow to determine the sign and the magnitude of the anomalous and the topological Hall effect and, in particular, how they evolve when the spin polarization changes. Experimentally, the latter may be realized by substitutional doping of Fe or Co into MnSi, where excellent agreement between theory and experiment has been observed as shown in Figs.~\ref{figure7}(c) and \ref{figure7}(d)~\cite{2014:Franz:PhysRevLett}. These results provide the quantitative microscopic underpinning that, while the anomalous Hall effect is due to the reciprocal-space Berry curvature~\cite{2010:Nagaosa:RevModPhys}, the topological Hall effect originates in real-space Berry phases. As a theoretical prediction that awaits further confirmation even  contributions arising from mixed phase-space Berry phases have been proposed~\cite{2013:Ritz:PhysRevB, 2013:Freimuth:PhysRevB}.

The efficient coupling of spin currents to the magnetic structure, together with the exceptional long-range order of the skyrmion lattice~\cite{2011:Adams:PhysRevLett} and the resulting very weak collective pinning to defects, causes a sizeable response of the magnetic textures at ultra-low current densities. Above an exceptionally low threshold current density of the order of $j_{c} \sim 10^{6}$\,A/m$^{2}$ the skyrmion lattice unpins and begins to drift~\cite{2010:Jonietz:Science,2012:Yu:NatCommun}. Numerical simulations revealed that the skyrmion motion exhibits a universal current-velocity relation that is (on the scale of the study) unaffected by impurities and non-adiabatic effects~\cite{2013:Iwasaki:NatCommun}. Flexible shape-deformations of individual skyrmions and the skyrmion lattice permit to avoid pinning centers. 

Theoretically, the spin transfer torques in the cubic chiral magnets may be accounted for in the framework of a Landau-Lifshitz-Gilbert equation using the Thiele approach~\cite{1973:Thiele:PhysRevLett, 2011:Everschor:PhysRevB}. Here, a Magnus force perpendicular to the current direction and a dissipative drag force along it are balanced by pinning forces, e.g., due to defects. The Magnus force represents the effective Lorentz force arising from the emergent magnetic field $\bm{B}^{e}$ and leads to a certain angle between the current direction and the drift direction of the skyrmion lattice. According to Faraday's law of induction, a moving skyrmion, which supports exactly one quantum of emergent magnetic flux, may then induce an emergent electric field $\bm{E}^{e}$ that inherits the topological quantization~\cite{2011:Zang:PhysRevLett}. These electric fields have been observed directly~\cite{2012:Schulz:NaturePhys}. A scaling plot as depicted in Fig.~\ref{figure7}(e) reveals a universal relation between the current density, $j$, and the drift velocity of the skyrmion lattice, $v_{\mathrm{d}\parallel}$, where typical pinning velocities are of the order of 0.1\,mm/s, i.e., the drift velocity of conduction electrons.


\section{Conclusions and outlook}
\label{Conclusions}

\begin{figure}
\includegraphics[width=1.0\linewidth]{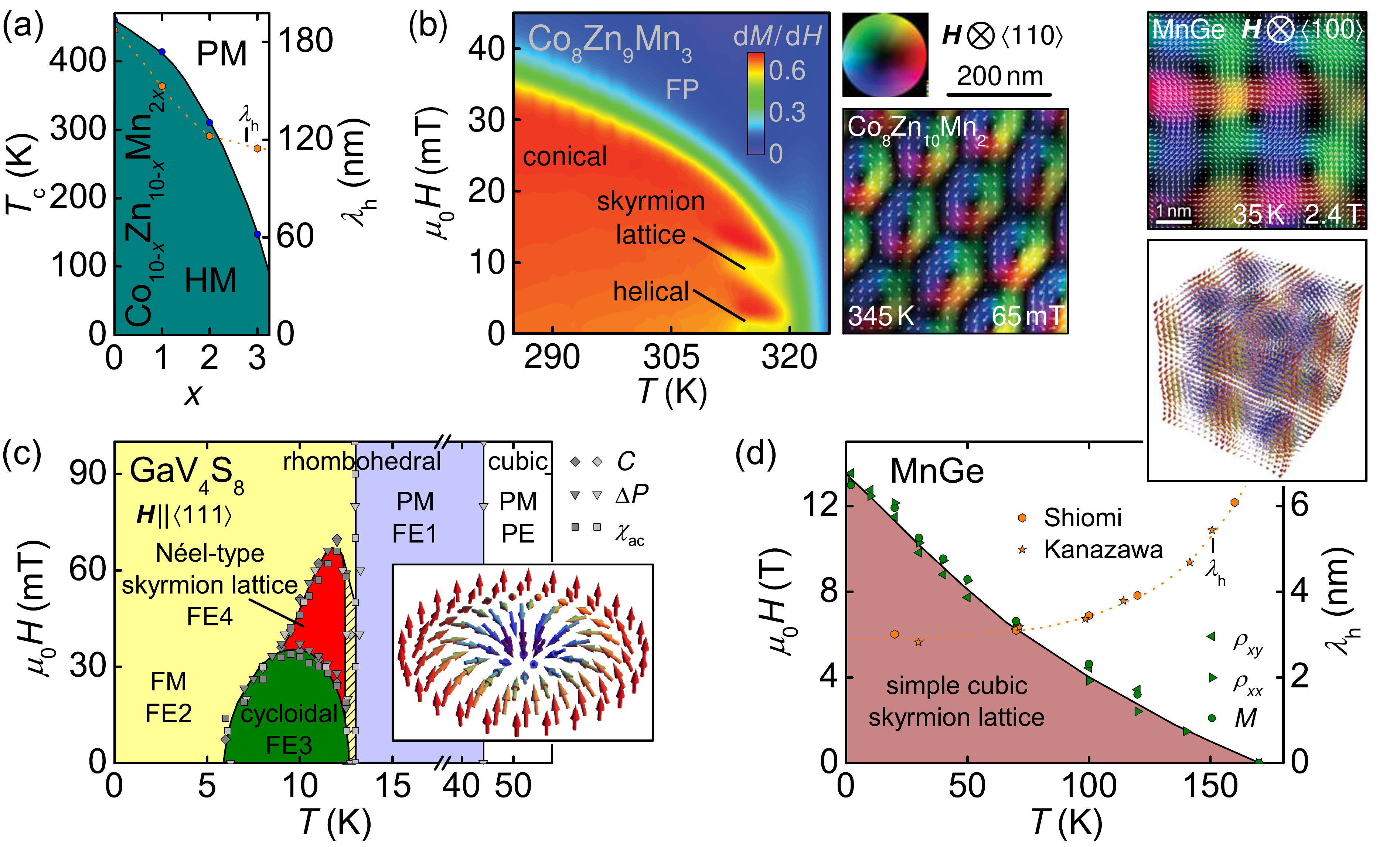}
\caption{Topologically non-trivial spin structures in further bulk materials. (a)~Part of the compositional phase diagram of the system Co$_{10-x}$Zn$_{10-y}$Mn$_{x+y}$. Long-wavelength helimagnetic order with transition temperatures exceeding room temperature has been reported~\cite{2015:Tokunaga:arXiv}. (b)~Colormap of the susceptibility of Co$_{8}$Zn$_{9}$Mn$_{3}$ revealing a skyrmion lattice state and corresponding real-space spin structure in Co$_{8}$Zn$_{10}$Mn$_{2}$ as obtained by LF-TEM~\cite{2015:Tokunaga:arXiv}. The behavior is highly reminiscent of the cubic chiral magnets. (c)~Magnetic phase diagram of GaV$_{4}$S$_{8}$ exhibiting a N\'{e}el-type skyrmion lattice and various types of ferroelectric order~\cite{2015:Kezsmarki:arXiv, 2015:Ruff:arXiv}. (d)~Magnetic phase diagram of MnGe~\cite{2011:Kanazawa:PhysRevLett, 2012:Kanazawa:PhysRevB, 2013:Shiomi:PhysRevB} giving rise to a simple cubic lattice of spin whirls as recently observed by LF-TEM~\cite{2015:Tanigaki:arXiv}.}
\label{figure8}
\end{figure}

Taken together, cubic chiral magnets with non-centrosymmetric space group $P2_{1}3$ represent a class of materials that share a universal magnetic phase diagram. The skyrmion lattice state occupies a single phase pocket and the entire magnetic phase diagram is well accounted for by a Ginzburg-Landau approach including the effects of thermal fluctuations. Depending on the specific material, key parameters such as the transition temperatures, critical fields, or the helix wavelength may be varied by two orders of magnitude. With compounds ranging from pure metals to magnetoelectric insulators, this material class provides well-understood model systems for experiments, theory, and simulations. In recent studies, for instance, aspects were addressed such as the topological unwinding at the transition to conventional helimagnetic order~\cite{2013:Milde:Science} or the collective excitations of the different spin structures~\cite{2010:Janoschek:PhysRevB, 2012:Koralek:PhysRevLett, 2012:Mochizuki:PhysRevLett, 2012:Onose:PhysRevLett}.

Current research activities on topologically non-trivial spin states, however, are not restricted to cubic chiral magnets. In thin films or monolayers, where the inversion symmetry is broken by the surface, skyrmions may be stabilized by the Dzyaloshinskii-Moriya interaction as combined with four-spin exchange interactions~\cite{2011:Heinze:NaturePhys, 2013:Romming:Science}. Another route towards skyrmionic textures may be long-range magnetodipolar interactions~\cite{2012:Yu:PNatlAcadSciUSA}. In such systems, it was already demonstrated that skyrmions may be created and annihilated individually using spin-polarized currents of a scanning tunneling microscope~\cite{2013:Romming:Science} or laser pulses~\cite{2013:Finazzi:PhysRevLett}. The creation, manipulation, and the dynamics of skyrmions in thin films, nanowires, and patterned nanostructures offer great potential for future applications, see for instance Refs.~\cite{2013:Sampaio:NatureNano, 2013:Yu:NanoLett, 2013:Iwasaki:NatureNano, 2013:Lin:PhysRevLett, 2013:Nagaosa:NatureNano, 2014:Iwasaki:NanoLett, 2014:Lin:PhysRevLett, 2015:Muller:PhysRevB, 2015:Zhang:SciRep}. The efficient gyromagnetic coupling, the topological stability, and the small size of the skyrmions promise devices for ultra-dense information storage and spintronics~\cite{2013:Fert:NatureNano}, while their unique collective excitations may be exploited for the design of conceptually new microwave devices~\cite{2013:Okamura:NatCommun, 2015:Schwarze:NatureMater, 2015:Ogawa:SciRep}.

In parallel, topologically non-trivial spin states have been identified in a rapidly growing number of bulk compounds suggesting that these complex magnetic structures may be in fact rather common. In Fig.~\ref{figure8} we summarize three recent examples. The first material, CoZn, crystallizes in the cubic space group $P4_{1}32$ or $P4_{3}32$, depending on the handedness, and orders magnetically well above room temperature~\cite{2013:Xie:InorgChem}. Doping manganese into the system, see Fig.~\ref{figure8}(a), reduces the transition temperature. Fig.~\ref{figure8}(b) shows the magnetic phase diagram of Co$_{8}$Zn$_{9}$Mn$_{3}$ extracted from the magnetic susceptibility. It is highly reminiscent to that of the cubic chiral magnets including a pocket of skyrmion lattice state as identified by LF-TEM and SANS measurements~\cite{2015:Tokunaga:arXiv}. Hence, the material system Co$_{10-x}$Zn$_{10-y}$Mn$_{x+y}$ is not only the first bulk compound with a space group other than $P2_{1}3$ that exhibits a skyrmion lattice state, but also the first compound stabilizing skyrmions above room temperature. 

Another important example is shown in Fig.~\ref{figure8}(c), which depicts the magnetic phase diagram of the lacunar spinel GaV$_{4}$S$_{8}$. This system crystallizes in the cubic space group $F\bar{4}3m$ at room temperature. At $T_{\mathrm{JT}} = 44$\,K GaV$_{4}$S$_{8}$ shows a structural phase transition~\cite{2010:Pocha:ChemMater} into the rhombohedral space group $R3m$ driven by Jahn-Teller orbital order and accompanied by an onset of ferroelectricity (FE). The structural transition creates a multi-domain state with submicron-thick sheets of the four different rhombohedral domains~\cite{2015:Kezsmarki:arXiv}. Below $T_{C} = 13$\,K magnetic order sets in~\cite{2008:Yadav:PhysicaB} and as a function of temperature and field a rich magnetic phase diagram unfolds with various magnetically ordered states of multiferroic nature~\cite{2015:Ruff:arXiv}. This phase diagram hosts a pocket of ferroelectric spin vortices forming a hexagonal skyrmion lattice as identified by means of force microscopy and SANS~\cite{2015:Kezsmarki:arXiv}. However, in contrast to the cubic chiral magnets or Co$_{10-x}$Zn$_{10-y}$Mn$_{x+y}$ where Bloch-type chiral skyrmions are described in terms of spin helices, in GaV$_{4}$S$_{8}$ N\'{e}el-type non-chiral skyrmions are addressed in form of a superposition of spin cycloids. Moreover, while in the cubic chiral magnets the skyrmion lines are always essentially parallel to the applied magnetic field, in GaV$_{4}$S$_{8}$ the vortex cores are confined along an $\langle111\rangle$ axis. In combination with the multiferroic nature of this polar magnetic semiconductor new ways of controlling and manipulating skyrmions may be possible.

Last but not least, we return to MnGe which is isostructural to the cubic chiral magnets with a magnetic phase diagram that differs from the ones described so far. In this compound, measurements of the topological Hall effect~\cite{2011:Kanazawa:PhysRevLett} and the topological Nernst effect~\cite{2013:Shiomi:PhysRevB} as well as data from SANS~\cite{2012:Kanazawa:PhysRevB} and LF-TEM~\cite{2015:Tanigaki:arXiv} consistently suggest the formation of a simple cubic lattice of spin whirls in zero and finite field. The magnetic lattice vectors are oriented along the $\langle100\rangle$ axes of the crystal lattice. The resulting magnetic phase diagram is depicted in Fig.~\ref{figure8}(c), where the inset schematically shows the spin structure and the upper panel shows the in-plane distribution of magnetic moments as obtained from LF-TEM. Compared to the cubic chiral magnets the corresponding lattice period is relatively small and exhibits a strong increase from 3\,nm at low temperatures to 6\,nm close to $T_{c} = 170\,$K. To what extent this marks the starting point of a new generic understanding of complex spin textures remains to be seen.

We are deeply indebted to our co-workers, in particular: T.~Adams, R.~Bamler, G.~Benka, H.~Berger, S.~Bl\"{u}gel, P.~B\"{o}ni, G.~Brandl, S.~Buhrandt, A.~Chacon, C.~Duvinage, H.-M.~Eiter, L.~M.~Eng, K.~Everschor, C.~Franz, F.~Freimuth, M.~Gangl, M.~Garst, R.~Georgii, D.~Grundler, R.~Hackl, M.~Halder, F.~Haslbeck, W.~H\"{a}u{\ss}ler, T.~Hesjedal, J.~P.~Hinton, C.~Hugenschmidt, M.~Janoschek, F.~Jarzembeck, P.~Jaschke, F.~Jonietz, J.~Kindervater, D.~K\"{o}hler, J.~D.~Koralek, P.~Krautscheid, M.~Kugler, A.~Kusmartseva, P.~Lemmens, N.~Martin, S.~Mayr, D.~Meier, M.~Meven, P.~Milde, Y.~Mokrousov, S.~M\"{u}hlbauer, J.~Orenstein, S.~A.~Parameswaran, B.~Pedersen, R.~Ramesh, T.~Reimann, M.~Reiner, R.~Ritz, A.~Rosch, F.~Rucker, S.~S\"{a}ubert, C.~Schnarr, R.~W.~Schoenlein, T.~Schr\"{o}der, S.~Schulte, T.~Schulz, C.~Sch\"{u}tte, T.~Schwarze, K.~Seemann, J.~Seidel, A.~Senyshyn, I.~Stasinopoulos, A. Vishwanath, M.~Wagner, J.~Waizner, T.~Weber, S.~Weichselbaumer, B.~Wiedemann, S.~Zhang, and the team of FRM~II. Financial support through DFG FOR960, DFG TRR80, and ERC advanced grant 291079 (TOPFIT) is gratefully acknowledged.

The final publication is available at Springer via \href{http://dx.doi.org/10.1007/978-3-319-25301-5_1}{http://dx.doi.org/10.1007/978-3-319-25301-5\textunderscore 1}.


\bibliography{BauerPfleiderer_SkyrmionsChiralMagnets_arXiv}

\begin{thebibliography}{147}%
\makeatletter
\providecommand \@ifxundefined [1]{%
 \@ifx{#1\undefined}
}%
\providecommand \@ifnum [1]{%
 \ifnum #1\expandafter \@firstoftwo
 \else \expandafter \@secondoftwo
 \fi
}%
\providecommand \@ifx [1]{%
 \ifx #1\expandafter \@firstoftwo
 \else \expandafter \@secondoftwo
 \fi
}%
\providecommand \natexlab [1]{#1}%
\providecommand \enquote  [1]{``#1''}%
\providecommand \bibnamefont  [1]{#1}%
\providecommand \bibfnamefont [1]{#1}%
\providecommand \citenamefont [1]{#1}%
\providecommand \href@noop [0]{\@secondoftwo}%
\providecommand \href [0]{\begingroup \@sanitize@url \@href}%
\providecommand \@href[1]{\@@startlink{#1}\@@href}%
\providecommand \@@href[1]{\endgroup#1\@@endlink}%
\providecommand \@sanitize@url [0]{\catcode `\\12\catcode `\$12\catcode
  `\&12\catcode `\#12\catcode `\^12\catcode `\_12\catcode `\%12\relax}%
\providecommand \@@startlink[1]{}%
\providecommand \@@endlink[0]{}%
\providecommand \url  [0]{\begingroup\@sanitize@url \@url }%
\providecommand \@url [1]{\endgroup\@href {#1}{\urlprefix }}%
\providecommand \urlprefix  [0]{URL }%
\providecommand \Eprint [0]{\href }%
\providecommand \doibase [0]{http://dx.doi.org/}%
\providecommand \selectlanguage [0]{\@gobble}%
\providecommand \bibinfo  [0]{\@secondoftwo}%
\providecommand \bibfield  [0]{\@secondoftwo}%
\providecommand \translation [1]{[#1]}%
\providecommand \BibitemOpen [0]{}%
\providecommand \bibitemStop [0]{}%
\providecommand \bibitemNoStop [0]{.\EOS\space}%
\providecommand \EOS [0]{\spacefactor3000\relax}%
\providecommand \BibitemShut  [1]{\csname bibitem#1\endcsname}%
\let\auto@bib@innerbib\@empty
\bibitem [{\citenamefont
  {Skyrme}(1961{\natexlab{a}})}]{1961:Skyrme:ProcRSocLondA}%
  \BibitemOpen
  \bibfield  {author} {\bibinfo {author} {\bibfnamefont {T.~H.~R.}\
  \bibnamefont {Skyrme}},\ }\bibfield  {title} {\enquote {\bibinfo {title} {{A
  Non-Linear Field Theory}},}\ }\href {\doibase 10.1098/rspa.1961.0018}
  {\bibfield  {journal} {\bibinfo  {journal} {Proc. R. Soc. Lond. A}\ }\textbf
  {\bibinfo {volume} {260}},\ \bibinfo {pages} {127} (\bibinfo {year}
  {1961}{\natexlab{a}})}\BibitemShut {NoStop}%
\bibitem [{\citenamefont
  {Skyrme}(1961{\natexlab{b}})}]{1961:Skyrme:ProcRSocLondA2}%
  \BibitemOpen
  \bibfield  {author} {\bibinfo {author} {\bibfnamefont {T.~H.~R.}\
  \bibnamefont {Skyrme}},\ }\bibfield  {title} {\enquote {\bibinfo {title}
  {{Particle States of a Quantized Meson Field}},}\ }\href {\doibase
  10.1098/rspa.1961.0115} {\bibfield  {journal} {\bibinfo  {journal} {Proc. R.
  Soc. Lond. A}\ }\textbf {\bibinfo {volume} {262}},\ \bibinfo {pages} {237}
  (\bibinfo {year} {1961}{\natexlab{b}})}\BibitemShut {NoStop}%
\bibitem [{\citenamefont {Skyrme}(1962)}]{1962:Skyrme:NuclPhys}%
  \BibitemOpen
  \bibfield  {author} {\bibinfo {author} {\bibfnamefont {T.~H.~R.}\
  \bibnamefont {Skyrme}},\ }\bibfield  {title} {\enquote {\bibinfo {title} {{A
  unified field theory of mesons and baryons}},}\ }\href {\doibase
  10.1016/0029-5582(62)90775-7} {\bibfield  {journal} {\bibinfo  {journal}
  {Nucl. Phys.}\ }\textbf {\bibinfo {volume} {31}},\ \bibinfo {pages} {556}
  (\bibinfo {year} {1962})}\BibitemShut {NoStop}%
\bibitem [{\citenamefont {Adkins}\ \emph {et~al.}(1983)\citenamefont {Adkins},
  \citenamefont {Nappi},\ and\ \citenamefont {Witten}}]{1983:Adkins:NuclPhysB}%
  \BibitemOpen
  \bibfield  {author} {\bibinfo {author} {\bibfnamefont {G.~S.}\ \bibnamefont
  {Adkins}}, \bibinfo {author} {\bibfnamefont {C.~R.}\ \bibnamefont {Nappi}}, \
  and\ \bibinfo {author} {\bibfnamefont {E.}~\bibnamefont {Witten}},\
  }\bibfield  {title} {\enquote {\bibinfo {title} {{Static properties of
  nucleons in the Skyrme model}},}\ }\href {\doibase
  10.1016/0550-3213(83)90559-X} {\bibfield  {journal} {\bibinfo  {journal}
  {Nucl. Phys. B}\ }\textbf {\bibinfo {volume} {228}},\ \bibinfo {pages} {552}
  (\bibinfo {year} {1983})}\BibitemShut {NoStop}%
\bibitem [{\citenamefont {Zahed}\ and\ \citenamefont
  {Brown}(1986)}]{1986:Zahed:PhysRep}%
  \BibitemOpen
  \bibfield  {author} {\bibinfo {author} {\bibfnamefont {I.}~\bibnamefont
  {Zahed}}\ and\ \bibinfo {author} {\bibfnamefont {G.~E.}\ \bibnamefont
  {Brown}},\ }\bibfield  {title} {\enquote {\bibinfo {title} {{The Skyrme
  model}},}\ }\href {\doibase 10.1016/0370-1573(86)90142-0} {\bibfield
  {journal} {\bibinfo  {journal} {Phys. Rep.}\ }\textbf {\bibinfo {volume}
  {142}},\ \bibinfo {pages} {1} (\bibinfo {year} {1986})}\BibitemShut {NoStop}%
\bibitem [{\citenamefont {Diakonov}\ \emph {et~al.}(1997)\citenamefont
  {Diakonov}, \citenamefont {Petrov},\ and\ \citenamefont
  {Polyakov}}]{1997:Diakonov:ZPhysA}%
  \BibitemOpen
  \bibfield  {author} {\bibinfo {author} {\bibfnamefont {D.}~\bibnamefont
  {Diakonov}}, \bibinfo {author} {\bibfnamefont {V.}~\bibnamefont {Petrov}}, \
  and\ \bibinfo {author} {\bibfnamefont {M.}~\bibnamefont {Polyakov}},\
  }\bibfield  {title} {\enquote {\bibinfo {title} {{Exotic anti-decuplet of
  baryons: prediction from chiral solitons}},}\ }\href {\doibase
  10.1007/s002180050406} {\bibfield  {journal} {\bibinfo  {journal} {Z. Phys.
  A}\ }\textbf {\bibinfo {volume} {359}},\ \bibinfo {pages} {305} (\bibinfo
  {year} {1997})}\BibitemShut {NoStop}%
\bibitem [{\citenamefont {Chabanat}\ \emph {et~al.}(1997)\citenamefont
  {Chabanat}, \citenamefont {Bonche}, \citenamefont {Haensel}, \citenamefont
  {Meyer},\ and\ \citenamefont {Schaeffer}}]{1997:Chabanat:NuclPhysA}%
  \BibitemOpen
  \bibfield  {author} {\bibinfo {author} {\bibfnamefont {E.}~\bibnamefont
  {Chabanat}}, \bibinfo {author} {\bibfnamefont {P.}~\bibnamefont {Bonche}},
  \bibinfo {author} {\bibfnamefont {P.}~\bibnamefont {Haensel}}, \bibinfo
  {author} {\bibfnamefont {J.}~\bibnamefont {Meyer}}, \ and\ \bibinfo {author}
  {\bibfnamefont {R.}~\bibnamefont {Schaeffer}},\ }\bibfield  {title} {\enquote
  {\bibinfo {title} {{A Skyrme parametrization from subnuclear to neutron star
  densitie}},}\ }\href {\doibase 10.1016/S0375-9474(97)00596-4} {\bibfield
  {journal} {\bibinfo  {journal} {Nucl. Phys. A}\ }\textbf {\bibinfo {volume}
  {627}},\ \bibinfo {pages} {710} (\bibinfo {year} {1997})}\BibitemShut
  {NoStop}%
\bibitem [{\citenamefont {Chabanat}\ \emph {et~al.}(1998)\citenamefont
  {Chabanat}, \citenamefont {Bonche}, \citenamefont {Haensel}, \citenamefont
  {Meyer},\ and\ \citenamefont {Schaeffer}}]{1998:Chabanat:NuclPhysA}%
  \BibitemOpen
  \bibfield  {author} {\bibinfo {author} {\bibfnamefont {E.}~\bibnamefont
  {Chabanat}}, \bibinfo {author} {\bibfnamefont {P.}~\bibnamefont {Bonche}},
  \bibinfo {author} {\bibfnamefont {P.}~\bibnamefont {Haensel}}, \bibinfo
  {author} {\bibfnamefont {J.}~\bibnamefont {Meyer}}, \ and\ \bibinfo {author}
  {\bibfnamefont {R.}~\bibnamefont {Schaeffer}},\ }\bibfield  {title} {\enquote
  {\bibinfo {title} {{A Skyrme parametrization from subnuclear to neutron star
  densities Part II. Nuclei far from stabilities}},}\ }\href {\doibase
  10.1016/S0375-9474(98)00180-8} {\bibfield  {journal} {\bibinfo  {journal}
  {Nucl. Phys. A}\ }\textbf {\bibinfo {volume} {635}},\ \bibinfo {pages} {231}
  (\bibinfo {year} {1998})}\BibitemShut {NoStop}%
\bibitem [{\citenamefont {Sondhi}\ \emph {et~al.}(1993)\citenamefont {Sondhi},
  \citenamefont {Karlhede}, \citenamefont {Kivelson},\ and\ \citenamefont
  {Rezayi}}]{1993:Sondhi:PhysRevB}%
  \BibitemOpen
  \bibfield  {author} {\bibinfo {author} {\bibfnamefont {S.~L.}\ \bibnamefont
  {Sondhi}}, \bibinfo {author} {\bibfnamefont {A.}~\bibnamefont {Karlhede}},
  \bibinfo {author} {\bibfnamefont {S.~A.}\ \bibnamefont {Kivelson}}, \ and\
  \bibinfo {author} {\bibfnamefont {E.~H.}\ \bibnamefont {Rezayi}},\ }\bibfield
   {title} {\enquote {\bibinfo {title} {{Skyrmions and the crossover from the
  integer to fractional quantum Hall effect at small Zeeman energies}},}\
  }\href {\doibase 10.1103/PhysRevB.47.16419} {\bibfield  {journal} {\bibinfo
  {journal} {Phys. Rev. B}\ }\textbf {\bibinfo {volume} {47}},\ \bibinfo
  {pages} {16419} (\bibinfo {year} {1993})}\BibitemShut {NoStop}%
\bibitem [{\citenamefont {Schmeller}\ \emph {et~al.}(1995)\citenamefont
  {Schmeller}, \citenamefont {Eisenstein}, \citenamefont {Pfeiffer},\ and\
  \citenamefont {West}}]{1995:Schmeller:PhysRevLett}%
  \BibitemOpen
  \bibfield  {author} {\bibinfo {author} {\bibfnamefont {A.}~\bibnamefont
  {Schmeller}}, \bibinfo {author} {\bibfnamefont {J.~P.}\ \bibnamefont
  {Eisenstein}}, \bibinfo {author} {\bibfnamefont {L.~N.}\ \bibnamefont
  {Pfeiffer}}, \ and\ \bibinfo {author} {\bibfnamefont {K.~W.}\ \bibnamefont
  {West}},\ }\bibfield  {title} {\enquote {\bibinfo {title} {{Evidence for
  Skyrmions and Single Spin Flips in the Integer Quantized Hall Effect}},}\
  }\href {\doibase 10.1103/PhysRevLett.75.4290} {\bibfield  {journal} {\bibinfo
   {journal} {Phys. Rev. Lett.}\ }\textbf {\bibinfo {volume} {75}},\ \bibinfo
  {pages} {4290} (\bibinfo {year} {1995})}\BibitemShut {NoStop}%
\bibitem [{\citenamefont {Yang}\ \emph {et~al.}(2006)\citenamefont {Yang},
  \citenamefont {Das~Sarma},\ and\ \citenamefont
  {MacDonald}}]{2006:Yang:PhysRevB}%
  \BibitemOpen
  \bibfield  {author} {\bibinfo {author} {\bibfnamefont {K.}~\bibnamefont
  {Yang}}, \bibinfo {author} {\bibfnamefont {S.}~\bibnamefont {Das~Sarma}}, \
  and\ \bibinfo {author} {\bibfnamefont {A.~H.}\ \bibnamefont {MacDonald}},\
  }\bibfield  {title} {\enquote {\bibinfo {title} {{Collective modes and
  skyrmion excitations in graphene $SU(4)$ quantum Hall ferromagnets}},}\
  }\href {\doibase 10.1103/PhysRevB.74.075423} {\bibfield  {journal} {\bibinfo
  {journal} {Phys. Rev. B}\ }\textbf {\bibinfo {volume} {74}},\ \bibinfo
  {pages} {075423} (\bibinfo {year} {2006})}\BibitemShut {NoStop}%
\bibitem [{\citenamefont {Ho}(1998)}]{1998:Ho:PhysRevLett}%
  \BibitemOpen
  \bibfield  {author} {\bibinfo {author} {\bibfnamefont {T.-L.}\ \bibnamefont
  {Ho}},\ }\bibfield  {title} {\enquote {\bibinfo {title} {{Spinor Bose
  Condensates in Optical Traps}},}\ }\href {\doibase
  10.1103/PhysRevLett.81.742} {\bibfield  {journal} {\bibinfo  {journal} {Phys.
  Rev. Lett.}\ }\textbf {\bibinfo {volume} {181}},\ \bibinfo {pages} {742}
  (\bibinfo {year} {1998})}\BibitemShut {NoStop}%
\bibitem [{\citenamefont {Khawaja}\ and\ \citenamefont
  {Stoof}(2001)}]{2001:Khawaja:Nature}%
  \BibitemOpen
  \bibfield  {author} {\bibinfo {author} {\bibfnamefont {U.~A.}\ \bibnamefont
  {Khawaja}}\ and\ \bibinfo {author} {\bibfnamefont {H.}~\bibnamefont
  {Stoof}},\ }\bibfield  {title} {\enquote {\bibinfo {title} {{Skyrmions in a
  ferromagnetic Bose-Einstein condensate}},}\ }\href {\doibase
  10.1038/35082010} {\bibfield  {journal} {\bibinfo  {journal} {Nature
  (London)}\ }\textbf {\bibinfo {volume} {411}},\ \bibinfo {pages} {918}
  (\bibinfo {year} {2001})}\BibitemShut {NoStop}%
\bibitem [{\citenamefont {Leslie}\ \emph {et~al.}(2009)\citenamefont {Leslie},
  \citenamefont {Hansen}, \citenamefont {Wright}, \citenamefont {Deutsch},\
  and\ \citenamefont {Bigelow}}]{2009:Leslie:PhysRevLett}%
  \BibitemOpen
  \bibfield  {author} {\bibinfo {author} {\bibfnamefont {L.~S.}\ \bibnamefont
  {Leslie}}, \bibinfo {author} {\bibfnamefont {A.}~\bibnamefont {Hansen}},
  \bibinfo {author} {\bibfnamefont {K.~C.}\ \bibnamefont {Wright}}, \bibinfo
  {author} {\bibfnamefont {B.~M.}\ \bibnamefont {Deutsch}}, \ and\ \bibinfo
  {author} {\bibfnamefont {N.~P.}\ \bibnamefont {Bigelow}},\ }\bibfield
  {title} {\enquote {\bibinfo {title} {{Creation and Detection of Skyrmions in
  a Bose-Einstein Condensate}},}\ }\href {\doibase
  10.1103/PhysRevLett.103.250401} {\bibfield  {journal} {\bibinfo  {journal}
  {Phys. Rev. Lett.}\ }\textbf {\bibinfo {volume} {103}},\ \bibinfo {pages}
  {250401} (\bibinfo {year} {2009})}\BibitemShut {NoStop}%
\bibitem [{\citenamefont {Fukuda}\ and\ \citenamefont
  {\v{Z}umer}(2011)}]{2011:Fukuda:NatCommun}%
  \BibitemOpen
  \bibfield  {author} {\bibinfo {author} {\bibfnamefont {J.}~\bibnamefont
  {Fukuda}}\ and\ \bibinfo {author} {\bibfnamefont {S.}~\bibnamefont
  {\v{Z}umer}},\ }\bibfield  {title} {\enquote {\bibinfo {title}
  {{Quasi-two-dimensional Skyrmion lattices in a chiral nematic liquid
  crystal}},}\ }\href {\doibase 10.1038/ncomms1250} {\bibfield  {journal}
  {\bibinfo  {journal} {Nat. Commun.}\ }\textbf {\bibinfo {volume} {2}},\
  \bibinfo {pages} {246} (\bibinfo {year} {2011})}\BibitemShut {NoStop}%
\bibitem [{\citenamefont {Bogdanov}\ and\ \citenamefont
  {Yablonskii}(1989)}]{1989:Bogdanov:SovPhysJETP}%
  \BibitemOpen
  \bibfield  {author} {\bibinfo {author} {\bibfnamefont {A.~N.}\ \bibnamefont
  {Bogdanov}}\ and\ \bibinfo {author} {\bibfnamefont {D.~A.}\ \bibnamefont
  {Yablonskii}},\ }\bibfield  {title} {\enquote {\bibinfo {title}
  {{Thermodynamically stable "vortices" in magnetically ordered crystals. The
  mixed state of magnets}},}\ }\href
  {http://jetp.ac.ru/cgi-bin/dn/e_068_01_0101.pdf} {\bibfield  {journal}
  {\bibinfo  {journal} {Sov. Phys. JETP}\ }\textbf {\bibinfo {volume} {95}},\
  \bibinfo {pages} {178} (\bibinfo {year} {1989})}\BibitemShut {NoStop}%
\bibitem [{\citenamefont {Bogdanov}\ and\ \citenamefont
  {Hubert}(1994)}]{1994:Bogdanov:JMagnMagnMater}%
  \BibitemOpen
  \bibfield  {author} {\bibinfo {author} {\bibfnamefont {A.}~\bibnamefont
  {Bogdanov}}\ and\ \bibinfo {author} {\bibfnamefont {A.}~\bibnamefont
  {Hubert}},\ }\bibfield  {title} {\enquote {\bibinfo {title}
  {{Thermodynamically stable magnetic vortex states in magnetic crystals}},}\
  }\href {\doibase 10.1016/0304-8853(94)90046-9} {\bibfield  {journal}
  {\bibinfo  {journal} {J. Magn. Magn. Mater.}\ }\textbf {\bibinfo {volume}
  {138}},\ \bibinfo {pages} {255} (\bibinfo {year} {1994})}\BibitemShut
  {NoStop}%
\bibitem [{\citenamefont {M\"{u}hlbauer}\ \emph {et~al.}(2009)\citenamefont
  {M\"{u}hlbauer}, \citenamefont {Binz}, \citenamefont {Jonietz}, \citenamefont
  {Pfleiderer}, \citenamefont {Rosch}, \citenamefont {Neubauer}, \citenamefont
  {Georgii},\ and\ \citenamefont {B\"{o}ni}}]{2009:Muhlbauer:Science}%
  \BibitemOpen
  \bibfield  {author} {\bibinfo {author} {\bibfnamefont {S.}~\bibnamefont
  {M\"{u}hlbauer}}, \bibinfo {author} {\bibfnamefont {B.}~\bibnamefont {Binz}},
  \bibinfo {author} {\bibfnamefont {F.}~\bibnamefont {Jonietz}}, \bibinfo
  {author} {\bibfnamefont {C.}~\bibnamefont {Pfleiderer}}, \bibinfo {author}
  {\bibfnamefont {A.}~\bibnamefont {Rosch}}, \bibinfo {author} {\bibfnamefont
  {A.}~\bibnamefont {Neubauer}}, \bibinfo {author} {\bibfnamefont
  {R.}~\bibnamefont {Georgii}}, \ and\ \bibinfo {author} {\bibfnamefont
  {P.}~\bibnamefont {B\"{o}ni}},\ }\bibfield  {title} {\enquote {\bibinfo
  {title} {{Skyrmion Lattice in a Chiral Magnet}},}\ }\href {\doibase
  10.1126/science.1166767} {\bibfield  {journal} {\bibinfo  {journal}
  {Science}\ }\textbf {\bibinfo {volume} {323}},\ \bibinfo {pages} {915}
  (\bibinfo {year} {2009})}\BibitemShut {NoStop}%
\bibitem [{\citenamefont {M\"{u}nzer}\ \emph {et~al.}(2010)\citenamefont
  {M\"{u}nzer}, \citenamefont {Neubauer}, \citenamefont {Adams}, \citenamefont
  {M\"{u}hlbauer}, \citenamefont {Franz}, \citenamefont {Jonietz},
  \citenamefont {Georgii}, \citenamefont {B\"{o}ni}, \citenamefont {Pedersen},
  \citenamefont {Schmidt}, \citenamefont {Rosch},\ and\ \citenamefont
  {Pfleiderer}}]{2010:Munzer:PhysRevB}%
  \BibitemOpen
  \bibfield  {author} {\bibinfo {author} {\bibfnamefont {W.}~\bibnamefont
  {M\"{u}nzer}}, \bibinfo {author} {\bibfnamefont {A.}~\bibnamefont
  {Neubauer}}, \bibinfo {author} {\bibfnamefont {T.}~\bibnamefont {Adams}},
  \bibinfo {author} {\bibfnamefont {S.}~\bibnamefont {M\"{u}hlbauer}}, \bibinfo
  {author} {\bibfnamefont {C.}~\bibnamefont {Franz}}, \bibinfo {author}
  {\bibfnamefont {F.}~\bibnamefont {Jonietz}}, \bibinfo {author} {\bibfnamefont
  {R.}~\bibnamefont {Georgii}}, \bibinfo {author} {\bibfnamefont
  {P.}~\bibnamefont {B\"{o}ni}}, \bibinfo {author} {\bibfnamefont
  {B.}~\bibnamefont {Pedersen}}, \bibinfo {author} {\bibfnamefont
  {M.}~\bibnamefont {Schmidt}}, \bibinfo {author} {\bibfnamefont
  {A.}~\bibnamefont {Rosch}}, \ and\ \bibinfo {author} {\bibfnamefont
  {C.}~\bibnamefont {Pfleiderer}},\ }\bibfield  {title} {\enquote {\bibinfo
  {title} {{Skyrmion lattice in the doped semiconductor
  Fe$_{1-x}$Co$_{x}$Si}},}\ }\href {\doibase 10.1103/PhysRevB.81.041203}
  {\bibfield  {journal} {\bibinfo  {journal} {Phys. Rev. B}\ }\textbf {\bibinfo
  {volume} {81}},\ \bibinfo {pages} {041203 (R)} (\bibinfo {year}
  {2010})}\BibitemShut {NoStop}%
\bibitem [{\citenamefont {Pfleiderer}\ \emph {et~al.}(2001)\citenamefont
  {Pfleiderer}, \citenamefont {Julian},\ and\ \citenamefont
  {Lonzarich}}]{2001:Pfleiderer:Nature}%
  \BibitemOpen
  \bibfield  {author} {\bibinfo {author} {\bibfnamefont {C.}~\bibnamefont
  {Pfleiderer}}, \bibinfo {author} {\bibfnamefont {S.~R.}\ \bibnamefont
  {Julian}}, \ and\ \bibinfo {author} {\bibfnamefont {G.~G.}\ \bibnamefont
  {Lonzarich}},\ }\bibfield  {title} {\enquote {\bibinfo {title}
  {{Non-Fermi-liquid nature of the normal state of itinerant-electron
  ferromagnets}},}\ }\href {\doibase 10.1038/35106527} {\bibfield  {journal}
  {\bibinfo  {journal} {Nature (London)}\ }\textbf {\bibinfo {volume} {414}},\
  \bibinfo {pages} {427} (\bibinfo {year} {2001})}\BibitemShut {NoStop}%
\bibitem [{\citenamefont {Pfleiderer}\ \emph {et~al.}(2004)\citenamefont
  {Pfleiderer}, \citenamefont {Reznik}, \citenamefont {Pintschovius},
  \citenamefont {v.~L\"{o}hneysen}, \citenamefont {Garst},\ and\ \citenamefont
  {Rosch}}]{2004:Pfleiderer:Nature}%
  \BibitemOpen
  \bibfield  {author} {\bibinfo {author} {\bibfnamefont {C.}~\bibnamefont
  {Pfleiderer}}, \bibinfo {author} {\bibfnamefont {D.}~\bibnamefont {Reznik}},
  \bibinfo {author} {\bibfnamefont {L.}~\bibnamefont {Pintschovius}}, \bibinfo
  {author} {\bibfnamefont {H.}~\bibnamefont {v.~L\"{o}hneysen}}, \bibinfo
  {author} {\bibfnamefont {M.}~\bibnamefont {Garst}}, \ and\ \bibinfo {author}
  {\bibfnamefont {A.}~\bibnamefont {Rosch}},\ }\bibfield  {title} {\enquote
  {\bibinfo {title} {{Partial order in the non-Fermi-liquid phase of MnSi}},}\
  }\href {\doibase 10.1038/nature02232} {\bibfield  {journal} {\bibinfo
  {journal} {Nature (London)}\ }\textbf {\bibinfo {volume} {427}},\ \bibinfo
  {pages} {227} (\bibinfo {year} {2004})}\BibitemShut {NoStop}%
\bibitem [{\citenamefont {Ritz}\ \emph
  {et~al.}(2013{\natexlab{a}})\citenamefont {Ritz}, \citenamefont {Halder},
  \citenamefont {Wagner}, \citenamefont {Franz}, \citenamefont {Bauer},\ and\
  \citenamefont {Pfleiderer}}]{2013:Ritz:Nature}%
  \BibitemOpen
  \bibfield  {author} {\bibinfo {author} {\bibfnamefont {R.}~\bibnamefont
  {Ritz}}, \bibinfo {author} {\bibfnamefont {M.}~\bibnamefont {Halder}},
  \bibinfo {author} {\bibfnamefont {M.}~\bibnamefont {Wagner}}, \bibinfo
  {author} {\bibfnamefont {C.}~\bibnamefont {Franz}}, \bibinfo {author}
  {\bibfnamefont {A.}~\bibnamefont {Bauer}}, \ and\ \bibinfo {author}
  {\bibfnamefont {C.}~\bibnamefont {Pfleiderer}},\ }\bibfield  {title}
  {\enquote {\bibinfo {title} {{Formation of a topological non-Fermi liquid in
  MnSi}},}\ }\href {\doibase 10.1038/nature12023} {\bibfield  {journal}
  {\bibinfo  {journal} {Nature (London)}\ }\textbf {\bibinfo {volume} {497}},\
  \bibinfo {pages} {231} (\bibinfo {year} {2013}{\natexlab{a}})}\BibitemShut
  {NoStop}%
\bibitem [{\citenamefont {Nagaosa}\ and\ \citenamefont
  {Tokura}(2013)}]{2013:Nagaosa:NatureNano}%
  \BibitemOpen
  \bibfield  {author} {\bibinfo {author} {\bibfnamefont {N.}~\bibnamefont
  {Nagaosa}}\ and\ \bibinfo {author} {\bibfnamefont {Y.}~\bibnamefont
  {Tokura}},\ }\bibfield  {title} {\enquote {\bibinfo {title} {{Topological
  properties and dynamics of magnetic skyrmions}},}\ }\href {\doibase
  10.1038/nnano.2013.243} {\bibfield  {journal} {\bibinfo  {journal} {Nature
  Nano.}\ }\textbf {\bibinfo {volume} {8}},\ \bibinfo {pages} {899} (\bibinfo
  {year} {2013})}\BibitemShut {NoStop}%
\bibitem [{\citenamefont {Jonietz}\ \emph {et~al.}(2010)\citenamefont
  {Jonietz}, \citenamefont {M\"{u}hlbauer}, \citenamefont {Pfleiderer},
  \citenamefont {Neubauer}, \citenamefont {M\"{u}nzer}, \citenamefont {Bauer},
  \citenamefont {Adams}, \citenamefont {Georgii}, \citenamefont {B\"{o}ni},
  \citenamefont {Duine}, \citenamefont {Everschor}, \citenamefont {Garst},\
  and\ \citenamefont {Rosch}}]{2010:Jonietz:Science}%
  \BibitemOpen
  \bibfield  {author} {\bibinfo {author} {\bibfnamefont {F.}~\bibnamefont
  {Jonietz}}, \bibinfo {author} {\bibfnamefont {S.}~\bibnamefont
  {M\"{u}hlbauer}}, \bibinfo {author} {\bibfnamefont {C.}~\bibnamefont
  {Pfleiderer}}, \bibinfo {author} {\bibfnamefont {A.}~\bibnamefont
  {Neubauer}}, \bibinfo {author} {\bibfnamefont {W.}~\bibnamefont
  {M\"{u}nzer}}, \bibinfo {author} {\bibfnamefont {A.}~\bibnamefont {Bauer}},
  \bibinfo {author} {\bibfnamefont {T.}~\bibnamefont {Adams}}, \bibinfo
  {author} {\bibfnamefont {R.}~\bibnamefont {Georgii}}, \bibinfo {author}
  {\bibfnamefont {P.}~\bibnamefont {B\"{o}ni}}, \bibinfo {author}
  {\bibfnamefont {R.~A.}\ \bibnamefont {Duine}}, \bibinfo {author}
  {\bibfnamefont {K.}~\bibnamefont {Everschor}}, \bibinfo {author}
  {\bibfnamefont {M.}~\bibnamefont {Garst}}, \ and\ \bibinfo {author}
  {\bibfnamefont {A.}~\bibnamefont {Rosch}},\ }\bibfield  {title} {\enquote
  {\bibinfo {title} {{Spin Transfer Torques in MnSi at Ultralow Current
  Densities}},}\ }\href {\doibase 10.1126/science.1195709} {\bibfield
  {journal} {\bibinfo  {journal} {Science}\ }\textbf {\bibinfo {volume}
  {330}},\ \bibinfo {pages} {1648} (\bibinfo {year} {2010})}\BibitemShut
  {NoStop}%
\bibitem [{\citenamefont {Fert}\ \emph {et~al.}(2013)\citenamefont {Fert},
  \citenamefont {Cros},\ and\ \citenamefont {Sampaio}}]{2013:Fert:NatureNano}%
  \BibitemOpen
  \bibfield  {author} {\bibinfo {author} {\bibfnamefont {A.}~\bibnamefont
  {Fert}}, \bibinfo {author} {\bibfnamefont {V.}~\bibnamefont {Cros}}, \ and\
  \bibinfo {author} {\bibfnamefont {J.}~\bibnamefont {Sampaio}},\ }\bibfield
  {title} {\enquote {\bibinfo {title} {{Skyrmions on the track}},}\ }\href
  {\doibase 10.1038/nnano.2013.29} {\bibfield  {journal} {\bibinfo  {journal}
  {Nature Nano.}\ }\textbf {\bibinfo {volume} {8}},\ \bibinfo {pages} {152}
  (\bibinfo {year} {2013})}\BibitemShut {NoStop}%
\bibitem [{\citenamefont {Seki}\ \emph
  {et~al.}(2012{\natexlab{a}})\citenamefont {Seki}, \citenamefont {Yu},
  \citenamefont {Ishiwata},\ and\ \citenamefont {Tokura}}]{2012:Seki:Science}%
  \BibitemOpen
  \bibfield  {author} {\bibinfo {author} {\bibfnamefont {S.}~\bibnamefont
  {Seki}}, \bibinfo {author} {\bibfnamefont {X.~Z.}\ \bibnamefont {Yu}},
  \bibinfo {author} {\bibfnamefont {S.}~\bibnamefont {Ishiwata}}, \ and\
  \bibinfo {author} {\bibfnamefont {Y.}~\bibnamefont {Tokura}},\ }\bibfield
  {title} {\enquote {\bibinfo {title} {{Observation of Skyrmions in a
  Multiferroic Material}},}\ }\href {\doibase 10.1126/science.1214143}
  {\bibfield  {journal} {\bibinfo  {journal} {Science}\ }\textbf {\bibinfo
  {volume} {336}},\ \bibinfo {pages} {198} (\bibinfo {year}
  {2012}{\natexlab{a}})}\BibitemShut {NoStop}%
\bibitem [{\citenamefont {Mochizuki}(2012)}]{2012:Mochizuki:PhysRevLett}%
  \BibitemOpen
  \bibfield  {author} {\bibinfo {author} {\bibfnamefont {M.}~\bibnamefont
  {Mochizuki}},\ }\bibfield  {title} {\enquote {\bibinfo {title} {{Spin-Wave
  Modes and Their Intense Excitation Effects in Skyrmion Crystals}},}\ }\href
  {\doibase 10.1103/PhysRevLett.108.017601} {\bibfield  {journal} {\bibinfo
  {journal} {Phys. Rev. Lett.}\ }\textbf {\bibinfo {volume} {108}},\ \bibinfo
  {pages} {017601} (\bibinfo {year} {2012})}\BibitemShut {NoStop}%
\bibitem [{\citenamefont {Okamura}\ \emph {et~al.}(2013)\citenamefont
  {Okamura}, \citenamefont {Kagawa}, \citenamefont {Mochizuki}, \citenamefont
  {Kubota}, \citenamefont {Seki}, \citenamefont {Ishiwata}, \citenamefont
  {Kawasaki}, \citenamefont {Onose},\ and\ \citenamefont
  {Tokura}}]{2013:Okamura:NatCommun}%
  \BibitemOpen
  \bibfield  {author} {\bibinfo {author} {\bibfnamefont {Y.}~\bibnamefont
  {Okamura}}, \bibinfo {author} {\bibfnamefont {F.}~\bibnamefont {Kagawa}},
  \bibinfo {author} {\bibfnamefont {M.}~\bibnamefont {Mochizuki}}, \bibinfo
  {author} {\bibfnamefont {M.}~\bibnamefont {Kubota}}, \bibinfo {author}
  {\bibfnamefont {S.}~\bibnamefont {Seki}}, \bibinfo {author} {\bibfnamefont
  {S.}~\bibnamefont {Ishiwata}}, \bibinfo {author} {\bibfnamefont
  {M.}~\bibnamefont {Kawasaki}}, \bibinfo {author} {\bibfnamefont
  {Y.}~\bibnamefont {Onose}}, \ and\ \bibinfo {author} {\bibfnamefont
  {Y.}~\bibnamefont {Tokura}},\ }\bibfield  {title} {\enquote {\bibinfo {title}
  {{Microwave magnetoelectric effect via skyrmion resonance modes in a
  helimagnetic multiferroic}},}\ }\href {\doibase 10.1038/ncomms3391}
  {\bibfield  {journal} {\bibinfo  {journal} {Nat. Commun.}\ }\textbf {\bibinfo
  {volume} {4}},\ \bibinfo {pages} {2391} (\bibinfo {year} {2013})}\BibitemShut
  {NoStop}%
\bibitem [{\citenamefont {Schwarze}\ \emph {et~al.}(2015)\citenamefont
  {Schwarze}, \citenamefont {Waizner}, \citenamefont {Garst}, \citenamefont
  {Bauer}, \citenamefont {Stasinopoulos}, \citenamefont {Berger}, \citenamefont
  {Rosch}, \citenamefont {Pfleiderer},\ and\ \citenamefont
  {Grundler}}]{2015:Schwarze:NatureMater}%
  \BibitemOpen
  \bibfield  {author} {\bibinfo {author} {\bibfnamefont {T.}~\bibnamefont
  {Schwarze}}, \bibinfo {author} {\bibfnamefont {J.}~\bibnamefont {Waizner}},
  \bibinfo {author} {\bibfnamefont {M.}~\bibnamefont {Garst}}, \bibinfo
  {author} {\bibfnamefont {A.}~\bibnamefont {Bauer}}, \bibinfo {author}
  {\bibfnamefont {I.}~\bibnamefont {Stasinopoulos}}, \bibinfo {author}
  {\bibfnamefont {H.}~\bibnamefont {Berger}}, \bibinfo {author} {\bibfnamefont
  {A.}~\bibnamefont {Rosch}}, \bibinfo {author} {\bibfnamefont
  {C.}~\bibnamefont {Pfleiderer}}, \ and\ \bibinfo {author} {\bibfnamefont
  {D.}~\bibnamefont {Grundler}},\ }\bibfield  {title} {\enquote {\bibinfo
  {title} {{Universal helimagnon and skyrmion excitations in metallic,
  semiconducting and insulating chiral magnets}},}\ }\href {\doibase
  10.1038/nmat4223} {\bibfield  {journal} {\bibinfo  {journal} {Nature Mater.}\
  }\textbf {\bibinfo {volume} {14}},\ \bibinfo {pages} {478} (\bibinfo {year}
  {2015})}\BibitemShut {NoStop}%
\bibitem [{\citenamefont {K\'{e}zsm\'{a}rki}\ \emph {et~al.}(2015)\citenamefont
  {K\'{e}zsm\'{a}rki}, \citenamefont {Bord\'{a}cs}, \citenamefont {Milde},
  \citenamefont {Neuber}, \citenamefont {Eng}, \citenamefont {White},
  \citenamefont {R{\o}nnow}, \citenamefont {Dewhurst}, \citenamefont
  {Mochizuki}, \citenamefont {Yanai}, \citenamefont {Nakamura}, \citenamefont
  {Ehlers}, \citenamefont {Tsurkan},\ and\ \citenamefont
  {Loidl}}]{2015:Kezsmarki:arXiv}%
  \BibitemOpen
  \bibfield  {author} {\bibinfo {author} {\bibfnamefont {I.}~\bibnamefont
  {K\'{e}zsm\'{a}rki}}, \bibinfo {author} {\bibfnamefont {S.}~\bibnamefont
  {Bord\'{a}cs}}, \bibinfo {author} {\bibfnamefont {P.}~\bibnamefont {Milde}},
  \bibinfo {author} {\bibfnamefont {E.}~\bibnamefont {Neuber}}, \bibinfo
  {author} {\bibfnamefont {L.~M.}\ \bibnamefont {Eng}}, \bibinfo {author}
  {\bibfnamefont {J.~S.}\ \bibnamefont {White}}, \bibinfo {author}
  {\bibfnamefont {H.~M.}\ \bibnamefont {R{\o}nnow}}, \bibinfo {author}
  {\bibfnamefont {C.~D.}\ \bibnamefont {Dewhurst}}, \bibinfo {author}
  {\bibfnamefont {M.}~\bibnamefont {Mochizuki}}, \bibinfo {author}
  {\bibfnamefont {K.}~\bibnamefont {Yanai}}, \bibinfo {author} {\bibfnamefont
  {H.}~\bibnamefont {Nakamura}}, \bibinfo {author} {\bibfnamefont
  {D.}~\bibnamefont {Ehlers}}, \bibinfo {author} {\bibfnamefont
  {V.}~\bibnamefont {Tsurkan}}, \ and\ \bibinfo {author} {\bibfnamefont
  {A.}~\bibnamefont {Loidl}},\ }\bibfield  {title} {\enquote {\bibinfo {title}
  {{N\'{e}el-type Skyrmion Lattice with Confined Orientation in the Polar
  Magnetic Semiconductor GaV$_{4}$S$_{8}$}},}\ }\href
  {http://arxiv.org/abs/1502.08049} {\bibfield  {journal} {\bibinfo  {journal}
  {arXiv:1502.08049}\ } (\bibinfo {year} {2015})}\BibitemShut {NoStop}%
\bibitem [{\citenamefont {Milde}\ \emph {et~al.}(2013)\citenamefont {Milde},
  \citenamefont {K\"{o}hler}, \citenamefont {Seidel}, \citenamefont {Eng},
  \citenamefont {Bauer}, \citenamefont {Chacon}, \citenamefont {Kindervater},
  \citenamefont {M\"{u}hlbauer}, \citenamefont {Pfleiderer}, \citenamefont
  {Buhrandt}, \citenamefont {Sch\"{u}tte},\ and\ \citenamefont
  {Rosch}}]{2013:Milde:Science}%
  \BibitemOpen
  \bibfield  {author} {\bibinfo {author} {\bibfnamefont {P.}~\bibnamefont
  {Milde}}, \bibinfo {author} {\bibfnamefont {D.}~\bibnamefont {K\"{o}hler}},
  \bibinfo {author} {\bibfnamefont {J.}~\bibnamefont {Seidel}}, \bibinfo
  {author} {\bibfnamefont {L.~M.}\ \bibnamefont {Eng}}, \bibinfo {author}
  {\bibfnamefont {A.}~\bibnamefont {Bauer}}, \bibinfo {author} {\bibfnamefont
  {A.}~\bibnamefont {Chacon}}, \bibinfo {author} {\bibfnamefont
  {J.}~\bibnamefont {Kindervater}}, \bibinfo {author} {\bibfnamefont
  {S.}~\bibnamefont {M\"{u}hlbauer}}, \bibinfo {author} {\bibfnamefont
  {C.}~\bibnamefont {Pfleiderer}}, \bibinfo {author} {\bibfnamefont
  {S.}~\bibnamefont {Buhrandt}}, \bibinfo {author} {\bibfnamefont
  {C.}~\bibnamefont {Sch\"{u}tte}}, \ and\ \bibinfo {author} {\bibfnamefont
  {A.}~\bibnamefont {Rosch}},\ }\bibfield  {title} {\enquote {\bibinfo {title}
  {{Unwinding of a Skyrmion Lattice by Magnetic Monopoles}},}\ }\href {\doibase
  10.1126/science.1234657} {\bibfield  {journal} {\bibinfo  {journal}
  {Science}\ }\textbf {\bibinfo {volume} {340}},\ \bibinfo {pages} {1076}
  (\bibinfo {year} {2013})}\BibitemShut {NoStop}%
\bibitem [{\citenamefont {Landau}\ and\ \citenamefont
  {Lifshitz}(1980)}]{1980:Landau:Book}%
  \BibitemOpen
  \bibfield  {author} {\bibinfo {author} {\bibfnamefont {L.~D.}\ \bibnamefont
  {Landau}}\ and\ \bibinfo {author} {\bibfnamefont {E.~M.}\ \bibnamefont
  {Lifshitz}},\ }\href@noop {} {\emph {\bibinfo {title} {{Course of Theoretical
  Physics}}}}\ (\bibinfo  {publisher} {Pergamon Press, New York},\ \bibinfo
  {year} {1980})\BibitemShut {NoStop}%
\bibitem [{\citenamefont
  {Dzyaloshinsky}(1957)}]{1957:Dzyaloshinsky:SovPhysJETP}%
  \BibitemOpen
  \bibfield  {author} {\bibinfo {author} {\bibfnamefont {I.~E.}\ \bibnamefont
  {Dzyaloshinsky}},\ }\bibfield  {title} {\enquote {\bibinfo {title}
  {{Thermodynamic theory of weak ferromagnetism in antiferromagnetic
  substances}},}\ }\href@noop {} {\bibfield  {journal} {\bibinfo  {journal}
  {Sov. Phys. JETP}\ }\textbf {\bibinfo {volume} {5}},\ \bibinfo {pages} {1259}
  (\bibinfo {year} {1957})}\BibitemShut {NoStop}%
\bibitem [{\citenamefont {Moriya}(1960)}]{1960:Moriya:PhysRev}%
  \BibitemOpen
  \bibfield  {author} {\bibinfo {author} {\bibfnamefont {T.}~\bibnamefont
  {Moriya}},\ }\bibfield  {title} {\enquote {\bibinfo {title} {{Anisotropic
  Superexchange Interaction and Weak Ferromagnetism}},}\ }\href {\doibase
  10.1103/PhysRev.120.91} {\bibfield  {journal} {\bibinfo  {journal} {Phys.
  Rev.}\ }\textbf {\bibinfo {volume} {120}},\ \bibinfo {pages} {91} (\bibinfo
  {year} {1960})}\BibitemShut {NoStop}%
\bibitem [{\citenamefont
  {Dzyaloshinsky}(1964)}]{1964:Dzyaloshinsky:SovPhysJETP}%
  \BibitemOpen
  \bibfield  {author} {\bibinfo {author} {\bibfnamefont {I.~E.}\ \bibnamefont
  {Dzyaloshinsky}},\ }\bibfield  {title} {\enquote {\bibinfo {title} {{Theory
  of helicoidal structures in antiferromagnets}},}\ }\href@noop {} {\bibfield
  {journal} {\bibinfo  {journal} {Sov. Phys. JETP}\ }\textbf {\bibinfo {volume}
  {19}},\ \bibinfo {pages} {960} (\bibinfo {year} {1964})}\BibitemShut
  {NoStop}%
\bibitem [{\citenamefont {Ishikawa}\ \emph {et~al.}(1976)\citenamefont
  {Ishikawa}, \citenamefont {Tajima}, \citenamefont {Bloch},\ and\
  \citenamefont {Roth}}]{1976:Ishikawa:SolidStateCommun}%
  \BibitemOpen
  \bibfield  {author} {\bibinfo {author} {\bibfnamefont {Y.}~\bibnamefont
  {Ishikawa}}, \bibinfo {author} {\bibfnamefont {K.}~\bibnamefont {Tajima}},
  \bibinfo {author} {\bibfnamefont {D.}~\bibnamefont {Bloch}}, \ and\ \bibinfo
  {author} {\bibfnamefont {M.}~\bibnamefont {Roth}},\ }\bibfield  {title}
  {\enquote {\bibinfo {title} {{Helical spin structure in manganese silicide
  MnSi}},}\ }\href {\doibase 10.1016/0038-1098(76)90057-0} {\bibfield
  {journal} {\bibinfo  {journal} {Solid State Commun.}\ }\textbf {\bibinfo
  {volume} {19}},\ \bibinfo {pages} {525} (\bibinfo {year} {1976})}\BibitemShut
  {NoStop}%
\bibitem [{\citenamefont {Motoya}\ \emph {et~al.}(1976)\citenamefont {Motoya},
  \citenamefont {Yasuoka}, \citenamefont {Nakamura},\ and\ \citenamefont
  {Wernick}}]{1976:Motoya:SolidStateCommun}%
  \BibitemOpen
  \bibfield  {author} {\bibinfo {author} {\bibfnamefont {K.}~\bibnamefont
  {Motoya}}, \bibinfo {author} {\bibfnamefont {H.}~\bibnamefont {Yasuoka}},
  \bibinfo {author} {\bibfnamefont {Y.}~\bibnamefont {Nakamura}}, \ and\
  \bibinfo {author} {\bibfnamefont {J.H.}\ \bibnamefont {Wernick}},\ }\bibfield
   {title} {\enquote {\bibinfo {title} {{Helical spin structure in MnSi-NMR
  studies}},}\ }\href {\doibase 10.1016/0038-1098(76)90058-2} {\bibfield
  {journal} {\bibinfo  {journal} {Solid State Commun.}\ }\textbf {\bibinfo
  {volume} {19}},\ \bibinfo {pages} {529} (\bibinfo {year} {1976})}\BibitemShut
  {NoStop}%
\bibitem [{\citenamefont {Shirane}\ \emph {et~al.}(1983)\citenamefont
  {Shirane}, \citenamefont {Cowley}, \citenamefont {Majkrzak}, \citenamefont
  {Sokoloff}, \citenamefont {Pagonis}, \citenamefont {Perry},\ and\
  \citenamefont {Ishikawa}}]{1983:Shirane:PhysRevB}%
  \BibitemOpen
  \bibfield  {author} {\bibinfo {author} {\bibfnamefont {G.}~\bibnamefont
  {Shirane}}, \bibinfo {author} {\bibfnamefont {R.}~\bibnamefont {Cowley}},
  \bibinfo {author} {\bibfnamefont {C.}~\bibnamefont {Majkrzak}}, \bibinfo
  {author} {\bibfnamefont {J.~B.}\ \bibnamefont {Sokoloff}}, \bibinfo {author}
  {\bibfnamefont {B.}~\bibnamefont {Pagonis}}, \bibinfo {author} {\bibfnamefont
  {C.~H.}\ \bibnamefont {Perry}}, \ and\ \bibinfo {author} {\bibfnamefont
  {Y.}~\bibnamefont {Ishikawa}},\ }\bibfield  {title} {\enquote {\bibinfo
  {title} {{Spiral magnetic correlation in cubic MnSi}},}\ }\href {\doibase
  10.1103/PhysRevB.28.6251} {\bibfield  {journal} {\bibinfo  {journal} {Phys.
  Rev. B}\ }\textbf {\bibinfo {volume} {28}},\ \bibinfo {pages} {6251}
  (\bibinfo {year} {1983})}\BibitemShut {NoStop}%
\bibitem [{\citenamefont {Ishida}\ \emph {et~al.}(1985)\citenamefont {Ishida},
  \citenamefont {Endoh}, \citenamefont {Mitsuda}, \citenamefont {Ishikawa},\
  and\ \citenamefont {Tanaka}}]{1985:Ishida:JPhysSocJpn}%
  \BibitemOpen
  \bibfield  {author} {\bibinfo {author} {\bibfnamefont {M.}~\bibnamefont
  {Ishida}}, \bibinfo {author} {\bibfnamefont {Y.}~\bibnamefont {Endoh}},
  \bibinfo {author} {\bibfnamefont {S.}~\bibnamefont {Mitsuda}}, \bibinfo
  {author} {\bibfnamefont {Y.}~\bibnamefont {Ishikawa}}, \ and\ \bibinfo
  {author} {\bibfnamefont {M.}~\bibnamefont {Tanaka}},\ }\bibfield  {title}
  {\enquote {\bibinfo {title} {{Crystal Chirality and Helicity of the Helical
  Spin Density Wave in MnSi. II. Polarized Neutron Diffraction}},}\ }\href
  {\doibase 10.1143/JPSJ.54.2975} {\bibfield  {journal} {\bibinfo  {journal}
  {J. Phys. Soc. Jpn.}\ }\textbf {\bibinfo {volume} {54}},\ \bibinfo {pages}
  {2975} (\bibinfo {year} {1985})}\BibitemShut {NoStop}%
\bibitem [{\citenamefont {Bak}\ and\ \citenamefont
  {Jensen}(1980)}]{1980:Bak:JPhysCSolidState}%
  \BibitemOpen
  \bibfield  {author} {\bibinfo {author} {\bibfnamefont {P.}~\bibnamefont
  {Bak}}\ and\ \bibinfo {author} {\bibfnamefont {M.~H.}\ \bibnamefont
  {Jensen}},\ }\bibfield  {title} {\enquote {\bibinfo {title} {{Theory of
  helical magnetic structures and phase transitions in MnSi and FeGe}},}\
  }\href {\doibase 10.1088/0022-3719/13/31/002} {\bibfield  {journal} {\bibinfo
   {journal} {J. Phys. C: Solid State}\ }\textbf {\bibinfo {volume} {13}},\
  \bibinfo {pages} {L881} (\bibinfo {year} {1980})}\BibitemShut {NoStop}%
\bibitem [{\citenamefont {Ishikawa}\ \emph {et~al.}(1985)\citenamefont
  {Ishikawa}, \citenamefont {Noda}, \citenamefont {Uemura}, \citenamefont
  {Majkrzak},\ and\ \citenamefont {Shirane}}]{1985:Ishikawa:PhysRevB}%
  \BibitemOpen
  \bibfield  {author} {\bibinfo {author} {\bibfnamefont {Y.}~\bibnamefont
  {Ishikawa}}, \bibinfo {author} {\bibfnamefont {Y.}~\bibnamefont {Noda}},
  \bibinfo {author} {\bibfnamefont {Y.~J.}\ \bibnamefont {Uemura}}, \bibinfo
  {author} {\bibfnamefont {C.~F.}\ \bibnamefont {Majkrzak}}, \ and\ \bibinfo
  {author} {\bibfnamefont {G.}~\bibnamefont {Shirane}},\ }\bibfield  {title}
  {\enquote {\bibinfo {title} {{Paramagnetic spin fluctuations in the weak
  itinerant-electron ferromagnet MnSi}},}\ }\href {\doibase
  10.1103/PhysRevB.31.5884} {\bibfield  {journal} {\bibinfo  {journal} {Phys.
  Rev. B}\ }\textbf {\bibinfo {volume} {31}},\ \bibinfo {pages} {5884}
  (\bibinfo {year} {1985})}\BibitemShut {NoStop}%
\bibitem [{\citenamefont {Lebech}(1993)}]{1993:Lebech:Book}%
  \BibitemOpen
  \bibfield  {author} {\bibinfo {author} {\bibfnamefont {B.}~\bibnamefont
  {Lebech}},\ }\href@noop {} {\emph {\bibinfo {title} {{Recent Advances in
  Magnetism of Transition Metal Compounds}}}}\ (\bibinfo  {publisher} {World
  Scientific, Singapore},\ \bibinfo {year} {1993})\ p.\ \bibinfo {pages}
  {167}\BibitemShut {NoStop}%
\bibitem [{\citenamefont {Bloch}\ \emph {et~al.}(1975)\citenamefont {Bloch},
  \citenamefont {Voiron}, \citenamefont {Jaccarino},\ and\ \citenamefont
  {Wernick}}]{1975:Bloch:PhysLettA}%
  \BibitemOpen
  \bibfield  {author} {\bibinfo {author} {\bibfnamefont {D.}~\bibnamefont
  {Bloch}}, \bibinfo {author} {\bibfnamefont {J.}~\bibnamefont {Voiron}},
  \bibinfo {author} {\bibfnamefont {V.}~\bibnamefont {Jaccarino}}, \ and\
  \bibinfo {author} {\bibfnamefont {J.~H.}\ \bibnamefont {Wernick}},\
  }\bibfield  {title} {\enquote {\bibinfo {title} {{The high field-high
  pressure magnetic properties of MnSi}},}\ }\href {\doibase
  http://dx.doi.org/10.1016/0375-9601(75)90438-7} {\bibfield  {journal}
  {\bibinfo  {journal} {Phys. Lett. A}\ }\textbf {\bibinfo {volume} {51}},\
  \bibinfo {pages} {259} (\bibinfo {year} {1975})}\BibitemShut {NoStop}%
\bibitem [{\citenamefont {Fawcett}\ \emph {et~al.}(1970)\citenamefont
  {Fawcett}, \citenamefont {Maita},\ and\ \citenamefont
  {Wernick}}]{1970:Fawcett:InternJMagnetism}%
  \BibitemOpen
  \bibfield  {author} {\bibinfo {author} {\bibfnamefont {E.}~\bibnamefont
  {Fawcett}}, \bibinfo {author} {\bibfnamefont {J.~P.}\ \bibnamefont {Maita}},
  \ and\ \bibinfo {author} {\bibfnamefont {J.~H.}\ \bibnamefont {Wernick}},\
  }\bibfield  {title} {\enquote {\bibinfo {title} {{Magnetoelastic and thermal
  properties of MnSi}},}\ }\href@noop {} {\bibfield  {journal} {\bibinfo
  {journal} {Int. J. Magn.}\ }\textbf {\bibinfo {volume} {1}},\ \bibinfo
  {pages} {29} (\bibinfo {year} {1970})}\BibitemShut {NoStop}%
\bibitem [{\citenamefont {Kusaka}\ \emph {et~al.}(1976)\citenamefont {Kusaka},
  \citenamefont {Yamamoto}, \citenamefont {Komatsubara},\ and\ \citenamefont
  {Ishikawa}}]{1976:Kusaka:SolidStateCommun}%
  \BibitemOpen
  \bibfield  {author} {\bibinfo {author} {\bibfnamefont {S.}~\bibnamefont
  {Kusaka}}, \bibinfo {author} {\bibfnamefont {K.}~\bibnamefont {Yamamoto}},
  \bibinfo {author} {\bibfnamefont {T.}~\bibnamefont {Komatsubara}}, \ and\
  \bibinfo {author} {\bibfnamefont {Y.}~\bibnamefont {Ishikawa}},\ }\bibfield
  {title} {\enquote {\bibinfo {title} {{Ultrasonic study of magnetic phase
  diagram of MnSi}},}\ }\href {\doibase 10.1016/0038-1098(76)91307-7}
  {\bibfield  {journal} {\bibinfo  {journal} {Solid State Commun.}\ }\textbf
  {\bibinfo {volume} {20}},\ \bibinfo {pages} {925} (\bibinfo {year}
  {1976})}\BibitemShut {NoStop}%
\bibitem [{\citenamefont {Adams}\ \emph {et~al.}(2011)\citenamefont {Adams},
  \citenamefont {M\"{u}hlbauer}, \citenamefont {Pfleiderer}, \citenamefont
  {Jonietz}, \citenamefont {Bauer}, \citenamefont {Neubauer}, \citenamefont
  {Georgii}, \citenamefont {B\"{o}ni}, \citenamefont {Keiderling},
  \citenamefont {Everschor}, \citenamefont {Garst},\ and\ \citenamefont
  {Rosch}}]{2011:Adams:PhysRevLett}%
  \BibitemOpen
  \bibfield  {author} {\bibinfo {author} {\bibfnamefont {T.}~\bibnamefont
  {Adams}}, \bibinfo {author} {\bibfnamefont {S.}~\bibnamefont
  {M\"{u}hlbauer}}, \bibinfo {author} {\bibfnamefont {C.}~\bibnamefont
  {Pfleiderer}}, \bibinfo {author} {\bibfnamefont {F.}~\bibnamefont {Jonietz}},
  \bibinfo {author} {\bibfnamefont {A.}~\bibnamefont {Bauer}}, \bibinfo
  {author} {\bibfnamefont {A.}~\bibnamefont {Neubauer}}, \bibinfo {author}
  {\bibfnamefont {R.}~\bibnamefont {Georgii}}, \bibinfo {author} {\bibfnamefont
  {P.}~\bibnamefont {B\"{o}ni}}, \bibinfo {author} {\bibfnamefont
  {U.}~\bibnamefont {Keiderling}}, \bibinfo {author} {\bibfnamefont
  {K.}~\bibnamefont {Everschor}}, \bibinfo {author} {\bibfnamefont
  {M.}~\bibnamefont {Garst}}, \ and\ \bibinfo {author} {\bibfnamefont
  {A.}~\bibnamefont {Rosch}},\ }\bibfield  {title} {\enquote {\bibinfo {title}
  {{Long-Range Crystalline Nature of the Skyrmion Lattice in MnSi}},}\ }\href
  {\doibase 10.1103/PhysRevLett.107.217206} {\bibfield  {journal} {\bibinfo
  {journal} {Phys. Rev. Lett.}\ }\textbf {\bibinfo {volume} {107}},\ \bibinfo
  {pages} {217206} (\bibinfo {year} {2011})}\BibitemShut {NoStop}%
\bibitem [{\citenamefont {Yu}\ \emph {et~al.}(2010)\citenamefont {Yu},
  \citenamefont {Onose}, \citenamefont {Kanazawa}, \citenamefont {Park},
  \citenamefont {Han}, \citenamefont {Matsui}, \citenamefont {Nagaosa},\ and\
  \citenamefont {Tokura}}]{2010:Yu:Nature}%
  \BibitemOpen
  \bibfield  {author} {\bibinfo {author} {\bibfnamefont {X.~Z.}\ \bibnamefont
  {Yu}}, \bibinfo {author} {\bibfnamefont {Y.}~\bibnamefont {Onose}}, \bibinfo
  {author} {\bibfnamefont {N.}~\bibnamefont {Kanazawa}}, \bibinfo {author}
  {\bibfnamefont {J.~H.}\ \bibnamefont {Park}}, \bibinfo {author}
  {\bibfnamefont {J.~H.}\ \bibnamefont {Han}}, \bibinfo {author} {\bibfnamefont
  {Y.}~\bibnamefont {Matsui}}, \bibinfo {author} {\bibfnamefont
  {N.}~\bibnamefont {Nagaosa}}, \ and\ \bibinfo {author} {\bibfnamefont
  {Y.}~\bibnamefont {Tokura}},\ }\bibfield  {title} {\enquote {\bibinfo {title}
  {{Real-space observation of a two-dimensional skyrmion crystal}},}\ }\href
  {\doibase 10.1038/nature09124} {\bibfield  {journal} {\bibinfo  {journal}
  {Nature (London)}\ }\textbf {\bibinfo {volume} {465}},\ \bibinfo {pages}
  {901} (\bibinfo {year} {2010})}\BibitemShut {NoStop}%
\bibitem [{\citenamefont {Pfleiderer}\ \emph {et~al.}(2010)\citenamefont
  {Pfleiderer}, \citenamefont {Adams}, \citenamefont {Bauer}, \citenamefont
  {Biberacher}, \citenamefont {Binz}, \citenamefont {Birkelbach}, \citenamefont
  {B\"{o}ni}, \citenamefont {Franz}, \citenamefont {Georgii}, \citenamefont
  {Janoschek}, \citenamefont {Jonietz}, \citenamefont {Keller}, \citenamefont
  {Ritz}, \citenamefont {M\"{u}hlbauer}, \citenamefont {M\"{u}nzer},
  \citenamefont {Neubauer}, \citenamefont {Pedersen},\ and\ \citenamefont
  {Rosch}}]{2010:Pfleiderer:JPhysCondensMatter}%
  \BibitemOpen
  \bibfield  {author} {\bibinfo {author} {\bibfnamefont {C.}~\bibnamefont
  {Pfleiderer}}, \bibinfo {author} {\bibfnamefont {T.}~\bibnamefont {Adams}},
  \bibinfo {author} {\bibfnamefont {A.}~\bibnamefont {Bauer}}, \bibinfo
  {author} {\bibfnamefont {W.}~\bibnamefont {Biberacher}}, \bibinfo {author}
  {\bibfnamefont {B.}~\bibnamefont {Binz}}, \bibinfo {author} {\bibfnamefont
  {F.}~\bibnamefont {Birkelbach}}, \bibinfo {author} {\bibfnamefont
  {P.}~\bibnamefont {B\"{o}ni}}, \bibinfo {author} {\bibfnamefont
  {C.}~\bibnamefont {Franz}}, \bibinfo {author} {\bibfnamefont
  {R.}~\bibnamefont {Georgii}}, \bibinfo {author} {\bibfnamefont
  {M.}~\bibnamefont {Janoschek}}, \bibinfo {author} {\bibfnamefont
  {F.}~\bibnamefont {Jonietz}}, \bibinfo {author} {\bibfnamefont
  {T.}~\bibnamefont {Keller}}, \bibinfo {author} {\bibfnamefont
  {R.}~\bibnamefont {Ritz}}, \bibinfo {author} {\bibfnamefont {S.}~\bibnamefont
  {M\"{u}hlbauer}}, \bibinfo {author} {\bibfnamefont {W.}~\bibnamefont
  {M\"{u}nzer}}, \bibinfo {author} {\bibfnamefont {A.}~\bibnamefont
  {Neubauer}}, \bibinfo {author} {\bibfnamefont {B.}~\bibnamefont {Pedersen}},
  \ and\ \bibinfo {author} {\bibfnamefont {A.}~\bibnamefont {Rosch}},\
  }\bibfield  {title} {\enquote {\bibinfo {title} {{Skyrmion lattices in
  metallic and semiconducting B20 transition metal compounds}},}\ }\href
  {\doibase 10.1088/0953-8984/22/16/164207} {\bibfield  {journal} {\bibinfo
  {journal} {J. Phys.: Condens. Matter}\ }\textbf {\bibinfo {volume} {22}},\
  \bibinfo {pages} {164207} (\bibinfo {year} {2010})}\BibitemShut {NoStop}%
\bibitem [{\citenamefont {Adams}\ \emph {et~al.}(2012)\citenamefont {Adams},
  \citenamefont {Chacon}, \citenamefont {Wagner}, \citenamefont {Bauer},
  \citenamefont {Brandl}, \citenamefont {Pedersen}, \citenamefont {Berger},
  \citenamefont {Lemmens},\ and\ \citenamefont
  {Pfleiderer}}]{2012:Adams:PhysRevLett}%
  \BibitemOpen
  \bibfield  {author} {\bibinfo {author} {\bibfnamefont {T.}~\bibnamefont
  {Adams}}, \bibinfo {author} {\bibfnamefont {A.}~\bibnamefont {Chacon}},
  \bibinfo {author} {\bibfnamefont {M.}~\bibnamefont {Wagner}}, \bibinfo
  {author} {\bibfnamefont {A.}~\bibnamefont {Bauer}}, \bibinfo {author}
  {\bibfnamefont {G.}~\bibnamefont {Brandl}}, \bibinfo {author} {\bibfnamefont
  {B.}~\bibnamefont {Pedersen}}, \bibinfo {author} {\bibfnamefont
  {H.}~\bibnamefont {Berger}}, \bibinfo {author} {\bibfnamefont
  {P.}~\bibnamefont {Lemmens}}, \ and\ \bibinfo {author} {\bibfnamefont
  {C.}~\bibnamefont {Pfleiderer}},\ }\bibfield  {title} {\enquote {\bibinfo
  {title} {{Long-Wavelength Helimagnetic Order and Skyrmion Lattice Phase in
  Cu$_{2}$OSeO$_{3}$}},}\ }\href {\doibase 10.1103/PhysRevLett.108.237204}
  {\bibfield  {journal} {\bibinfo  {journal} {Phys. Rev. Lett.}\ }\textbf
  {\bibinfo {volume} {108}},\ \bibinfo {pages} {237204} (\bibinfo {year}
  {2012})}\BibitemShut {NoStop}%
\bibitem [{\citenamefont {Moskvin}\ \emph {et~al.}(2013)\citenamefont
  {Moskvin}, \citenamefont {Grigoriev}, \citenamefont {Dyadkin}, \citenamefont
  {Eckerlebe}, \citenamefont {Baenitz}, \citenamefont {Schmidt},\ and\
  \citenamefont {Wilhelm}}]{2013:Moskvin:PhysRevLett}%
  \BibitemOpen
  \bibfield  {author} {\bibinfo {author} {\bibfnamefont {E.}~\bibnamefont
  {Moskvin}}, \bibinfo {author} {\bibfnamefont {S.}~\bibnamefont {Grigoriev}},
  \bibinfo {author} {\bibfnamefont {V.}~\bibnamefont {Dyadkin}}, \bibinfo
  {author} {\bibfnamefont {H.}~\bibnamefont {Eckerlebe}}, \bibinfo {author}
  {\bibfnamefont {M.}~\bibnamefont {Baenitz}}, \bibinfo {author} {\bibfnamefont
  {M.}~\bibnamefont {Schmidt}}, \ and\ \bibinfo {author} {\bibfnamefont
  {H.}~\bibnamefont {Wilhelm}},\ }\bibfield  {title} {\enquote {\bibinfo
  {title} {{Complex Chiral Modulations in FeGe Close to Magnetic Ordering}},}\
  }\href {\doibase 10.1103/PhysRevLett.110.077207} {\bibfield  {journal}
  {\bibinfo  {journal} {Phys. Rev. Lett.}\ }\textbf {\bibinfo {volume} {110}},\
  \bibinfo {pages} {077207} (\bibinfo {year} {2013})}\BibitemShut {NoStop}%
\bibitem [{\citenamefont {Yu}\ \emph {et~al.}(2011)\citenamefont {Yu},
  \citenamefont {Kanazawa}, \citenamefont {Onose}, \citenamefont {Kimoto},
  \citenamefont {Zhang}, \citenamefont {Ishiwata}, \citenamefont {Matsui},\
  and\ \citenamefont {Tokura}}]{2011:Yu:NatureMater}%
  \BibitemOpen
  \bibfield  {author} {\bibinfo {author} {\bibfnamefont {X.~Z.}\ \bibnamefont
  {Yu}}, \bibinfo {author} {\bibfnamefont {N.}~\bibnamefont {Kanazawa}},
  \bibinfo {author} {\bibfnamefont {Y.}~\bibnamefont {Onose}}, \bibinfo
  {author} {\bibfnamefont {K.}~\bibnamefont {Kimoto}}, \bibinfo {author}
  {\bibfnamefont {W.~Z.}\ \bibnamefont {Zhang}}, \bibinfo {author}
  {\bibfnamefont {S.}~\bibnamefont {Ishiwata}}, \bibinfo {author}
  {\bibfnamefont {Y.}~\bibnamefont {Matsui}}, \ and\ \bibinfo {author}
  {\bibfnamefont {Y.}~\bibnamefont {Tokura}},\ }\bibfield  {title} {\enquote
  {\bibinfo {title} {{Near room-temperature formation of a skyrmion crystal in
  thin-films of the helimagnet FeGe}},}\ }\href {\doibase 10.1038/nmat2916}
  {\bibfield  {journal} {\bibinfo  {journal} {Nature Mater.}\ }\textbf
  {\bibinfo {volume} {10}},\ \bibinfo {pages} {106} (\bibinfo {year}
  {2011})}\BibitemShut {NoStop}%
\bibitem [{\citenamefont {Tonomura}\ \emph {et~al.}(2012)\citenamefont
  {Tonomura}, \citenamefont {Yu}, \citenamefont {Yanagisawa}, \citenamefont
  {Matsuda}, \citenamefont {Onose}, \citenamefont {Kanazawa}, \citenamefont
  {Park},\ and\ \citenamefont {Tokura}}]{2012:Tonomura:NanoLett}%
  \BibitemOpen
  \bibfield  {author} {\bibinfo {author} {\bibfnamefont {A.}~\bibnamefont
  {Tonomura}}, \bibinfo {author} {\bibfnamefont {X.}~\bibnamefont {Yu}},
  \bibinfo {author} {\bibfnamefont {K.}~\bibnamefont {Yanagisawa}}, \bibinfo
  {author} {\bibfnamefont {T.}~\bibnamefont {Matsuda}}, \bibinfo {author}
  {\bibfnamefont {Y.}~\bibnamefont {Onose}}, \bibinfo {author} {\bibfnamefont
  {N.}~\bibnamefont {Kanazawa}}, \bibinfo {author} {\bibfnamefont {H.~S.}\
  \bibnamefont {Park}}, \ and\ \bibinfo {author} {\bibfnamefont
  {Y.}~\bibnamefont {Tokura}},\ }\bibfield  {title} {\enquote {\bibinfo {title}
  {{Real-Space Observation of Skyrmion Lattice in Helimagnet MnSi Thin
  Samples}},}\ }\href {\doibase 10.1021/nl300073m} {\bibfield  {journal}
  {\bibinfo  {journal} {Nano Lett.}\ }\textbf {\bibinfo {volume} {12}},\
  \bibinfo {pages} {1673} (\bibinfo {year} {2012})}\BibitemShut {NoStop}%
\bibitem [{\citenamefont {Park}\ \emph {et~al.}(2014)\citenamefont {Park},
  \citenamefont {Yu}, \citenamefont {Aizawa}, \citenamefont {Tanigaki},
  \citenamefont {Akashi}, \citenamefont {Takahashi}, \citenamefont {Matsuda},
  \citenamefont {Kanazawa}, \citenamefont {Onose}, \citenamefont {Shindo},
  \citenamefont {Tonomura},\ and\ \citenamefont
  {Tokura}}]{2014:Park:NatureNano}%
  \BibitemOpen
  \bibfield  {author} {\bibinfo {author} {\bibfnamefont {H.~S.}\ \bibnamefont
  {Park}}, \bibinfo {author} {\bibfnamefont {X.}~\bibnamefont {Yu}}, \bibinfo
  {author} {\bibfnamefont {S.}~\bibnamefont {Aizawa}}, \bibinfo {author}
  {\bibfnamefont {T.}~\bibnamefont {Tanigaki}}, \bibinfo {author}
  {\bibfnamefont {T.}~\bibnamefont {Akashi}}, \bibinfo {author} {\bibfnamefont
  {Y.}~\bibnamefont {Takahashi}}, \bibinfo {author} {\bibfnamefont
  {T.}~\bibnamefont {Matsuda}}, \bibinfo {author} {\bibfnamefont
  {N.}~\bibnamefont {Kanazawa}}, \bibinfo {author} {\bibfnamefont
  {Y.}~\bibnamefont {Onose}}, \bibinfo {author} {\bibfnamefont
  {D.}~\bibnamefont {Shindo}}, \bibinfo {author} {\bibfnamefont
  {A.}~\bibnamefont {Tonomura}}, \ and\ \bibinfo {author} {\bibfnamefont
  {Y.}~\bibnamefont {Tokura}},\ }\bibfield  {title} {\enquote {\bibinfo {title}
  {{Observation of the magnetic flux and three-dimensional structure of
  skyrmion lattices by electron holography}},}\ }\href {\doibase
  10.1038/nnano.2014.52} {\bibfield  {journal} {\bibinfo  {journal} {Nature
  Nano.}\ }\textbf {\bibinfo {volume} {9}},\ \bibinfo {pages} {337} (\bibinfo
  {year} {2014})}\BibitemShut {NoStop}%
\bibitem [{\citenamefont {Uchida}\ \emph {et~al.}(2006)\citenamefont {Uchida},
  \citenamefont {Onose}, \citenamefont {Matsui},\ and\ \citenamefont
  {Tokura}}]{2006:Uchida:Science}%
  \BibitemOpen
  \bibfield  {author} {\bibinfo {author} {\bibfnamefont {M.}~\bibnamefont
  {Uchida}}, \bibinfo {author} {\bibfnamefont {Y.}~\bibnamefont {Onose}},
  \bibinfo {author} {\bibfnamefont {Y.}~\bibnamefont {Matsui}}, \ and\ \bibinfo
  {author} {\bibfnamefont {Y.}~\bibnamefont {Tokura}},\ }\bibfield  {title}
  {\enquote {\bibinfo {title} {{Real-Space Observation of Helical Spin
  Order}},}\ }\href {\doibase 10.1126/science.1120639} {\bibfield  {journal}
  {\bibinfo  {journal} {Science}\ }\textbf {\bibinfo {volume} {311}},\ \bibinfo
  {pages} {359} (\bibinfo {year} {2006})}\BibitemShut {NoStop}%
\bibitem [{\citenamefont {Ishikawa}\ and\ \citenamefont
  {Arai}(1984)}]{1984:Ishikawa:JPhysSocJpn}%
  \BibitemOpen
  \bibfield  {author} {\bibinfo {author} {\bibfnamefont {Y.}~\bibnamefont
  {Ishikawa}}\ and\ \bibinfo {author} {\bibfnamefont {M.}~\bibnamefont
  {Arai}},\ }\bibfield  {title} {\enquote {\bibinfo {title} {{Magnetic Phase
  Diagram of MnSi near Critical Temperature Studied by Neutron Small Angle
  Scattering}},}\ }\href {\doibase 10.1143/JPSJ.53.2726} {\bibfield  {journal}
  {\bibinfo  {journal} {J. Phys. Soc. Jpn.}\ }\textbf {\bibinfo {volume}
  {53}},\ \bibinfo {pages} {2726} (\bibinfo {year} {1984})}\BibitemShut
  {NoStop}%
\bibitem [{\citenamefont {Lebech}\ \emph {et~al.}(1995)\citenamefont {Lebech},
  \citenamefont {Harris}, \citenamefont {Pedersen}, \citenamefont {Mortensen},
  \citenamefont {Gregory}, \citenamefont {Bernhoeft}, \citenamefont {Jermy},\
  and\ \citenamefont {Brown}}]{1995:Lebech:JMagnMagnMater}%
  \BibitemOpen
  \bibfield  {author} {\bibinfo {author} {\bibfnamefont {B.}~\bibnamefont
  {Lebech}}, \bibinfo {author} {\bibfnamefont {P.}~\bibnamefont {Harris}},
  \bibinfo {author} {\bibfnamefont {J.~S.}\ \bibnamefont {Pedersen}}, \bibinfo
  {author} {\bibfnamefont {K.}~\bibnamefont {Mortensen}}, \bibinfo {author}
  {\bibfnamefont {C.~I.}\ \bibnamefont {Gregory}}, \bibinfo {author}
  {\bibfnamefont {N.~R.}\ \bibnamefont {Bernhoeft}}, \bibinfo {author}
  {\bibfnamefont {M.}~\bibnamefont {Jermy}}, \ and\ \bibinfo {author}
  {\bibfnamefont {S.~A.}\ \bibnamefont {Brown}},\ }\bibfield  {title} {\enquote
  {\bibinfo {title} {{Magnetic phase diagram of MnSi}},}\ }\href {\doibase
  10.1016/0304-8853(94)01115-X} {\bibfield  {journal} {\bibinfo  {journal} {J.
  Magn. Magn. Mater.}\ }\textbf {\bibinfo {volume} {140--144}},\ \bibinfo
  {pages} {119} (\bibinfo {year} {1995})}\BibitemShut {NoStop}%
\bibitem [{\citenamefont {Grigoriev}\ \emph {et~al.}(2006)\citenamefont
  {Grigoriev}, \citenamefont {Maleyev}, \citenamefont {Okorokov}, \citenamefont
  {Chetverikov},\ and\ \citenamefont {Eckerlebe}}]{2006:Grigoriev:PhysRevB}%
  \BibitemOpen
  \bibfield  {author} {\bibinfo {author} {\bibfnamefont {S.~V.}\ \bibnamefont
  {Grigoriev}}, \bibinfo {author} {\bibfnamefont {S.~V.}\ \bibnamefont
  {Maleyev}}, \bibinfo {author} {\bibfnamefont {A.~I.}\ \bibnamefont
  {Okorokov}}, \bibinfo {author} {\bibfnamefont {Yu.~O.}\ \bibnamefont
  {Chetverikov}}, \ and\ \bibinfo {author} {\bibfnamefont {H.}~\bibnamefont
  {Eckerlebe}},\ }\bibfield  {title} {\enquote {\bibinfo {title}
  {{Field-induced reorientation of the spin helix in MnSi near $T_{c}$}},}\
  }\href {\doibase 10.1103/PhysRevB.73.224440} {\bibfield  {journal} {\bibinfo
  {journal} {Phys. Rev. B}\ }\textbf {\bibinfo {volume} {73}},\ \bibinfo
  {pages} {224440} (\bibinfo {year} {2006})}\BibitemShut {NoStop}%
\bibitem [{\citenamefont {Huang}\ and\ \citenamefont
  {Chien}(2012)}]{2012:Huang:PhysRevLett}%
  \BibitemOpen
  \bibfield  {author} {\bibinfo {author} {\bibfnamefont {S.~X.}\ \bibnamefont
  {Huang}}\ and\ \bibinfo {author} {\bibfnamefont {C.~L.}\ \bibnamefont
  {Chien}},\ }\bibfield  {title} {\enquote {\bibinfo {title} {{Extended
  Skyrmion Phase in Epitaxial FeGe$(111)$ Thin Films}},}\ }\href {\doibase
  10.1103/PhysRevLett.108.267201} {\bibfield  {journal} {\bibinfo  {journal}
  {Phys. Rev. Lett.}\ }\textbf {\bibinfo {volume} {108}},\ \bibinfo {pages}
  {267201} (\bibinfo {year} {2012})}\BibitemShut {NoStop}%
\bibitem [{\citenamefont {Li}\ \emph {et~al.}(2013)\citenamefont {Li},
  \citenamefont {Kanazawa}, \citenamefont {Yu}, \citenamefont {Tsukazaki},
  \citenamefont {Kawasaki}, \citenamefont {Ichikawa}, \citenamefont {Jin},
  \citenamefont {Kagawa},\ and\ \citenamefont {Tokura}}]{2013:Li:PhysRevLett}%
  \BibitemOpen
  \bibfield  {author} {\bibinfo {author} {\bibfnamefont {Y.}~\bibnamefont
  {Li}}, \bibinfo {author} {\bibfnamefont {N.}~\bibnamefont {Kanazawa}},
  \bibinfo {author} {\bibfnamefont {X.~Z.}\ \bibnamefont {Yu}}, \bibinfo
  {author} {\bibfnamefont {A.}~\bibnamefont {Tsukazaki}}, \bibinfo {author}
  {\bibfnamefont {M.}~\bibnamefont {Kawasaki}}, \bibinfo {author}
  {\bibfnamefont {M.}~\bibnamefont {Ichikawa}}, \bibinfo {author}
  {\bibfnamefont {X.~F.}\ \bibnamefont {Jin}}, \bibinfo {author} {\bibfnamefont
  {F.}~\bibnamefont {Kagawa}}, \ and\ \bibinfo {author} {\bibfnamefont
  {Y.}~\bibnamefont {Tokura}},\ }\bibfield  {title} {\enquote {\bibinfo {title}
  {{Robust Formation of Skyrmions and Topological Hall Effect Anomaly in
  Epitaxial Thin Films of MnSi}},}\ }\href {\doibase
  10.1103/PhysRevLett.110.117202} {\bibfield  {journal} {\bibinfo  {journal}
  {Phys. Rev. Lett.}\ }\textbf {\bibinfo {volume} {110}},\ \bibinfo {pages}
  {117202} (\bibinfo {year} {2013})}\BibitemShut {NoStop}%
\bibitem [{\citenamefont {Sinha}\ \emph {et~al.}(2014)\citenamefont {Sinha},
  \citenamefont {Porter},\ and\ \citenamefont {Marrows}}]{2014:Sinha:PhysRevB}%
  \BibitemOpen
  \bibfield  {author} {\bibinfo {author} {\bibfnamefont {P.}~\bibnamefont
  {Sinha}}, \bibinfo {author} {\bibfnamefont {N.~A.}\ \bibnamefont {Porter}}, \
  and\ \bibinfo {author} {\bibfnamefont {C.~H.}\ \bibnamefont {Marrows}},\
  }\bibfield  {title} {\enquote {\bibinfo {title} {{Strain-induced effects on
  the magnetic and electronic properties of epitaxial Fe$_{1-x}$Co$_{x}$Si thin
  films}},}\ }\href {\doibase 10.1103/PhysRevB.89.134426} {\bibfield  {journal}
  {\bibinfo  {journal} {Phys. Rev. B}\ }\textbf {\bibinfo {volume} {89}},\
  \bibinfo {pages} {134426} (\bibinfo {year} {2014})}\BibitemShut {NoStop}%
\bibitem [{\citenamefont {Wilson}\ \emph {et~al.}(2014)\citenamefont {Wilson},
  \citenamefont {Butenko}, \citenamefont {Bogdanov},\ and\ \citenamefont
  {Monchesky}}]{2014:Wilson:PhysRevB}%
  \BibitemOpen
  \bibfield  {author} {\bibinfo {author} {\bibfnamefont {M.~N.}\ \bibnamefont
  {Wilson}}, \bibinfo {author} {\bibfnamefont {A.~B.}\ \bibnamefont {Butenko}},
  \bibinfo {author} {\bibfnamefont {A.~N.}\ \bibnamefont {Bogdanov}}, \ and\
  \bibinfo {author} {\bibfnamefont {T.~L.}\ \bibnamefont {Monchesky}},\
  }\bibfield  {title} {\enquote {\bibinfo {title} {{Chiral skyrmions in cubic
  helimagnet films: The role of uniaxial anisotropy}},}\ }\href {\doibase
  10.1103/PhysRevB.89.094411} {\bibfield  {journal} {\bibinfo  {journal} {Phys.
  Rev. B}\ }\textbf {\bibinfo {volume} {89}},\ \bibinfo {pages} {094411}
  (\bibinfo {year} {2014})}\BibitemShut {NoStop}%
\bibitem [{\citenamefont {Brazovskii}(1975)}]{1975:Brazovskii:SovPhysJETP}%
  \BibitemOpen
  \bibfield  {author} {\bibinfo {author} {\bibfnamefont {S.~A.}\ \bibnamefont
  {Brazovskii}},\ }\bibfield  {title} {\enquote {\bibinfo {title} {{Phase
  transition of an isotropic system to a nonuniform state}},}\ }\href
  {http://www.jetp.ac.ru/cgi-bin/e/index/e/41/1/p85?a=list} {\bibfield
  {journal} {\bibinfo  {journal} {Sov. Phys. JETP}\ }\textbf {\bibinfo {volume}
  {41}},\ \bibinfo {pages} {85} (\bibinfo {year} {1975})}\BibitemShut {NoStop}%
\bibitem [{\citenamefont {Janoschek}\ \emph {et~al.}(2013)\citenamefont
  {Janoschek}, \citenamefont {Garst}, \citenamefont {Bauer}, \citenamefont
  {Krautscheid}, \citenamefont {Georgii}, \citenamefont {B\"{o}ni},\ and\
  \citenamefont {Pfleiderer}}]{2013:Janoschek:PhysRevB}%
  \BibitemOpen
  \bibfield  {author} {\bibinfo {author} {\bibfnamefont {M.}~\bibnamefont
  {Janoschek}}, \bibinfo {author} {\bibfnamefont {M.}~\bibnamefont {Garst}},
  \bibinfo {author} {\bibfnamefont {A.}~\bibnamefont {Bauer}}, \bibinfo
  {author} {\bibfnamefont {P.}~\bibnamefont {Krautscheid}}, \bibinfo {author}
  {\bibfnamefont {R.}~\bibnamefont {Georgii}}, \bibinfo {author} {\bibfnamefont
  {P.}~\bibnamefont {B\"{o}ni}}, \ and\ \bibinfo {author} {\bibfnamefont
  {C.}~\bibnamefont {Pfleiderer}},\ }\bibfield  {title} {\enquote {\bibinfo
  {title} {{Fluctuation-induced first-order phase transition in
  Dzyaloshinskii-Moriya helimagnets}},}\ }\href {\doibase
  10.1103/PhysRevB.87.134407} {\bibfield  {journal} {\bibinfo  {journal} {Phys.
  Rev. B}\ }\textbf {\bibinfo {volume} {87}},\ \bibinfo {pages} {134407}
  (\bibinfo {year} {2013})}\BibitemShut {NoStop}%
\bibitem [{\citenamefont {Grigoriev}\ \emph {et~al.}(2010)\citenamefont
  {Grigoriev}, \citenamefont {Maleyev}, \citenamefont {Moskvin}, \citenamefont
  {Dyadkin}, \citenamefont {Fouquet},\ and\ \citenamefont
  {Eckerlebe}}]{2010:Grigoriev:PhysRevB2}%
  \BibitemOpen
  \bibfield  {author} {\bibinfo {author} {\bibfnamefont {S.~V.}\ \bibnamefont
  {Grigoriev}}, \bibinfo {author} {\bibfnamefont {S.~V.}\ \bibnamefont
  {Maleyev}}, \bibinfo {author} {\bibfnamefont {E.~V.}\ \bibnamefont
  {Moskvin}}, \bibinfo {author} {\bibfnamefont {V.~A.}\ \bibnamefont
  {Dyadkin}}, \bibinfo {author} {\bibfnamefont {P.}~\bibnamefont {Fouquet}}, \
  and\ \bibinfo {author} {\bibfnamefont {H.}~\bibnamefont {Eckerlebe}},\
  }\bibfield  {title} {\enquote {\bibinfo {title} {{Crossover behavior of
  critical helix fluctuations in MnSi}},}\ }\href {\doibase
  10.1103/PhysRevB.81.144413} {\bibfield  {journal} {\bibinfo  {journal} {Phys.
  Rev. B}\ }\textbf {\bibinfo {volume} {81}},\ \bibinfo {pages} {144413}
  (\bibinfo {year} {2010})}\BibitemShut {NoStop}%
\bibitem [{\citenamefont {Grigoriev}\ \emph {et~al.}(2011)\citenamefont
  {Grigoriev}, \citenamefont {Moskvin}, \citenamefont {Dyadkin}, \citenamefont
  {Lamago}, \citenamefont {Wolf}, \citenamefont {Eckerlebe},\ and\
  \citenamefont {Maleyev}}]{2011:Grigoriev:PhysRevB}%
  \BibitemOpen
  \bibfield  {author} {\bibinfo {author} {\bibfnamefont {S.~V.}\ \bibnamefont
  {Grigoriev}}, \bibinfo {author} {\bibfnamefont {E.~V.}\ \bibnamefont
  {Moskvin}}, \bibinfo {author} {\bibfnamefont {V.~A.}\ \bibnamefont
  {Dyadkin}}, \bibinfo {author} {\bibfnamefont {D.}~\bibnamefont {Lamago}},
  \bibinfo {author} {\bibfnamefont {T.}~\bibnamefont {Wolf}}, \bibinfo {author}
  {\bibfnamefont {H.}~\bibnamefont {Eckerlebe}}, \ and\ \bibinfo {author}
  {\bibfnamefont {S.~V.}\ \bibnamefont {Maleyev}},\ }\bibfield  {title}
  {\enquote {\bibinfo {title} {{Chiral criticality in the doped helimagnets
  Mn$_{1-y}$Fe$_{y}$Si}},}\ }\href {\doibase 10.1103/PhysRevB.83.224411}
  {\bibfield  {journal} {\bibinfo  {journal} {Phys. Rev. B}\ }\textbf {\bibinfo
  {volume} {83}},\ \bibinfo {pages} {224411} (\bibinfo {year}
  {2011})}\BibitemShut {NoStop}%
\bibitem [{\citenamefont {\v{Z}ivkovi\'{c}}\ \emph {et~al.}(2014)\citenamefont
  {\v{Z}ivkovi\'{c}}, \citenamefont {White}, \citenamefont {R{\o}nnow},
  \citenamefont {Pr\v{s}a},\ and\ \citenamefont
  {Berger}}]{2014:Zivkovic:PhysRevB}%
  \BibitemOpen
  \bibfield  {author} {\bibinfo {author} {\bibfnamefont {I.}~\bibnamefont
  {\v{Z}ivkovi\'{c}}}, \bibinfo {author} {\bibfnamefont {J.~S.}\ \bibnamefont
  {White}}, \bibinfo {author} {\bibfnamefont {H.~M.}\ \bibnamefont
  {R{\o}nnow}}, \bibinfo {author} {\bibfnamefont {K.}~\bibnamefont {Pr\v{s}a}},
  \ and\ \bibinfo {author} {\bibfnamefont {H.}~\bibnamefont {Berger}},\
  }\bibfield  {title} {\enquote {\bibinfo {title} {{Critical scaling in the
  cubic helimagnet Cu$_{2}$OSeO$_{3}$}},}\ }\href {\doibase
  10.1103/PhysRevB.89.060401} {\bibfield  {journal} {\bibinfo  {journal} {Phys.
  Rev. B}\ }\textbf {\bibinfo {volume} {89}},\ \bibinfo {pages} {060401}
  (\bibinfo {year} {2014})}\BibitemShut {NoStop}%
\bibitem [{\citenamefont {Buhrandt}\ and\ \citenamefont
  {Fritz}(2013)}]{2013:Buhrandt:PhysRevB}%
  \BibitemOpen
  \bibfield  {author} {\bibinfo {author} {\bibfnamefont {S.}~\bibnamefont
  {Buhrandt}}\ and\ \bibinfo {author} {\bibfnamefont {L.}~\bibnamefont
  {Fritz}},\ }\bibfield  {title} {\enquote {\bibinfo {title} {{Skyrmion lattice
  phase in three-dimensional chiral magnets from Monte Carlo simulations}},}\
  }\href {\doibase 10.1103/PhysRevB.88.195137} {\bibfield  {journal} {\bibinfo
  {journal} {Phys. Rev. B}\ }\textbf {\bibinfo {volume} {88}},\ \bibinfo
  {pages} {195137} (\bibinfo {year} {2013})}\BibitemShut {NoStop}%
\bibitem [{\citenamefont {Bauer}\ and\ \citenamefont
  {Pfleiderer}(2012)}]{2012:Bauer:PhysRevB}%
  \BibitemOpen
  \bibfield  {author} {\bibinfo {author} {\bibfnamefont {A.}~\bibnamefont
  {Bauer}}\ and\ \bibinfo {author} {\bibfnamefont {C.}~\bibnamefont
  {Pfleiderer}},\ }\bibfield  {title} {\enquote {\bibinfo {title} {{Magnetic
  phase diagram of MnSi inferred from magnetization and ac susceptibility}},}\
  }\href {\doibase 10.1103/PhysRevB.85.214418} {\bibfield  {journal} {\bibinfo
  {journal} {Phys. Rev. B}\ }\textbf {\bibinfo {volume} {85}},\ \bibinfo
  {pages} {214418} (\bibinfo {year} {2012})}\BibitemShut {NoStop}%
\bibitem [{\citenamefont {Seki}\ \emph
  {et~al.}(2012{\natexlab{b}})\citenamefont {Seki}, \citenamefont {Ishiwata},\
  and\ \citenamefont {Tokura}}]{2012:Seki:PhysRevB2}%
  \BibitemOpen
  \bibfield  {author} {\bibinfo {author} {\bibfnamefont {S.}~\bibnamefont
  {Seki}}, \bibinfo {author} {\bibfnamefont {S.}~\bibnamefont {Ishiwata}}, \
  and\ \bibinfo {author} {\bibfnamefont {Y.}~\bibnamefont {Tokura}},\
  }\bibfield  {title} {\enquote {\bibinfo {title} {{Magnetoelectric nature of
  skyrmions in a chiral magnetic insulator Cu$_{2}$OSeO$_{3}$}},}\ }\href
  {\doibase 10.1103/PhysRevB.86.060403} {\bibfield  {journal} {\bibinfo
  {journal} {Phys. Rev. B}\ }\textbf {\bibinfo {volume} {86}},\ \bibinfo
  {pages} {060403} (\bibinfo {year} {2012}{\natexlab{b}})}\BibitemShut
  {NoStop}%
\bibitem [{\citenamefont {R\"{o}{\ss}ler}\ \emph {et~al.}(2006)\citenamefont
  {R\"{o}{\ss}ler}, \citenamefont {Bogdanov},\ and\ \citenamefont
  {Pfleiderer}}]{2006:Roessler:Nature}%
  \BibitemOpen
  \bibfield  {author} {\bibinfo {author} {\bibfnamefont {U.~K.}\ \bibnamefont
  {R\"{o}{\ss}ler}}, \bibinfo {author} {\bibfnamefont {A.~N.}\ \bibnamefont
  {Bogdanov}}, \ and\ \bibinfo {author} {\bibfnamefont {C.}~\bibnamefont
  {Pfleiderer}},\ }\bibfield  {title} {\enquote {\bibinfo {title} {{Spontaneous
  skyrmion ground states in magnetic metals}},}\ }\href {\doibase
  doi:10.1038/nature05056} {\bibfield  {journal} {\bibinfo  {journal} {Nature
  (London)}\ }\textbf {\bibinfo {volume} {442}},\ \bibinfo {pages} {797}
  (\bibinfo {year} {2006})}\BibitemShut {NoStop}%
\bibitem [{\citenamefont {Butenko}\ \emph {et~al.}(2010)\citenamefont
  {Butenko}, \citenamefont {Leonov}, \citenamefont {R\"{o}{\ss}ler},\ and\
  \citenamefont {Bogdanov}}]{2010:Butenko:PhysRevB}%
  \BibitemOpen
  \bibfield  {author} {\bibinfo {author} {\bibfnamefont {A.~B.}\ \bibnamefont
  {Butenko}}, \bibinfo {author} {\bibfnamefont {A.~A.}\ \bibnamefont {Leonov}},
  \bibinfo {author} {\bibfnamefont {U.~K.}\ \bibnamefont {R\"{o}{\ss}ler}}, \
  and\ \bibinfo {author} {\bibfnamefont {A.~N.}\ \bibnamefont {Bogdanov}},\
  }\bibfield  {title} {\enquote {\bibinfo {title} {{Stabilization of skyrmion
  textures by uniaxial distortions in noncentrosymmetric cubic helimagnets}},}\
  }\href {\doibase 10.1103/PhysRevB.82.052403} {\bibfield  {journal} {\bibinfo
  {journal} {Phys. Rev. B}\ }\textbf {\bibinfo {volume} {82}},\ \bibinfo
  {pages} {052403} (\bibinfo {year} {2010})}\BibitemShut {NoStop}%
\bibitem [{\citenamefont {Wilhelm}\ \emph {et~al.}(2011)\citenamefont
  {Wilhelm}, \citenamefont {Baenitz}, \citenamefont {Schmidt}, \citenamefont
  {R\"{o}{\ss}ler}, \citenamefont {Leonov},\ and\ \citenamefont
  {Bogdanov}}]{2011:Wilhelm:PhysRevLett}%
  \BibitemOpen
  \bibfield  {author} {\bibinfo {author} {\bibfnamefont {H.}~\bibnamefont
  {Wilhelm}}, \bibinfo {author} {\bibfnamefont {M.}~\bibnamefont {Baenitz}},
  \bibinfo {author} {\bibfnamefont {M.}~\bibnamefont {Schmidt}}, \bibinfo
  {author} {\bibfnamefont {U.~K.}\ \bibnamefont {R\"{o}{\ss}ler}}, \bibinfo
  {author} {\bibfnamefont {A.~A.}\ \bibnamefont {Leonov}}, \ and\ \bibinfo
  {author} {\bibfnamefont {A.~N.}\ \bibnamefont {Bogdanov}},\ }\bibfield
  {title} {\enquote {\bibinfo {title} {{Precursor Phenomena at the Magnetic
  Ordering of the Cubic Helimagnet FeGe}},}\ }\href {\doibase
  10.1103/PhysRevLett.107.127203} {\bibfield  {journal} {\bibinfo  {journal}
  {Phys. Rev. Lett.}\ }\textbf {\bibinfo {volume} {107}},\ \bibinfo {pages}
  {127203} (\bibinfo {year} {2011})}\BibitemShut {NoStop}%
\bibitem [{\citenamefont {Wilhelm}\ \emph {et~al.}(2012)\citenamefont
  {Wilhelm}, \citenamefont {Baenitz}, \citenamefont {Schmidt}, \citenamefont
  {Naylor}, \citenamefont {Lortz}, \citenamefont {R\"{o}{\ss}ler},
  \citenamefont {Leonov},\ and\ \citenamefont
  {Bogdanov}}]{2012:Wilhelm:JPhysCondensMatter}%
  \BibitemOpen
  \bibfield  {author} {\bibinfo {author} {\bibfnamefont {H.}~\bibnamefont
  {Wilhelm}}, \bibinfo {author} {\bibfnamefont {M.}~\bibnamefont {Baenitz}},
  \bibinfo {author} {\bibfnamefont {M.}~\bibnamefont {Schmidt}}, \bibinfo
  {author} {\bibfnamefont {C.}~\bibnamefont {Naylor}}, \bibinfo {author}
  {\bibfnamefont {R.}~\bibnamefont {Lortz}}, \bibinfo {author} {\bibfnamefont
  {U.~K.}\ \bibnamefont {R\"{o}{\ss}ler}}, \bibinfo {author} {\bibfnamefont
  {A.~A.}\ \bibnamefont {Leonov}}, \ and\ \bibinfo {author} {\bibfnamefont
  {A.~N.}\ \bibnamefont {Bogdanov}},\ }\bibfield  {title} {\enquote {\bibinfo
  {title} {{Confinement of chiral magnetic modulations in the precursor region
  of FeGe}},}\ }\href {\doibase 10.1088/0953-8984/24/29/294204} {\bibfield
  {journal} {\bibinfo  {journal} {J. Phys.: Condens. Matter}\ }\textbf
  {\bibinfo {volume} {24}},\ \bibinfo {pages} {294204} (\bibinfo {year}
  {2012})}\BibitemShut {NoStop}%
\bibitem [{\citenamefont {Cevey}\ \emph {et~al.}(2013)\citenamefont {Cevey},
  \citenamefont {Wilhelm}, \citenamefont {Schmidt},\ and\ \citenamefont
  {Lortz}}]{2013:Cevey:PhysStatusSolidiB}%
  \BibitemOpen
  \bibfield  {author} {\bibinfo {author} {\bibfnamefont {L.}~\bibnamefont
  {Cevey}}, \bibinfo {author} {\bibfnamefont {H.}~\bibnamefont {Wilhelm}},
  \bibinfo {author} {\bibfnamefont {M.}~\bibnamefont {Schmidt}}, \ and\
  \bibinfo {author} {\bibfnamefont {R.}~\bibnamefont {Lortz}},\ }\bibfield
  {title} {\enquote {\bibinfo {title} {{Thermodynamic investigations in the
  precursor region of FeGe}},}\ }\href {\doibase 10.1002/pssb.201200632}
  {\bibfield  {journal} {\bibinfo  {journal} {Phys. Status Solidi B}\ }\textbf
  {\bibinfo {volume} {250}},\ \bibinfo {pages} {650} (\bibinfo {year}
  {2013})}\BibitemShut {NoStop}%
\bibitem [{\citenamefont {Bauer}\ \emph {et~al.}(2010)\citenamefont {Bauer},
  \citenamefont {Neubauer}, \citenamefont {Franz}, \citenamefont {M\"{u}nzer},
  \citenamefont {Garst},\ and\ \citenamefont
  {Pfleiderer}}]{2010:Bauer:PhysRevB}%
  \BibitemOpen
  \bibfield  {author} {\bibinfo {author} {\bibfnamefont {A.}~\bibnamefont
  {Bauer}}, \bibinfo {author} {\bibfnamefont {A.}~\bibnamefont {Neubauer}},
  \bibinfo {author} {\bibfnamefont {C.}~\bibnamefont {Franz}}, \bibinfo
  {author} {\bibfnamefont {W.}~\bibnamefont {M\"{u}nzer}}, \bibinfo {author}
  {\bibfnamefont {M.}~\bibnamefont {Garst}}, \ and\ \bibinfo {author}
  {\bibfnamefont {C.}~\bibnamefont {Pfleiderer}},\ }\bibfield  {title}
  {\enquote {\bibinfo {title} {{Quantum phase transitions in single-crystal
  Mn$_{1-x}$Fe$_{x}$Si and Mn$_{1-x}$Co$_{x}$Si: Crystal growth, magnetization,
  ac susceptibility, and specific heat}},}\ }\href {\doibase
  10.1103/PhysRevB.82.064404} {\bibfield  {journal} {\bibinfo  {journal} {Phys.
  Rev. B}\ }\textbf {\bibinfo {volume} {82}},\ \bibinfo {pages} {064404}
  (\bibinfo {year} {2010})}\BibitemShut {NoStop}%
\bibitem [{\citenamefont {Bauer}\ \emph {et~al.}(2013)\citenamefont {Bauer},
  \citenamefont {Garst},\ and\ \citenamefont
  {Pfleiderer}}]{2013:Bauer:PhysRevLett}%
  \BibitemOpen
  \bibfield  {author} {\bibinfo {author} {\bibfnamefont {A.}~\bibnamefont
  {Bauer}}, \bibinfo {author} {\bibfnamefont {M.}~\bibnamefont {Garst}}, \ and\
  \bibinfo {author} {\bibfnamefont {C.}~\bibnamefont {Pfleiderer}},\ }\bibfield
   {title} {\enquote {\bibinfo {title} {{Specific Heat of the Skyrmion Lattice
  Phase and Field-Induced Tricritical Point in MnSi}},}\ }\href {\doibase
  10.1103/PhysRevLett.110.177207} {\bibfield  {journal} {\bibinfo  {journal}
  {Phys. Rev. Lett.}\ }\textbf {\bibinfo {volume} {110}},\ \bibinfo {pages}
  {177207} (\bibinfo {year} {2013})}\BibitemShut {NoStop}%
\bibitem [{\citenamefont {Levati\'{c}}\ \emph {et~al.}(2014)\citenamefont
  {Levati\'{c}}, \citenamefont {\v{S}urija}, \citenamefont {Berger},\ and\
  \citenamefont {\v{Z}ivkovi\'{c}}}]{2014:Levatic:PhysRevB}%
  \BibitemOpen
  \bibfield  {author} {\bibinfo {author} {\bibfnamefont {I.}~\bibnamefont
  {Levati\'{c}}}, \bibinfo {author} {\bibfnamefont {V.}~\bibnamefont
  {\v{S}urija}}, \bibinfo {author} {\bibfnamefont {H.}~\bibnamefont {Berger}},
  \ and\ \bibinfo {author} {\bibfnamefont {I.}~\bibnamefont
  {\v{Z}ivkovi\'{c}}},\ }\bibfield  {title} {\enquote {\bibinfo {title}
  {{Dissipation processes in the insulating skyrmion compound
  Cu$_{2}$OSeO$_{3}$}},}\ }\href {\doibase 10.1103/PhysRevB.90.224412}
  {\bibfield  {journal} {\bibinfo  {journal} {Phys. Rev. B}\ }\textbf {\bibinfo
  {volume} {90}},\ \bibinfo {pages} {224412} (\bibinfo {year}
  {2014})}\BibitemShut {NoStop}%
\bibitem [{\citenamefont {Vollhardt}(1997)}]{1997:Vollhardt:PhysRevLett}%
  \BibitemOpen
  \bibfield  {author} {\bibinfo {author} {\bibfnamefont {D.}~\bibnamefont
  {Vollhardt}},\ }\bibfield  {title} {\enquote {\bibinfo {title}
  {{Characteristic Crossing Points in Specific Heat Curves of Correlated
  Systems}},}\ }\href {\doibase 10.1103/PhysRevLett.78.1307} {\bibfield
  {journal} {\bibinfo  {journal} {Phys. Rev. Lett.}\ }\textbf {\bibinfo
  {volume} {78}},\ \bibinfo {pages} {1307} (\bibinfo {year}
  {1997})}\BibitemShut {NoStop}%
\bibitem [{\citenamefont {Aharoni}(1998)}]{1998:Aharoni:JApplPhys}%
  \BibitemOpen
  \bibfield  {author} {\bibinfo {author} {\bibfnamefont {A.}~\bibnamefont
  {Aharoni}},\ }\bibfield  {title} {\enquote {\bibinfo {title} {{Demagnetizing
  factors for rectangular ferromagnetic prisms}},}\ }\href {\doibase
  10.1063/1.367113} {\bibfield  {journal} {\bibinfo  {journal} {J. Appl.
  Phys.}\ }\textbf {\bibinfo {volume} {83}},\ \bibinfo {pages} {3432} (\bibinfo
  {year} {1998})}\BibitemShut {NoStop}%
\bibitem [{\citenamefont {Ludgren}\ \emph {et~al.}(1970)\citenamefont
  {Ludgren}, \citenamefont {Beckman}, \citenamefont {Attia}, \citenamefont
  {Bhattacheriee},\ and\ \citenamefont {Richardson}}]{1970:Lundgren:PhysScr}%
  \BibitemOpen
  \bibfield  {author} {\bibinfo {author} {\bibfnamefont {L.}~\bibnamefont
  {Ludgren}}, \bibinfo {author} {\bibfnamefont {O.}~\bibnamefont {Beckman}},
  \bibinfo {author} {\bibfnamefont {V.}~\bibnamefont {Attia}}, \bibinfo
  {author} {\bibfnamefont {S.~P.}\ \bibnamefont {Bhattacheriee}}, \ and\
  \bibinfo {author} {\bibfnamefont {M.}~\bibnamefont {Richardson}},\ }\bibfield
   {title} {\enquote {\bibinfo {title} {{Helical Spin Arrangement in Cubic
  FeGe}},}\ }\href {\doibase 10.1088/0031-8949/1/1/012} {\bibfield  {journal}
  {\bibinfo  {journal} {Phys. Scr.}\ }\textbf {\bibinfo {volume} {1}},\
  \bibinfo {pages} {69} (\bibinfo {year} {1970})}\BibitemShut {NoStop}%
\bibitem [{\citenamefont {Lebech}\ \emph {et~al.}(1989)\citenamefont {Lebech},
  \citenamefont {Bernhard},\ and\ \citenamefont
  {Freltoft}}]{1989:Lebech:JPhysCondensMatter}%
  \BibitemOpen
  \bibfield  {author} {\bibinfo {author} {\bibfnamefont {B.}~\bibnamefont
  {Lebech}}, \bibinfo {author} {\bibfnamefont {J.}~\bibnamefont {Bernhard}}, \
  and\ \bibinfo {author} {\bibfnamefont {T.}~\bibnamefont {Freltoft}},\
  }\bibfield  {title} {\enquote {\bibinfo {title} {{Magnetic structures of
  cubic FeGe studied by small-angle neutron scattering}},}\ }\href {\doibase
  10.1088/0953-8984/1/35/010} {\bibfield  {journal} {\bibinfo  {journal} {J.
  Phys.: Condens. Matter}\ }\textbf {\bibinfo {volume} {1}},\ \bibinfo {pages}
  {6105} (\bibinfo {year} {1989})}\BibitemShut {NoStop}%
\bibitem [{\citenamefont {Beille}\ \emph {et~al.}(1983)\citenamefont {Beille},
  \citenamefont {Voiron},\ and\ \citenamefont
  {Roth}}]{1983:Beille:SolidStateCommun}%
  \BibitemOpen
  \bibfield  {author} {\bibinfo {author} {\bibfnamefont {J.}~\bibnamefont
  {Beille}}, \bibinfo {author} {\bibfnamefont {J.}~\bibnamefont {Voiron}}, \
  and\ \bibinfo {author} {\bibfnamefont {M.}~\bibnamefont {Roth}},\ }\bibfield
  {title} {\enquote {\bibinfo {title} {{Long period helimagnetism in the cubic
  B20 Fe$_{x}$Co$_{1-x}$Si and Co$_{x}$Mn$_{1-x}$Si alloys}},}\ }\href
  {\doibase 10.1016/0038-1098(83)90928-6} {\bibfield  {journal} {\bibinfo
  {journal} {Solid State Commun.}\ }\textbf {\bibinfo {volume} {47}},\ \bibinfo
  {pages} {399} (\bibinfo {year} {1983})}\BibitemShut {NoStop}%
\bibitem [{\citenamefont {Ishimoto}\ \emph {et~al.}(1995)\citenamefont
  {Ishimoto}, \citenamefont {Yamaguchi}, \citenamefont {Suzuki}, \citenamefont
  {Arai}, \citenamefont {Furusaka},\ and\ \citenamefont
  {Endoh}}]{1995:Ishimoto:PhysicaB}%
  \BibitemOpen
  \bibfield  {author} {\bibinfo {author} {\bibfnamefont {K.}~\bibnamefont
  {Ishimoto}}, \bibinfo {author} {\bibfnamefont {Y.}~\bibnamefont {Yamaguchi}},
  \bibinfo {author} {\bibfnamefont {J.}~\bibnamefont {Suzuki}}, \bibinfo
  {author} {\bibfnamefont {M.}~\bibnamefont {Arai}}, \bibinfo {author}
  {\bibfnamefont {M.}~\bibnamefont {Furusaka}}, \ and\ \bibinfo {author}
  {\bibfnamefont {Y.}~\bibnamefont {Endoh}},\ }\bibfield  {title} {\enquote
  {\bibinfo {title} {{Small-angle neutron diffraction from the helical magnet
  Fe$_{0.8}$Co$_{0.2}$Si}},}\ }\href {\doibase 10.1016/0921-4526(95)00163-4}
  {\bibfield  {journal} {\bibinfo  {journal} {Physica B}\ }\textbf {\bibinfo
  {volume} {213--214}},\ \bibinfo {pages} {381} (\bibinfo {year}
  {1995})}\BibitemShut {NoStop}%
\bibitem [{\citenamefont {Chernikov}\ \emph {et~al.}(1997)\citenamefont
  {Chernikov}, \citenamefont {Degiorgi}, \citenamefont {Felder}, \citenamefont
  {Paschen}, \citenamefont {Bianchi}, \citenamefont {Ott}, \citenamefont
  {Sarrao}, \citenamefont {Fisk},\ and\ \citenamefont
  {Mandrus}}]{1997:Chernikov:PhysRevB}%
  \BibitemOpen
  \bibfield  {author} {\bibinfo {author} {\bibfnamefont {M.~A.}\ \bibnamefont
  {Chernikov}}, \bibinfo {author} {\bibfnamefont {L.}~\bibnamefont {Degiorgi}},
  \bibinfo {author} {\bibfnamefont {E.}~\bibnamefont {Felder}}, \bibinfo
  {author} {\bibfnamefont {S.}~\bibnamefont {Paschen}}, \bibinfo {author}
  {\bibfnamefont {A.~D.}\ \bibnamefont {Bianchi}}, \bibinfo {author}
  {\bibfnamefont {H.~R.}\ \bibnamefont {Ott}}, \bibinfo {author} {\bibfnamefont
  {J.~L.}\ \bibnamefont {Sarrao}}, \bibinfo {author} {\bibfnamefont
  {Z.}~\bibnamefont {Fisk}}, \ and\ \bibinfo {author} {\bibfnamefont
  {D.}~\bibnamefont {Mandrus}},\ }\bibfield  {title} {\enquote {\bibinfo
  {title} {{Low-temperature transport, optical, magnetic and thermodynamic
  properties of Fe$_{1-x}$Co$_{x}$Si}},}\ }\href {\doibase
  10.1103/PhysRevB.56.1366} {\bibfield  {journal} {\bibinfo  {journal} {Phys.
  Rev. B}\ }\textbf {\bibinfo {volume} {56}},\ \bibinfo {pages} {1366}
  (\bibinfo {year} {1997})}\BibitemShut {NoStop}%
\bibitem [{\citenamefont {Manyala}\ \emph {et~al.}(2000)\citenamefont
  {Manyala}, \citenamefont {Sidis}, \citenamefont {DiTusa}, \citenamefont
  {Aeppli}, \citenamefont {Young},\ and\ \citenamefont
  {Fisk}}]{2000:Manyala:Nature}%
  \BibitemOpen
  \bibfield  {author} {\bibinfo {author} {\bibfnamefont {N.}~\bibnamefont
  {Manyala}}, \bibinfo {author} {\bibfnamefont {Y.}~\bibnamefont {Sidis}},
  \bibinfo {author} {\bibfnamefont {J.~F.}\ \bibnamefont {DiTusa}}, \bibinfo
  {author} {\bibfnamefont {G.}~\bibnamefont {Aeppli}}, \bibinfo {author}
  {\bibfnamefont {D.~P.}\ \bibnamefont {Young}}, \ and\ \bibinfo {author}
  {\bibfnamefont {Z.}~\bibnamefont {Fisk}},\ }\bibfield  {title} {\enquote
  {\bibinfo {title} {{Magnetoresistance from quantum interference effects in
  ferromagnets}},}\ }\href {\doibase 10.1038/35007030} {\bibfield  {journal}
  {\bibinfo  {journal} {Nature (London)}\ }\textbf {\bibinfo {volume} {404}},\
  \bibinfo {pages} {581} (\bibinfo {year} {2000})}\BibitemShut {NoStop}%
\bibitem [{\citenamefont {Onose}\ \emph {et~al.}(2005)\citenamefont {Onose},
  \citenamefont {Takeshita}, \citenamefont {Terakura}, \citenamefont {Takagi},\
  and\ \citenamefont {Tokura}}]{2005:Onose:PhysRevB}%
  \BibitemOpen
  \bibfield  {author} {\bibinfo {author} {\bibfnamefont {Y.}~\bibnamefont
  {Onose}}, \bibinfo {author} {\bibfnamefont {N.}~\bibnamefont {Takeshita}},
  \bibinfo {author} {\bibfnamefont {C.}~\bibnamefont {Terakura}}, \bibinfo
  {author} {\bibfnamefont {H.}~\bibnamefont {Takagi}}, \ and\ \bibinfo {author}
  {\bibfnamefont {Y.}~\bibnamefont {Tokura}},\ }\bibfield  {title} {\enquote
  {\bibinfo {title} {{Doping dependence of transport properties in
  Fe$_{1-x}$Co$_{x}$Si}},}\ }\href {\doibase 10.1103/PhysRevB.72.224431}
  {\bibfield  {journal} {\bibinfo  {journal} {Phys. Rev. B}\ }\textbf {\bibinfo
  {volume} {72}},\ \bibinfo {pages} {224431} (\bibinfo {year}
  {2005})}\BibitemShut {NoStop}%
\bibitem [{\citenamefont {Grigoriev}\ \emph {et~al.}(2007)\citenamefont
  {Grigoriev}, \citenamefont {Dyadkin}, \citenamefont {Menzel}, \citenamefont
  {Schoenes}, \citenamefont {Chetverikov}, \citenamefont {Okorokov},
  \citenamefont {Eckerlebe},\ and\ \citenamefont
  {Maleyev}}]{2007:Grigoriev:PhysRevB}%
  \BibitemOpen
  \bibfield  {author} {\bibinfo {author} {\bibfnamefont {S.~V.}\ \bibnamefont
  {Grigoriev}}, \bibinfo {author} {\bibfnamefont {V.~A.}\ \bibnamefont
  {Dyadkin}}, \bibinfo {author} {\bibfnamefont {D.}~\bibnamefont {Menzel}},
  \bibinfo {author} {\bibfnamefont {J.}~\bibnamefont {Schoenes}}, \bibinfo
  {author} {\bibfnamefont {Yu.~O.}\ \bibnamefont {Chetverikov}}, \bibinfo
  {author} {\bibfnamefont {A.~I.}\ \bibnamefont {Okorokov}}, \bibinfo {author}
  {\bibfnamefont {H.}~\bibnamefont {Eckerlebe}}, \ and\ \bibinfo {author}
  {\bibfnamefont {S.~V.}\ \bibnamefont {Maleyev}},\ }\bibfield  {title}
  {\enquote {\bibinfo {title} {{Magnetic structure of Fe$_{1-x}$Co$_{x}$Si in a
  magnetic field studied via small-angle polarized neutron diffraction}},}\
  }\href {\doibase 10.1103/PhysRevB.76.224424} {\bibfield  {journal} {\bibinfo
  {journal} {Phys. Rev. B}\ }\textbf {\bibinfo {volume} {76}},\ \bibinfo
  {pages} {224424} (\bibinfo {year} {2007})}\BibitemShut {NoStop}%
\bibitem [{\citenamefont {Franz}\ \emph {et~al.}(2014)\citenamefont {Franz},
  \citenamefont {Freimuth}, \citenamefont {Bauer}, \citenamefont {Ritz},
  \citenamefont {Schnarr}, \citenamefont {Duvinage}, \citenamefont {Adams},
  \citenamefont {Bl\"{u}gel}, \citenamefont {Rosch}, \citenamefont
  {Mokrousov},\ and\ \citenamefont {Pfleiderer}}]{2014:Franz:PhysRevLett}%
  \BibitemOpen
  \bibfield  {author} {\bibinfo {author} {\bibfnamefont {C.}~\bibnamefont
  {Franz}}, \bibinfo {author} {\bibfnamefont {F.}~\bibnamefont {Freimuth}},
  \bibinfo {author} {\bibfnamefont {A.}~\bibnamefont {Bauer}}, \bibinfo
  {author} {\bibfnamefont {R.}~\bibnamefont {Ritz}}, \bibinfo {author}
  {\bibfnamefont {C.}~\bibnamefont {Schnarr}}, \bibinfo {author} {\bibfnamefont
  {C.}~\bibnamefont {Duvinage}}, \bibinfo {author} {\bibfnamefont
  {T.}~\bibnamefont {Adams}}, \bibinfo {author} {\bibfnamefont
  {S.}~\bibnamefont {Bl\"{u}gel}}, \bibinfo {author} {\bibfnamefont
  {A.}~\bibnamefont {Rosch}}, \bibinfo {author} {\bibfnamefont
  {Y.}~\bibnamefont {Mokrousov}}, \ and\ \bibinfo {author} {\bibfnamefont
  {C.}~\bibnamefont {Pfleiderer}},\ }\bibfield  {title} {\enquote {\bibinfo
  {title} {{Real-Space and Reciprocal-Space Berry Phases in the Hall Effect of
  Mn$_{1-x}$Fe$_{x}$Si}},}\ }\href {\doibase 10.1103/PhysRevLett.112.186601}
  {\bibfield  {journal} {\bibinfo  {journal} {Phys. Rev. Lett.}\ }\textbf
  {\bibinfo {volume} {112}},\ \bibinfo {pages} {186601} (\bibinfo {year}
  {2014})}\BibitemShut {NoStop}%
\bibitem [{\citenamefont {Teyssier}\ \emph {et~al.}(2010)\citenamefont
  {Teyssier}, \citenamefont {Giannini}, \citenamefont {Guritanu}, \citenamefont
  {Viennois}, \citenamefont {van~der Marel}, \citenamefont {Amato},\ and\
  \citenamefont {Gvasaliya}}]{2010:Teyssier:PhysRevB}%
  \BibitemOpen
  \bibfield  {author} {\bibinfo {author} {\bibfnamefont {J.}~\bibnamefont
  {Teyssier}}, \bibinfo {author} {\bibfnamefont {E.}~\bibnamefont {Giannini}},
  \bibinfo {author} {\bibfnamefont {V.}~\bibnamefont {Guritanu}}, \bibinfo
  {author} {\bibfnamefont {R.}~\bibnamefont {Viennois}}, \bibinfo {author}
  {\bibfnamefont {D.}~\bibnamefont {van~der Marel}}, \bibinfo {author}
  {\bibfnamefont {A.}~\bibnamefont {Amato}}, \ and\ \bibinfo {author}
  {\bibfnamefont {S.~N.}\ \bibnamefont {Gvasaliya}},\ }\bibfield  {title}
  {\enquote {\bibinfo {title} {{Spin-glass ground state in
  Mn$_{1-x}$Co$_{x}$Si}},}\ }\href {\doibase 10.1103/PhysRevB.82.064417}
  {\bibfield  {journal} {\bibinfo  {journal} {Phys. Rev. B}\ }\textbf {\bibinfo
  {volume} {82}},\ \bibinfo {pages} {064417} (\bibinfo {year}
  {2010})}\BibitemShut {NoStop}%
\bibitem [{\citenamefont {Achu}\ \emph {et~al.}(1998)\citenamefont {Achu},
  \citenamefont {Al-Kanani}, \citenamefont {Booth}, \citenamefont {Costa},\
  and\ \citenamefont {Lebech}}]{1998:Achu:JMagnMagnMater}%
  \BibitemOpen
  \bibfield  {author} {\bibinfo {author} {\bibfnamefont {E.~W.}\ \bibnamefont
  {Achu}}, \bibinfo {author} {\bibfnamefont {H.~J.}\ \bibnamefont {Al-Kanani}},
  \bibinfo {author} {\bibfnamefont {J.~G.}\ \bibnamefont {Booth}}, \bibinfo
  {author} {\bibfnamefont {M.~M.~R.}\ \bibnamefont {Costa}}, \ and\ \bibinfo
  {author} {\bibfnamefont {B.}~\bibnamefont {Lebech}},\ }\bibfield  {title}
  {\enquote {\bibinfo {title} {{Studies of the incommensurate structures of B20
  alloys}},}\ }\href {\doibase 10.1016/S0304-8853(97)00277-1} {\bibfield
  {journal} {\bibinfo  {journal} {J. Magn. Magn. Mater.}\ }\textbf {\bibinfo
  {volume} {177--181}},\ \bibinfo {pages} {779} (\bibinfo {year}
  {1998})}\BibitemShut {NoStop}%
\bibitem [{\citenamefont {Motokawa}\ \emph {et~al.}(1987)\citenamefont
  {Motokawa}, \citenamefont {Kawarazaki}, \citenamefont {Nojiri},\ and\
  \citenamefont {Inoue}}]{1987:Motokawa:JMagnMagnMater}%
  \BibitemOpen
  \bibfield  {author} {\bibinfo {author} {\bibfnamefont {M.}~\bibnamefont
  {Motokawa}}, \bibinfo {author} {\bibfnamefont {S.}~\bibnamefont
  {Kawarazaki}}, \bibinfo {author} {\bibfnamefont {H.}~\bibnamefont {Nojiri}},
  \ and\ \bibinfo {author} {\bibfnamefont {T.}~\bibnamefont {Inoue}},\
  }\bibfield  {title} {\enquote {\bibinfo {title} {{Magnetization measurements
  of Fe$_{1-x}$Co$_{x}$Si}},}\ }\href {\doibase 10.1016/0304-8853(87)90425-2}
  {\bibfield  {journal} {\bibinfo  {journal} {J. Magn. Magn. Mater.}\ }\textbf
  {\bibinfo {volume} {70}},\ \bibinfo {pages} {245} (\bibinfo {year}
  {1987})}\BibitemShut {NoStop}%
\bibitem [{\citenamefont {Jaccarino}\ \emph {et~al.}(1967)\citenamefont
  {Jaccarino}, \citenamefont {Wertheim}, \citenamefont {Wernick}, \citenamefont
  {Walker},\ and\ \citenamefont {Arajs}}]{1967:Jaccarino:PhysRev}%
  \BibitemOpen
  \bibfield  {author} {\bibinfo {author} {\bibfnamefont {V.}~\bibnamefont
  {Jaccarino}}, \bibinfo {author} {\bibfnamefont {G.~K.}\ \bibnamefont
  {Wertheim}}, \bibinfo {author} {\bibfnamefont {J.~H.}\ \bibnamefont
  {Wernick}}, \bibinfo {author} {\bibfnamefont {L.~R.}\ \bibnamefont {Walker}},
  \ and\ \bibinfo {author} {\bibfnamefont {Sigurds}\ \bibnamefont {Arajs}},\
  }\bibfield  {title} {\enquote {\bibinfo {title} {{Paramagnetic Excited State
  of FeSi}},}\ }\href {\doibase 10.1103/PhysRev.160.476} {\bibfield  {journal}
  {\bibinfo  {journal} {Phys. Rev.}\ }\textbf {\bibinfo {volume} {160}},\
  \bibinfo {pages} {476} (\bibinfo {year} {1967})}\BibitemShut {NoStop}%
\bibitem [{\citenamefont {Shinoda}\ and\ \citenamefont
  {Asanabe}(1966)}]{1966:Shinoda:JPhysSocJpn}%
  \BibitemOpen
  \bibfield  {author} {\bibinfo {author} {\bibfnamefont {D.}~\bibnamefont
  {Shinoda}}\ and\ \bibinfo {author} {\bibfnamefont {S.}~\bibnamefont
  {Asanabe}},\ }\bibfield  {title} {\enquote {\bibinfo {title} {{Magnetic
  Properties of Silicides of Iron Group Transition Elements}},}\ }\href
  {\doibase 10.1143/JPSJ.21.555} {\bibfield  {journal} {\bibinfo  {journal} {J.
  Phys. Soc. Jpn.}\ }\textbf {\bibinfo {volume} {21}},\ \bibinfo {pages} {555}
  (\bibinfo {year} {1966})}\BibitemShut {NoStop}%
\bibitem [{\citenamefont {Arita}\ \emph {et~al.}(2008)\citenamefont {Arita},
  \citenamefont {Shimada}, \citenamefont {Takeda}, \citenamefont {Nakatake},
  \citenamefont {Namatame}, \citenamefont {Taniguchi}, \citenamefont {Negishi},
  \citenamefont {Oguchi}, \citenamefont {Saitoh}, \citenamefont {Fujimori},\
  and\ \citenamefont {Kanomata}}]{2008:Arita:PhysRevB}%
  \BibitemOpen
  \bibfield  {author} {\bibinfo {author} {\bibfnamefont {M.}~\bibnamefont
  {Arita}}, \bibinfo {author} {\bibfnamefont {K.}~\bibnamefont {Shimada}},
  \bibinfo {author} {\bibfnamefont {Y.}~\bibnamefont {Takeda}}, \bibinfo
  {author} {\bibfnamefont {M.}~\bibnamefont {Nakatake}}, \bibinfo {author}
  {\bibfnamefont {H.}~\bibnamefont {Namatame}}, \bibinfo {author}
  {\bibfnamefont {M.}~\bibnamefont {Taniguchi}}, \bibinfo {author}
  {\bibfnamefont {H.}~\bibnamefont {Negishi}}, \bibinfo {author} {\bibfnamefont
  {T.}~\bibnamefont {Oguchi}}, \bibinfo {author} {\bibfnamefont
  {T.}~\bibnamefont {Saitoh}}, \bibinfo {author} {\bibfnamefont
  {A.}~\bibnamefont {Fujimori}}, \ and\ \bibinfo {author} {\bibfnamefont
  {T.}~\bibnamefont {Kanomata}},\ }\bibfield  {title} {\enquote {\bibinfo
  {title} {{Angle-resolved photoemission study of the strongly correlated
  semiconductor FeSi}},}\ }\href {\doibase 10.1103/PhysRevB.77.205117}
  {\bibfield  {journal} {\bibinfo  {journal} {Phys. Rev. B}\ }\textbf {\bibinfo
  {volume} {77}},\ \bibinfo {pages} {205117} (\bibinfo {year}
  {2008})}\BibitemShut {NoStop}%
\bibitem [{\citenamefont {Manyala}\ \emph {et~al.}(2004)\citenamefont
  {Manyala}, \citenamefont {Sidis}, \citenamefont {DiTusa}, \citenamefont
  {Aeppli}, \citenamefont {Young},\ and\ \citenamefont
  {Fisk}}]{2004:Manyala:NatureMater}%
  \BibitemOpen
  \bibfield  {author} {\bibinfo {author} {\bibfnamefont {N.}~\bibnamefont
  {Manyala}}, \bibinfo {author} {\bibfnamefont {Y.}~\bibnamefont {Sidis}},
  \bibinfo {author} {\bibfnamefont {J.~F.}\ \bibnamefont {DiTusa}}, \bibinfo
  {author} {\bibfnamefont {G.}~\bibnamefont {Aeppli}}, \bibinfo {author}
  {\bibfnamefont {D.~P.}\ \bibnamefont {Young}}, \ and\ \bibinfo {author}
  {\bibfnamefont {Z.}~\bibnamefont {Fisk}},\ }\bibfield  {title} {\enquote
  {\bibinfo {title} {{Large anomalous Hall effect in a silicon-based magnetic
  semiconductor}},}\ }\href {\doibase 10.1038/nmat1103} {\bibfield  {journal}
  {\bibinfo  {journal} {Nature Mater.}\ }\textbf {\bibinfo {volume} {3}},\
  \bibinfo {pages} {255} (\bibinfo {year} {2004})}\BibitemShut {NoStop}%
\bibitem [{\citenamefont {Ritz}\ \emph
  {et~al.}(2013{\natexlab{b}})\citenamefont {Ritz}, \citenamefont {Halder},
  \citenamefont {Franz}, \citenamefont {Bauer}, \citenamefont {Wagner},
  \citenamefont {Bamler}, \citenamefont {Rosch},\ and\ \citenamefont
  {Pfleiderer}}]{2013:Ritz:PhysRevB}%
  \BibitemOpen
  \bibfield  {author} {\bibinfo {author} {\bibfnamefont {R.}~\bibnamefont
  {Ritz}}, \bibinfo {author} {\bibfnamefont {M.}~\bibnamefont {Halder}},
  \bibinfo {author} {\bibfnamefont {C.}~\bibnamefont {Franz}}, \bibinfo
  {author} {\bibfnamefont {A.}~\bibnamefont {Bauer}}, \bibinfo {author}
  {\bibfnamefont {M.}~\bibnamefont {Wagner}}, \bibinfo {author} {\bibfnamefont
  {R.}~\bibnamefont {Bamler}}, \bibinfo {author} {\bibfnamefont
  {A.}~\bibnamefont {Rosch}}, \ and\ \bibinfo {author} {\bibfnamefont
  {C.}~\bibnamefont {Pfleiderer}},\ }\bibfield  {title} {\enquote {\bibinfo
  {title} {{Giant generic topological Hall resistivity of MnSi under
  pressure}},}\ }\href {\doibase 10.1103/PhysRevB.87.134424} {\bibfield
  {journal} {\bibinfo  {journal} {Phys. Rev. B}\ }\textbf {\bibinfo {volume}
  {87}},\ \bibinfo {pages} {134424} (\bibinfo {year}
  {2013}{\natexlab{b}})}\BibitemShut {NoStop}%
\bibitem [{\citenamefont {Adams}\ \emph {et~al.}(2010)\citenamefont {Adams},
  \citenamefont {M\"{u}hlbauer}, \citenamefont {Neubauer}, \citenamefont
  {M\"{u}nzer}, \citenamefont {Jonietz}, \citenamefont {Georgii}, \citenamefont
  {Pedersen}, \citenamefont {B\"{o}ni}, \citenamefont {Rosch},\ and\
  \citenamefont {Pfleiderer}}]{2010:Adams:JPhysConfSer}%
  \BibitemOpen
  \bibfield  {author} {\bibinfo {author} {\bibfnamefont {T.}~\bibnamefont
  {Adams}}, \bibinfo {author} {\bibfnamefont {S.}~\bibnamefont
  {M\"{u}hlbauer}}, \bibinfo {author} {\bibfnamefont {A.}~\bibnamefont
  {Neubauer}}, \bibinfo {author} {\bibfnamefont {W.}~\bibnamefont
  {M\"{u}nzer}}, \bibinfo {author} {\bibfnamefont {F.}~\bibnamefont {Jonietz}},
  \bibinfo {author} {\bibfnamefont {R.}~\bibnamefont {Georgii}}, \bibinfo
  {author} {\bibfnamefont {B.}~\bibnamefont {Pedersen}}, \bibinfo {author}
  {\bibfnamefont {P.}~\bibnamefont {B\"{o}ni}}, \bibinfo {author}
  {\bibfnamefont {A.}~\bibnamefont {Rosch}}, \ and\ \bibinfo {author}
  {\bibfnamefont {C.}~\bibnamefont {Pfleiderer}},\ }\bibfield  {title}
  {\enquote {\bibinfo {title} {{Skyrmion Lattice Domains in
  Fe$_{1-x}$Co$_{x}$Si}},}\ }\href {\doibase 10.1088/1742-6596/200/3/032001}
  {\bibfield  {journal} {\bibinfo  {journal} {J. Phys.: Conf. Ser.}\ }\textbf
  {\bibinfo {volume} {200}},\ \bibinfo {pages} {032001} (\bibinfo {year}
  {2010})}\BibitemShut {NoStop}%
\bibitem [{\citenamefont {Meunier}\ and\ \citenamefont
  {Bertaud}(1976)}]{1976:Meunier:JApplCryst}%
  \BibitemOpen
  \bibfield  {author} {\bibinfo {author} {\bibfnamefont {G.}~\bibnamefont
  {Meunier}}\ and\ \bibinfo {author} {\bibfnamefont {M.}~\bibnamefont
  {Bertaud}},\ }\bibfield  {title} {\enquote {\bibinfo {title} {{Constantes
  cristallographiques de CuSe$_{2}$O$_{5}$, CuSeO$_{3}$ et
  Cu$_{2}$SeO$_{4}$}},}\ }\href {\doibase 10.1107/S0021889876011540} {\bibfield
   {journal} {\bibinfo  {journal} {J. Appl. Cryst.}\ }\textbf {\bibinfo
  {volume} {9}},\ \bibinfo {pages} {364} (\bibinfo {year} {1976})}\BibitemShut
  {NoStop}%
\bibitem [{\citenamefont {Kohn}(1977)}]{1977:Kohn:JPhysSocJpn}%
  \BibitemOpen
  \bibfield  {author} {\bibinfo {author} {\bibfnamefont {K.}~\bibnamefont
  {Kohn}},\ }\bibfield  {title} {\enquote {\bibinfo {title} {{A New Ferrimagnet
  Cu$_{2}$SeO$_{4}$}},}\ }\href {\doibase 10.1143/JPSJ.42.2065} {\bibfield
  {journal} {\bibinfo  {journal} {J. Phys. Soc. Jpn}\ }\textbf {\bibinfo
  {volume} {42}},\ \bibinfo {pages} {2065} (\bibinfo {year}
  {1977})}\BibitemShut {NoStop}%
\bibitem [{\citenamefont {Belesi}\ \emph {et~al.}(2010)\citenamefont {Belesi},
  \citenamefont {Rousochatzakis}, \citenamefont {Wu}, \citenamefont {Berger},
  \citenamefont {Shvets}, \citenamefont {Mila},\ and\ \citenamefont
  {Ansermet}}]{2010:Belesi:PhysRevB}%
  \BibitemOpen
  \bibfield  {author} {\bibinfo {author} {\bibfnamefont {M.}~\bibnamefont
  {Belesi}}, \bibinfo {author} {\bibfnamefont {I.}~\bibnamefont
  {Rousochatzakis}}, \bibinfo {author} {\bibfnamefont {H.~C.}\ \bibnamefont
  {Wu}}, \bibinfo {author} {\bibfnamefont {H.}~\bibnamefont {Berger}}, \bibinfo
  {author} {\bibfnamefont {I.~V.}\ \bibnamefont {Shvets}}, \bibinfo {author}
  {\bibfnamefont {F.}~\bibnamefont {Mila}}, \ and\ \bibinfo {author}
  {\bibfnamefont {J.~P}\ \bibnamefont {Ansermet}},\ }\bibfield  {title}
  {\enquote {\bibinfo {title} {{Ferrimagnetism of the magnetoelectric compound
  Cu$_{2}$OSeO$_{3}$ probed by $^{77}$Se NMR}},}\ }\href {\doibase
  10.1103/PhysRevB.82.094422} {\bibfield  {journal} {\bibinfo  {journal} {Phys.
  Rev. B}\ }\textbf {\bibinfo {volume} {82}},\ \bibinfo {pages} {094422}
  (\bibinfo {year} {2010})}\BibitemShut {NoStop}%
\bibitem [{\citenamefont {Huang}\ \emph {et~al.}(2011)\citenamefont {Huang},
  \citenamefont {Tseng}, \citenamefont {Chou}, \citenamefont {Mukherjee},
  \citenamefont {Her}, \citenamefont {Matsuda}, \citenamefont {Kindo},
  \citenamefont {Berger},\ and\ \citenamefont {Yang}}]{2011:Huang:PhysRevB}%
  \BibitemOpen
  \bibfield  {author} {\bibinfo {author} {\bibfnamefont {C.~L.}\ \bibnamefont
  {Huang}}, \bibinfo {author} {\bibfnamefont {K.~F.}\ \bibnamefont {Tseng}},
  \bibinfo {author} {\bibfnamefont {C.~C.}\ \bibnamefont {Chou}}, \bibinfo
  {author} {\bibfnamefont {S.}~\bibnamefont {Mukherjee}}, \bibinfo {author}
  {\bibfnamefont {J.~L.}\ \bibnamefont {Her}}, \bibinfo {author} {\bibfnamefont
  {Y.~H.}\ \bibnamefont {Matsuda}}, \bibinfo {author} {\bibfnamefont
  {K.}~\bibnamefont {Kindo}}, \bibinfo {author} {\bibfnamefont
  {H.}~\bibnamefont {Berger}}, \ and\ \bibinfo {author} {\bibfnamefont {H.~D.}\
  \bibnamefont {Yang}},\ }\bibfield  {title} {\enquote {\bibinfo {title}
  {{Observation of a second metastable spin-ordered state in ferrimagnet
  Cu$_{2}$OSeO$_{3}$}},}\ }\href {\doibase 10.1103/PhysRevB.83.052402}
  {\bibfield  {journal} {\bibinfo  {journal} {Phys. Rev. B}\ }\textbf {\bibinfo
  {volume} {83}},\ \bibinfo {pages} {052402} (\bibinfo {year}
  {2011})}\BibitemShut {NoStop}%
\bibitem [{\citenamefont {Seki}\ \emph
  {et~al.}(2012{\natexlab{c}})\citenamefont {Seki}, \citenamefont {Kim},
  \citenamefont {Inosov}, \citenamefont {Georgii}, \citenamefont {Keimer},
  \citenamefont {Ishiwata},\ and\ \citenamefont {Tokura}}]{2012:Seki:PhysRevB}%
  \BibitemOpen
  \bibfield  {author} {\bibinfo {author} {\bibfnamefont {S.}~\bibnamefont
  {Seki}}, \bibinfo {author} {\bibfnamefont {J.-H.}\ \bibnamefont {Kim}},
  \bibinfo {author} {\bibfnamefont {D.~S.}\ \bibnamefont {Inosov}}, \bibinfo
  {author} {\bibfnamefont {R.}~\bibnamefont {Georgii}}, \bibinfo {author}
  {\bibfnamefont {B.}~\bibnamefont {Keimer}}, \bibinfo {author} {\bibfnamefont
  {S.}~\bibnamefont {Ishiwata}}, \ and\ \bibinfo {author} {\bibfnamefont
  {Y.}~\bibnamefont {Tokura}},\ }\bibfield  {title} {\enquote {\bibinfo {title}
  {{Formation and rotation of skyrmion crystal in the chiral-lattice insulator
  Cu$_{2}$OSeO$_{3}$}},}\ }\href {\doibase 10.1103/PhysRevB.85.220406}
  {\bibfield  {journal} {\bibinfo  {journal} {Phys. Rev. B}\ }\textbf {\bibinfo
  {volume} {85}},\ \bibinfo {pages} {220406 (R)} (\bibinfo {year}
  {2012}{\natexlab{c}})}\BibitemShut {NoStop}%
\bibitem [{\citenamefont {Langner}\ \emph {et~al.}(2014)\citenamefont
  {Langner}, \citenamefont {Roy}, \citenamefont {Mishra}, \citenamefont {Lee},
  \citenamefont {Shi}, \citenamefont {Hossain}, \citenamefont {Chuang},
  \citenamefont {Seki}, \citenamefont {Tokura}, \citenamefont {Kevan},\ and\
  \citenamefont {Schoenlein}}]{2014:Langner:PhysRevLett}%
  \BibitemOpen
  \bibfield  {author} {\bibinfo {author} {\bibfnamefont {M.~C.}\ \bibnamefont
  {Langner}}, \bibinfo {author} {\bibfnamefont {S.}~\bibnamefont {Roy}},
  \bibinfo {author} {\bibfnamefont {S.~K.}\ \bibnamefont {Mishra}}, \bibinfo
  {author} {\bibfnamefont {J.~C.~T.}\ \bibnamefont {Lee}}, \bibinfo {author}
  {\bibfnamefont {X.~W.}\ \bibnamefont {Shi}}, \bibinfo {author} {\bibfnamefont
  {M.~A.}\ \bibnamefont {Hossain}}, \bibinfo {author} {\bibfnamefont {Y.-D.}\
  \bibnamefont {Chuang}}, \bibinfo {author} {\bibfnamefont {S.}~\bibnamefont
  {Seki}}, \bibinfo {author} {\bibfnamefont {Y.}~\bibnamefont {Tokura}},
  \bibinfo {author} {\bibfnamefont {S.~D.}\ \bibnamefont {Kevan}}, \ and\
  \bibinfo {author} {\bibfnamefont {R.~W.}\ \bibnamefont {Schoenlein}},\
  }\bibfield  {title} {\enquote {\bibinfo {title} {{Coupled Skyrmion
  Sublattices in Cu$_{2}$OSeO$_{3}$}},}\ }\href {\doibase
  10.1103/PhysRevLett.112.167202} {\bibfield  {journal} {\bibinfo  {journal}
  {Phys. Rev. Lett.}\ }\textbf {\bibinfo {volume} {112}},\ \bibinfo {pages}
  {167202} (\bibinfo {year} {2014})}\BibitemShut {NoStop}%
\bibitem [{\citenamefont {Jia}\ \emph {et~al.}(2006)\citenamefont {Jia},
  \citenamefont {Onoda}, \citenamefont {Nagaosa},\ and\ \citenamefont
  {Han}}]{2006:Jia:PhysRevB}%
  \BibitemOpen
  \bibfield  {author} {\bibinfo {author} {\bibfnamefont {C.}~\bibnamefont
  {Jia}}, \bibinfo {author} {\bibfnamefont {S.}~\bibnamefont {Onoda}}, \bibinfo
  {author} {\bibfnamefont {N.}~\bibnamefont {Nagaosa}}, \ and\ \bibinfo
  {author} {\bibfnamefont {J.~H.}\ \bibnamefont {Han}},\ }\bibfield  {title}
  {\enquote {\bibinfo {title} {{Bond electronic polarization induced by
  spin}},}\ }\href {\doibase 10.1103/PhysRevB.74.224444} {\bibfield  {journal}
  {\bibinfo  {journal} {Phys. Rev. B}\ }\textbf {\bibinfo {volume} {74}},\
  \bibinfo {pages} {224444} (\bibinfo {year} {2006})}\BibitemShut {NoStop}%
\bibitem [{\citenamefont {White}\ \emph {et~al.}(2012)\citenamefont {White},
  \citenamefont {Levati\'{c}}, \citenamefont {Omrani}, \citenamefont
  {Egetenmeyer}, \citenamefont {Pr\v{s}a}, \citenamefont {\v{Z}ivkovi\'{c}},
  \citenamefont {Gavilano}, \citenamefont {Kohlbrecher}, \citenamefont
  {Bartkowiak}, \citenamefont {Berger},\ and\ \citenamefont
  {R{\o}nnow}}]{2012:White:JPhysCondensMatter}%
  \BibitemOpen
  \bibfield  {author} {\bibinfo {author} {\bibfnamefont {J.~S.}\ \bibnamefont
  {White}}, \bibinfo {author} {\bibfnamefont {I.}~\bibnamefont {Levati\'{c}}},
  \bibinfo {author} {\bibfnamefont {A.~A.}\ \bibnamefont {Omrani}}, \bibinfo
  {author} {\bibfnamefont {N.}~\bibnamefont {Egetenmeyer}}, \bibinfo {author}
  {\bibfnamefont {K.}~\bibnamefont {Pr\v{s}a}}, \bibinfo {author}
  {\bibfnamefont {I.}~\bibnamefont {\v{Z}ivkovi\'{c}}}, \bibinfo {author}
  {\bibfnamefont {J.~L.}\ \bibnamefont {Gavilano}}, \bibinfo {author}
  {\bibfnamefont {J.}~\bibnamefont {Kohlbrecher}}, \bibinfo {author}
  {\bibfnamefont {M.}~\bibnamefont {Bartkowiak}}, \bibinfo {author}
  {\bibfnamefont {H.}~\bibnamefont {Berger}}, \ and\ \bibinfo {author}
  {\bibfnamefont {H.~M.}\ \bibnamefont {R{\o}nnow}},\ }\bibfield  {title}
  {\enquote {\bibinfo {title} {{Electric field control of the skyrmion lattice
  in Cu$_{2}$OSeO$_{3}$}},}\ }\href {\doibase 10.1088/0953-8984/24/43/432201}
  {\bibfield  {journal} {\bibinfo  {journal} {J. Phys.: Condens. Matter}\
  }\textbf {\bibinfo {volume} {24}},\ \bibinfo {pages} {432201} (\bibinfo
  {year} {2012})}\BibitemShut {NoStop}%
\bibitem [{\citenamefont {Mochizuki}\ and\ \citenamefont
  {Seki}(2013)}]{2013:Mochizuki:PhysRevB}%
  \BibitemOpen
  \bibfield  {author} {\bibinfo {author} {\bibfnamefont {M.}~\bibnamefont
  {Mochizuki}}\ and\ \bibinfo {author} {\bibfnamefont {S.}~\bibnamefont
  {Seki}},\ }\bibfield  {title} {\enquote {\bibinfo {title} {{Magnetoelectric
  resonances and predicted microwave diode effect of the skyrmion crystal in a
  multiferroic chiral-lattice magnet}},}\ }\href {\doibase
  10.1103/PhysRevB.87.134403} {\bibfield  {journal} {\bibinfo  {journal} {Phys.
  Rev. B}\ }\textbf {\bibinfo {volume} {87}},\ \bibinfo {pages} {134403}
  (\bibinfo {year} {2013})}\BibitemShut {NoStop}%
\bibitem [{\citenamefont {Mochizuki}\ \emph {et~al.}(2014)\citenamefont
  {Mochizuki}, \citenamefont {Yu}, \citenamefont {Seki}, \citenamefont
  {Kanazawa}, \citenamefont {Koshibae}, \citenamefont {Zang}, \citenamefont
  {Mostovoy}, \citenamefont {Tokura},\ and\ \citenamefont
  {Nagaosa}}]{2014:Mochizuki:NatureMater}%
  \BibitemOpen
  \bibfield  {author} {\bibinfo {author} {\bibfnamefont {M.}~\bibnamefont
  {Mochizuki}}, \bibinfo {author} {\bibfnamefont {X.~Z.}\ \bibnamefont {Yu}},
  \bibinfo {author} {\bibfnamefont {S.}~\bibnamefont {Seki}}, \bibinfo {author}
  {\bibfnamefont {N.}~\bibnamefont {Kanazawa}}, \bibinfo {author}
  {\bibfnamefont {W.}~\bibnamefont {Koshibae}}, \bibinfo {author}
  {\bibfnamefont {J.}~\bibnamefont {Zang}}, \bibinfo {author} {\bibfnamefont
  {M.}~\bibnamefont {Mostovoy}}, \bibinfo {author} {\bibfnamefont
  {Y.}~\bibnamefont {Tokura}}, \ and\ \bibinfo {author} {\bibfnamefont
  {N.}~\bibnamefont {Nagaosa}},\ }\bibfield  {title} {\enquote {\bibinfo
  {title} {{Thermally driven ratchet motion of a skyrmion microcrystal and
  topological magnon Hall effect}},}\ }\href {\doibase 10.1038/nmat3862}
  {\bibfield  {journal} {\bibinfo  {journal} {Nature Mater.}\ }\textbf
  {\bibinfo {volume} {13}},\ \bibinfo {pages} {241} (\bibinfo {year}
  {2014})}\BibitemShut {NoStop}%
\bibitem [{\citenamefont {Okamura}\ \emph {et~al.}(2015)\citenamefont
  {Okamura}, \citenamefont {Kagawa}, \citenamefont {Seki}, \citenamefont
  {Kubota}, \citenamefont {Kawasaki},\ and\ \citenamefont
  {Tokura}}]{2015:Okamura:PhysRevLett}%
  \BibitemOpen
  \bibfield  {author} {\bibinfo {author} {\bibfnamefont {Y.}~\bibnamefont
  {Okamura}}, \bibinfo {author} {\bibfnamefont {F.}~\bibnamefont {Kagawa}},
  \bibinfo {author} {\bibfnamefont {S.}~\bibnamefont {Seki}}, \bibinfo {author}
  {\bibfnamefont {M.}~\bibnamefont {Kubota}}, \bibinfo {author} {\bibfnamefont
  {M.}~\bibnamefont {Kawasaki}}, \ and\ \bibinfo {author} {\bibfnamefont
  {Y.}~\bibnamefont {Tokura}},\ }\bibfield  {title} {\enquote {\bibinfo {title}
  {{Microwave Magnetochiral Dichroism in the Chiral-Lattice Magnet
  Cu$_{2}$OSeO$_{3}$}},}\ }\href {\doibase 10.1103/PhysRevLett.114.197202}
  {\bibfield  {journal} {\bibinfo  {journal} {Phys. Rev. Lett}\ }\textbf
  {\bibinfo {volume} {114}},\ \bibinfo {pages} {197202} (\bibinfo {year}
  {2015})}\BibitemShut {NoStop}%
\bibitem [{\citenamefont {Neubauer}\ \emph {et~al.}(2009)\citenamefont
  {Neubauer}, \citenamefont {Pfleiderer}, \citenamefont {Binz}, \citenamefont
  {Rosch}, \citenamefont {Ritz}, \citenamefont {Niklowitz},\ and\ \citenamefont
  {B\"{o}ni}}]{2009:Neubauer:PhysRevLett}%
  \BibitemOpen
  \bibfield  {author} {\bibinfo {author} {\bibfnamefont {A.}~\bibnamefont
  {Neubauer}}, \bibinfo {author} {\bibfnamefont {C.}~\bibnamefont
  {Pfleiderer}}, \bibinfo {author} {\bibfnamefont {B.}~\bibnamefont {Binz}},
  \bibinfo {author} {\bibfnamefont {A.}~\bibnamefont {Rosch}}, \bibinfo
  {author} {\bibfnamefont {R.}~\bibnamefont {Ritz}}, \bibinfo {author}
  {\bibfnamefont {P.~G.}\ \bibnamefont {Niklowitz}}, \ and\ \bibinfo {author}
  {\bibfnamefont {P.}~\bibnamefont {B\"{o}ni}},\ }\bibfield  {title} {\enquote
  {\bibinfo {title} {{Topological Hall Effect in the $A$ Phase of MnSi}},}\
  }\href {\doibase 10.1103/PhysRevLett.102.186602} {\bibfield  {journal}
  {\bibinfo  {journal} {Phys. Rev. Lett.}\ }\textbf {\bibinfo {volume} {102}},\
  \bibinfo {pages} {186602} (\bibinfo {year} {2009})}\BibitemShut {NoStop}%
\bibitem [{\citenamefont {Pfleiderer}\ \emph {et~al.}(2007)\citenamefont
  {Pfleiderer}, \citenamefont {B\"{o}ni}, \citenamefont {Keller}, \citenamefont
  {R\"{o}{\ss}ler},\ and\ \citenamefont {Rosch}}]{2007:Pfleiderer:Science}%
  \BibitemOpen
  \bibfield  {author} {\bibinfo {author} {\bibfnamefont {C.}~\bibnamefont
  {Pfleiderer}}, \bibinfo {author} {\bibfnamefont {P.}~\bibnamefont
  {B\"{o}ni}}, \bibinfo {author} {\bibfnamefont {T.}~\bibnamefont {Keller}},
  \bibinfo {author} {\bibfnamefont {U.~K.}\ \bibnamefont {R\"{o}{\ss}ler}}, \
  and\ \bibinfo {author} {\bibfnamefont {A.}~\bibnamefont {Rosch}},\ }\bibfield
   {title} {\enquote {\bibinfo {title} {{Non-Fermi Liquid Metal Without Quantum
  Criticality}},}\ }\href {\doibase 10.1126/science.1142644} {\bibfield
  {journal} {\bibinfo  {journal} {Science}\ }\textbf {\bibinfo {volume}
  {316}},\ \bibinfo {pages} {1871} (\bibinfo {year} {2007})}\BibitemShut
  {NoStop}%
\bibitem [{\citenamefont {Schulz}\ \emph {et~al.}(2012)\citenamefont {Schulz},
  \citenamefont {Ritz}, \citenamefont {Bauer}, \citenamefont {Halder},
  \citenamefont {Wagner}, \citenamefont {Franz}, \citenamefont {Pfleiderer},
  \citenamefont {Everschor}, \citenamefont {Garst},\ and\ \citenamefont
  {Rosch}}]{2012:Schulz:NaturePhys}%
  \BibitemOpen
  \bibfield  {author} {\bibinfo {author} {\bibfnamefont {T.}~\bibnamefont
  {Schulz}}, \bibinfo {author} {\bibfnamefont {R.}~\bibnamefont {Ritz}},
  \bibinfo {author} {\bibfnamefont {A.}~\bibnamefont {Bauer}}, \bibinfo
  {author} {\bibfnamefont {M.}~\bibnamefont {Halder}}, \bibinfo {author}
  {\bibfnamefont {M.}~\bibnamefont {Wagner}}, \bibinfo {author} {\bibfnamefont
  {C.}~\bibnamefont {Franz}}, \bibinfo {author} {\bibfnamefont
  {C.}~\bibnamefont {Pfleiderer}}, \bibinfo {author} {\bibfnamefont
  {K.}~\bibnamefont {Everschor}}, \bibinfo {author} {\bibfnamefont
  {M.}~\bibnamefont {Garst}}, \ and\ \bibinfo {author} {\bibfnamefont
  {A.}~\bibnamefont {Rosch}},\ }\bibfield  {title} {\enquote {\bibinfo {title}
  {{Emergent electrodynamics of skyrmions in a chiral magnet}},}\ }\href
  {\doibase 10.1038/nphys2231} {\bibfield  {journal} {\bibinfo  {journal}
  {Nature Phys.}\ }\textbf {\bibinfo {volume} {8}},\ \bibinfo {pages} {301}
  (\bibinfo {year} {2012})}\BibitemShut {NoStop}%
\bibitem [{\citenamefont {Pfleiderer}\ \emph {et~al.}(1997)\citenamefont
  {Pfleiderer}, \citenamefont {McMullan}, \citenamefont {Julian},\ and\
  \citenamefont {Lonzarich}}]{1997:Pfleiderer:PhysRevB}%
  \BibitemOpen
  \bibfield  {author} {\bibinfo {author} {\bibfnamefont {C.}~\bibnamefont
  {Pfleiderer}}, \bibinfo {author} {\bibfnamefont {G.~J.}\ \bibnamefont
  {McMullan}}, \bibinfo {author} {\bibfnamefont {S.~R.}\ \bibnamefont
  {Julian}}, \ and\ \bibinfo {author} {\bibfnamefont {G.~G.}\ \bibnamefont
  {Lonzarich}},\ }\bibfield  {title} {\enquote {\bibinfo {title} {{Magnetic
  quantum phase transition in MnSi under hydrostatic pressure}},}\ }\href
  {\doibase 10.1103/PhysRevB.55.8330} {\bibfield  {journal} {\bibinfo
  {journal} {Phys. Rev. B}\ }\textbf {\bibinfo {volume} {55}},\ \bibinfo
  {pages} {8330} (\bibinfo {year} {1997})}\BibitemShut {NoStop}%
\bibitem [{\citenamefont {Thessieu}\ \emph {et~al.}(1997)\citenamefont
  {Thessieu}, \citenamefont {Pfleiderer}, \citenamefont {Stepanov},\ and\
  \citenamefont {Flouquet}}]{1997:Thessieu:JPhysCondensMatter}%
  \BibitemOpen
  \bibfield  {author} {\bibinfo {author} {\bibfnamefont {C.}~\bibnamefont
  {Thessieu}}, \bibinfo {author} {\bibfnamefont {C.}~\bibnamefont
  {Pfleiderer}}, \bibinfo {author} {\bibfnamefont {A.~N.}\ \bibnamefont
  {Stepanov}}, \ and\ \bibinfo {author} {\bibfnamefont {J.}~\bibnamefont
  {Flouquet}},\ }\bibfield  {title} {\enquote {\bibinfo {title} {{Field
  dependence of the magnetic quantum phase transition in MnSi}},}\ }\href
  {\doibase 10.1088/0953-8984/9/31/019} {\bibfield  {journal} {\bibinfo
  {journal} {J. Phys.: Condens. Matter}\ }\textbf {\bibinfo {volume} {9}},\
  \bibinfo {pages} {6677} (\bibinfo {year} {1997})}\BibitemShut {NoStop}%
\bibitem [{\citenamefont {Doiron-Leyraud}\ \emph {et~al.}(2003)\citenamefont
  {Doiron-Leyraud}, \citenamefont {Walker}, \citenamefont {Taillefer},
  \citenamefont {Steiner}, \citenamefont {Julian},\ and\ \citenamefont
  {Lonzarich}}]{2003:Doiron-Leyraud:Nature}%
  \BibitemOpen
  \bibfield  {author} {\bibinfo {author} {\bibfnamefont {N.}~\bibnamefont
  {Doiron-Leyraud}}, \bibinfo {author} {\bibfnamefont {I.~R.}\ \bibnamefont
  {Walker}}, \bibinfo {author} {\bibfnamefont {L.}~\bibnamefont {Taillefer}},
  \bibinfo {author} {\bibfnamefont {M.~J.}\ \bibnamefont {Steiner}}, \bibinfo
  {author} {\bibfnamefont {S.~R.}\ \bibnamefont {Julian}}, \ and\ \bibinfo
  {author} {\bibfnamefont {G.~G.}\ \bibnamefont {Lonzarich}},\ }\bibfield
  {title} {\enquote {\bibinfo {title} {{Fermi-liquid breakdown in the
  paramagnetic phase of a pure metal}},}\ }\href {\doibase 10.1038/nature01968}
  {\bibfield  {journal} {\bibinfo  {journal} {Nature (London)}\ }\textbf
  {\bibinfo {volume} {425}},\ \bibinfo {pages} {595} (\bibinfo {year}
  {2003})}\BibitemShut {NoStop}%
\bibitem [{\citenamefont {Uemura}\ \emph {et~al.}(2007)\citenamefont {Uemura},
  \citenamefont {Goko}, \citenamefont {Gat-Malureanu}, \citenamefont {Carlo},
  \citenamefont {Russo}, \citenamefont {Savici}, \citenamefont {Aczel},
  \citenamefont {MacDougall}, \citenamefont {Rodriguez}, \citenamefont {Luke},
  \citenamefont {Dunsiger}, \citenamefont {McCollam}, \citenamefont {Arai},
  \citenamefont {Pfleiderer}, \citenamefont {B\"{o}ni}, \citenamefont
  {Yoshimura}, \citenamefont {Baggio-Saitovitch}, \citenamefont {Fontes},
  \citenamefont {Larrea}, \citenamefont {Sushko},\ and\ \citenamefont
  {Sereni}}]{2007:Uemura:NaturePhys}%
  \BibitemOpen
  \bibfield  {author} {\bibinfo {author} {\bibfnamefont {Y.~J.}\ \bibnamefont
  {Uemura}}, \bibinfo {author} {\bibfnamefont {T.}~\bibnamefont {Goko}},
  \bibinfo {author} {\bibfnamefont {I.~M.}\ \bibnamefont {Gat-Malureanu}},
  \bibinfo {author} {\bibfnamefont {J.~P.}\ \bibnamefont {Carlo}}, \bibinfo
  {author} {\bibfnamefont {P.~L.}\ \bibnamefont {Russo}}, \bibinfo {author}
  {\bibfnamefont {A.~T.}\ \bibnamefont {Savici}}, \bibinfo {author}
  {\bibfnamefont {A.}~\bibnamefont {Aczel}}, \bibinfo {author} {\bibfnamefont
  {G.~J.}\ \bibnamefont {MacDougall}}, \bibinfo {author} {\bibfnamefont
  {J.~A.}\ \bibnamefont {Rodriguez}}, \bibinfo {author} {\bibfnamefont {G.~M.}\
  \bibnamefont {Luke}}, \bibinfo {author} {\bibfnamefont {S.~R.}\ \bibnamefont
  {Dunsiger}}, \bibinfo {author} {\bibfnamefont {A.}~\bibnamefont {McCollam}},
  \bibinfo {author} {\bibfnamefont {J.}~\bibnamefont {Arai}}, \bibinfo {author}
  {\bibfnamefont {Ch.}\ \bibnamefont {Pfleiderer}}, \bibinfo {author}
  {\bibfnamefont {P.}~\bibnamefont {B\"{o}ni}}, \bibinfo {author}
  {\bibfnamefont {K.}~\bibnamefont {Yoshimura}}, \bibinfo {author}
  {\bibfnamefont {E.}~\bibnamefont {Baggio-Saitovitch}}, \bibinfo {author}
  {\bibfnamefont {M.~B.}\ \bibnamefont {Fontes}}, \bibinfo {author}
  {\bibfnamefont {J.}~\bibnamefont {Larrea}}, \bibinfo {author} {\bibfnamefont
  {Y.~V.}\ \bibnamefont {Sushko}}, \ and\ \bibinfo {author} {\bibfnamefont
  {J.}~\bibnamefont {Sereni}},\ }\bibfield  {title} {\enquote {\bibinfo {title}
  {{Phase separation and suppression of critical dynamics at quantum phase
  transitions of MnSi and (Sr$_{1-x}$Ca$_{x}$)RuO$_{3}$}},}\ }\href {\doibase
  10.1038/nphys488} {\bibfield  {journal} {\bibinfo  {journal} {Nature Phys.}\
  }\textbf {\bibinfo {volume} {3}},\ \bibinfo {pages} {29} (\bibinfo {year}
  {2007})}\BibitemShut {NoStop}%
\bibitem [{\citenamefont {Nagaosa}\ \emph {et~al.}(2010)\citenamefont
  {Nagaosa}, \citenamefont {Sinova}, \citenamefont {Onoda}, \citenamefont
  {MacDonald},\ and\ \citenamefont {Ong}}]{2010:Nagaosa:RevModPhys}%
  \BibitemOpen
  \bibfield  {author} {\bibinfo {author} {\bibfnamefont {N.}~\bibnamefont
  {Nagaosa}}, \bibinfo {author} {\bibfnamefont {J.}~\bibnamefont {Sinova}},
  \bibinfo {author} {\bibfnamefont {S.}~\bibnamefont {Onoda}}, \bibinfo
  {author} {\bibfnamefont {A.~H.}\ \bibnamefont {MacDonald}}, \ and\ \bibinfo
  {author} {\bibfnamefont {N.~P.}\ \bibnamefont {Ong}},\ }\bibfield  {title}
  {\enquote {\bibinfo {title} {{Anomalous Hall effect}},}\ }\href {\doibase
  10.1103/RevModPhys.82.1539} {\bibfield  {journal} {\bibinfo  {journal} {Rev.
  Mod. Phys.}\ }\textbf {\bibinfo {volume} {82}},\ \bibinfo {pages} {1539}
  (\bibinfo {year} {2010})}\BibitemShut {NoStop}%
\bibitem [{\citenamefont {Freimuth}\ \emph {et~al.}(2013)\citenamefont
  {Freimuth}, \citenamefont {Bamler}, \citenamefont {Mokrousov},\ and\
  \citenamefont {Rosch}}]{2013:Freimuth:PhysRevB}%
  \BibitemOpen
  \bibfield  {author} {\bibinfo {author} {\bibfnamefont {F.}~\bibnamefont
  {Freimuth}}, \bibinfo {author} {\bibfnamefont {R.}~\bibnamefont {Bamler}},
  \bibinfo {author} {\bibfnamefont {Y.}~\bibnamefont {Mokrousov}}, \ and\
  \bibinfo {author} {\bibfnamefont {A.}~\bibnamefont {Rosch}},\ }\bibfield
  {title} {\enquote {\bibinfo {title} {{Phase-space Berry phases in chiral
  magnets: Dzyaloshinskii-Moriya interaction and the charge of skyrmions}},}\
  }\href {\doibase 10.1103/PhysRevB.88.214409} {\bibfield  {journal} {\bibinfo
  {journal} {Phys. Rev. B}\ }\textbf {\bibinfo {volume} {88}},\ \bibinfo
  {pages} {214409} (\bibinfo {year} {2013})}\BibitemShut {NoStop}%
\bibitem [{\citenamefont {Yu}\ \emph {et~al.}(2012{\natexlab{a}})\citenamefont
  {Yu}, \citenamefont {Kanazawa}, \citenamefont {Zhang}, \citenamefont {Nagai},
  \citenamefont {Hara}, \citenamefont {Kimoto}, \citenamefont {Matsui},
  \citenamefont {Onose},\ and\ \citenamefont {Tokura}}]{2012:Yu:NatCommun}%
  \BibitemOpen
  \bibfield  {author} {\bibinfo {author} {\bibfnamefont {X.~Z.}\ \bibnamefont
  {Yu}}, \bibinfo {author} {\bibfnamefont {N.}~\bibnamefont {Kanazawa}},
  \bibinfo {author} {\bibfnamefont {W.~Z.}\ \bibnamefont {Zhang}}, \bibinfo
  {author} {\bibfnamefont {T.}~\bibnamefont {Nagai}}, \bibinfo {author}
  {\bibfnamefont {T.}~\bibnamefont {Hara}}, \bibinfo {author} {\bibfnamefont
  {K.}~\bibnamefont {Kimoto}}, \bibinfo {author} {\bibfnamefont
  {Y.}~\bibnamefont {Matsui}}, \bibinfo {author} {\bibfnamefont
  {Y.}~\bibnamefont {Onose}}, \ and\ \bibinfo {author} {\bibfnamefont
  {Y.}~\bibnamefont {Tokura}},\ }\bibfield  {title} {\enquote {\bibinfo {title}
  {{Skyrmion flow near room temperature in an ultralow current density}},}\
  }\href {\doibase 10.1038/ncomms1990} {\bibfield  {journal} {\bibinfo
  {journal} {Nat. Commun.}\ }\textbf {\bibinfo {volume} {3}},\ \bibinfo {pages}
  {988} (\bibinfo {year} {2012}{\natexlab{a}})}\BibitemShut {NoStop}%
\bibitem [{\citenamefont {Iwasaki}\ \emph
  {et~al.}(2013{\natexlab{a}})\citenamefont {Iwasaki}, \citenamefont
  {Mochizuki},\ and\ \citenamefont {Nagaosa}}]{2013:Iwasaki:NatCommun}%
  \BibitemOpen
  \bibfield  {author} {\bibinfo {author} {\bibfnamefont {J.}~\bibnamefont
  {Iwasaki}}, \bibinfo {author} {\bibfnamefont {M.}~\bibnamefont {Mochizuki}},
  \ and\ \bibinfo {author} {\bibfnamefont {N.}~\bibnamefont {Nagaosa}},\
  }\bibfield  {title} {\enquote {\bibinfo {title} {{Universal current-velocity
  relation of skyrmion motion in chiral magnets}},}\ }\href {\doibase
  10.1038/ncomms2442} {\bibfield  {journal} {\bibinfo  {journal} {Nat.
  Commun.}\ }\textbf {\bibinfo {volume} {4}},\ \bibinfo {pages} {1463}
  (\bibinfo {year} {2013}{\natexlab{a}})}\BibitemShut {NoStop}%
\bibitem [{\citenamefont {Thiele}(1973)}]{1973:Thiele:PhysRevLett}%
  \BibitemOpen
  \bibfield  {author} {\bibinfo {author} {\bibfnamefont {A.~A.}\ \bibnamefont
  {Thiele}},\ }\bibfield  {title} {\enquote {\bibinfo {title} {{Steady-State
  Motion of Magnetic Domains}},}\ }\href {\doibase 10.1103/PhysRevLett.30.230}
  {\bibfield  {journal} {\bibinfo  {journal} {Phys. Rev. Lett.}\ }\textbf
  {\bibinfo {volume} {30}},\ \bibinfo {pages} {230} (\bibinfo {year}
  {1973})}\BibitemShut {NoStop}%
\bibitem [{\citenamefont {Everschor}\ \emph {et~al.}(2011)\citenamefont
  {Everschor}, \citenamefont {Garst}, \citenamefont {Duine},\ and\
  \citenamefont {Rosch}}]{2011:Everschor:PhysRevB}%
  \BibitemOpen
  \bibfield  {author} {\bibinfo {author} {\bibfnamefont {K.}~\bibnamefont
  {Everschor}}, \bibinfo {author} {\bibfnamefont {M.}~\bibnamefont {Garst}},
  \bibinfo {author} {\bibfnamefont {R.~A.}\ \bibnamefont {Duine}}, \ and\
  \bibinfo {author} {\bibfnamefont {A.}~\bibnamefont {Rosch}},\ }\bibfield
  {title} {\enquote {\bibinfo {title} {{Current-induced rotational torques in
  the skyrmion lattice phase of chiral magnets}},}\ }\href {\doibase
  10.1103/PhysRevB.84.064401} {\bibfield  {journal} {\bibinfo  {journal} {Phys.
  Rev. B}\ }\textbf {\bibinfo {volume} {84}},\ \bibinfo {pages} {064401}
  (\bibinfo {year} {2011})}\BibitemShut {NoStop}%
\bibitem [{\citenamefont {Zang}\ \emph {et~al.}(2011)\citenamefont {Zang},
  \citenamefont {Mostovoy}, \citenamefont {Han},\ and\ \citenamefont
  {Nagaosa}}]{2011:Zang:PhysRevLett}%
  \BibitemOpen
  \bibfield  {author} {\bibinfo {author} {\bibfnamefont {J.}~\bibnamefont
  {Zang}}, \bibinfo {author} {\bibfnamefont {M.}~\bibnamefont {Mostovoy}},
  \bibinfo {author} {\bibfnamefont {J.~H.}\ \bibnamefont {Han}}, \ and\
  \bibinfo {author} {\bibfnamefont {N.}~\bibnamefont {Nagaosa}},\ }\bibfield
  {title} {\enquote {\bibinfo {title} {{Dynamics of Skyrmion Crystals in
  Metallic Thin Films}},}\ }\href {\doibase 10.1103/PhysRevLett.107.136804}
  {\bibfield  {journal} {\bibinfo  {journal} {Phys. Rev. Lett.}\ }\textbf
  {\bibinfo {volume} {107}},\ \bibinfo {pages} {136804} (\bibinfo {year}
  {2011})}\BibitemShut {NoStop}%
\bibitem [{\citenamefont {Tokunaga}\ \emph {et~al.}(2015)\citenamefont
  {Tokunaga}, \citenamefont {Yu}, \citenamefont {White}, \citenamefont
  {R{\o}nnow}, \citenamefont {Morikawa}, \citenamefont {Taguchi},\ and\
  \citenamefont {Tokura}}]{2015:Tokunaga:arXiv}%
  \BibitemOpen
  \bibfield  {author} {\bibinfo {author} {\bibfnamefont {Y.}~\bibnamefont
  {Tokunaga}}, \bibinfo {author} {\bibfnamefont {X.~Z.}\ \bibnamefont {Yu}},
  \bibinfo {author} {\bibfnamefont {J.~S.}\ \bibnamefont {White}}, \bibinfo
  {author} {\bibfnamefont {H.~M.}\ \bibnamefont {R{\o}nnow}}, \bibinfo {author}
  {\bibfnamefont {D.}~\bibnamefont {Morikawa}}, \bibinfo {author}
  {\bibfnamefont {Y.}~\bibnamefont {Taguchi}}, \ and\ \bibinfo {author}
  {\bibfnamefont {Y.}~\bibnamefont {Tokura}},\ }\bibfield  {title} {\enquote
  {\bibinfo {title} {{A new class of chiral materials hosting magnetic
  skyrmions beyond room temperature}},}\ }\href
  {http://arxiv.org/abs/1503.05651} {\bibfield  {journal} {\bibinfo  {journal}
  {arXiv:1503.05651}\ } (\bibinfo {year} {2015})}\BibitemShut {NoStop}%
\bibitem [{\citenamefont {Ruff}\ \emph {et~al.}(2015)\citenamefont {Ruff},
  \citenamefont {Widmann}, \citenamefont {Lunkenheimer}, \citenamefont
  {Tsurkan}, \citenamefont {Bord\'{a}cs}, \citenamefont {K\'{e}zsm\'{a}rki},\
  and\ \citenamefont {Loidl}}]{2015:Ruff:arXiv}%
  \BibitemOpen
  \bibfield  {author} {\bibinfo {author} {\bibfnamefont {E.}~\bibnamefont
  {Ruff}}, \bibinfo {author} {\bibfnamefont {S.}~\bibnamefont {Widmann}},
  \bibinfo {author} {\bibfnamefont {P.}~\bibnamefont {Lunkenheimer}}, \bibinfo
  {author} {\bibfnamefont {V.}~\bibnamefont {Tsurkan}}, \bibinfo {author}
  {\bibfnamefont {S.}~\bibnamefont {Bord\'{a}cs}}, \bibinfo {author}
  {\bibfnamefont {I.}~\bibnamefont {K\'{e}zsm\'{a}rki}}, \ and\ \bibinfo
  {author} {\bibfnamefont {A.}~\bibnamefont {Loidl}},\ }\bibfield  {title}
  {\enquote {\bibinfo {title} {{Ferroelectric Skyrmions and a Zoo of
  Multiferroic Phases in GaV$_{4}$S$_{8}$}},}\ }\href
  {http://arxiv.org/abs/1504.00309} {\bibfield  {journal} {\bibinfo  {journal}
  {arXiv:1504.00309}\ } (\bibinfo {year} {2015})}\BibitemShut {NoStop}%
\bibitem [{\citenamefont {Kanazawa}\ \emph {et~al.}(2011)\citenamefont
  {Kanazawa}, \citenamefont {Onose}, \citenamefont {Arima}, \citenamefont
  {Okuyama}, \citenamefont {Ohoyama}, \citenamefont {Wakimoto}, \citenamefont
  {Kakurai}, \citenamefont {Ishiwata},\ and\ \citenamefont
  {Tokura}}]{2011:Kanazawa:PhysRevLett}%
  \BibitemOpen
  \bibfield  {author} {\bibinfo {author} {\bibfnamefont {N.}~\bibnamefont
  {Kanazawa}}, \bibinfo {author} {\bibfnamefont {Y.}~\bibnamefont {Onose}},
  \bibinfo {author} {\bibfnamefont {T.}~\bibnamefont {Arima}}, \bibinfo
  {author} {\bibfnamefont {D.}~\bibnamefont {Okuyama}}, \bibinfo {author}
  {\bibfnamefont {K.}~\bibnamefont {Ohoyama}}, \bibinfo {author} {\bibfnamefont
  {S.}~\bibnamefont {Wakimoto}}, \bibinfo {author} {\bibfnamefont
  {K.}~\bibnamefont {Kakurai}}, \bibinfo {author} {\bibfnamefont
  {S.}~\bibnamefont {Ishiwata}}, \ and\ \bibinfo {author} {\bibfnamefont
  {Y.}~\bibnamefont {Tokura}},\ }\bibfield  {title} {\enquote {\bibinfo {title}
  {{Large Topological Hall Effect in a Short-Period Helimagnet MnGe}},}\ }\href
  {\doibase 10.1103/PhysRevLett.106.156603} {\bibfield  {journal} {\bibinfo
  {journal} {Phys. Rev. Lett.}\ }\textbf {\bibinfo {volume} {106}},\ \bibinfo
  {pages} {156603} (\bibinfo {year} {2011})}\BibitemShut {NoStop}%
\bibitem [{\citenamefont {Kanazawa}\ \emph {et~al.}(2012)\citenamefont
  {Kanazawa}, \citenamefont {Kim}, \citenamefont {Inosov}, \citenamefont
  {White}, \citenamefont {Egetenmeyer}, \citenamefont {Gavilano}, \citenamefont
  {Ishiwata}, \citenamefont {Onose}, \citenamefont {Arima}, \citenamefont
  {Keimer},\ and\ \citenamefont {Tokura}}]{2012:Kanazawa:PhysRevB}%
  \BibitemOpen
  \bibfield  {author} {\bibinfo {author} {\bibfnamefont {N.}~\bibnamefont
  {Kanazawa}}, \bibinfo {author} {\bibfnamefont {J.-H.}\ \bibnamefont {Kim}},
  \bibinfo {author} {\bibfnamefont {D.~S.}\ \bibnamefont {Inosov}}, \bibinfo
  {author} {\bibfnamefont {J.~S.}\ \bibnamefont {White}}, \bibinfo {author}
  {\bibfnamefont {N.}~\bibnamefont {Egetenmeyer}}, \bibinfo {author}
  {\bibfnamefont {J.~L.}\ \bibnamefont {Gavilano}}, \bibinfo {author}
  {\bibfnamefont {S.}~\bibnamefont {Ishiwata}}, \bibinfo {author}
  {\bibfnamefont {Y.}~\bibnamefont {Onose}}, \bibinfo {author} {\bibfnamefont
  {T.}~\bibnamefont {Arima}}, \bibinfo {author} {\bibfnamefont
  {B.}~\bibnamefont {Keimer}}, \ and\ \bibinfo {author} {\bibfnamefont
  {Y.}~\bibnamefont {Tokura}},\ }\bibfield  {title} {\enquote {\bibinfo {title}
  {{Possible skyrmion-lattice ground state in the $B20$ chiral-lattice magnet
  MnGe as seen via small-angle neutron scattering}},}\ }\href {\doibase
  10.1103/PhysRevB.86.134425} {\bibfield  {journal} {\bibinfo  {journal} {Phys.
  Rev. B}\ }\textbf {\bibinfo {volume} {86}},\ \bibinfo {pages} {134425}
  (\bibinfo {year} {2012})}\BibitemShut {NoStop}%
\bibitem [{\citenamefont {Shiomi}\ \emph {et~al.}(2013)\citenamefont {Shiomi},
  \citenamefont {Kanazawa}, \citenamefont {Shibata}, \citenamefont {Onose},\
  and\ \citenamefont {Tokura}}]{2013:Shiomi:PhysRevB}%
  \BibitemOpen
  \bibfield  {author} {\bibinfo {author} {\bibfnamefont {Y.}~\bibnamefont
  {Shiomi}}, \bibinfo {author} {\bibfnamefont {N.}~\bibnamefont {Kanazawa}},
  \bibinfo {author} {\bibfnamefont {K.}~\bibnamefont {Shibata}}, \bibinfo
  {author} {\bibfnamefont {Y.}~\bibnamefont {Onose}}, \ and\ \bibinfo {author}
  {\bibfnamefont {Y.}~\bibnamefont {Tokura}},\ }\bibfield  {title} {\enquote
  {\bibinfo {title} {{Topological Nernst effect in a three-dimensional
  skyrmion-lattice phase}},}\ }\href {\doibase 10.1103/PhysRevB.88.064409}
  {\bibfield  {journal} {\bibinfo  {journal} {Phys. Rev. B}\ }\textbf {\bibinfo
  {volume} {88}},\ \bibinfo {pages} {064409} (\bibinfo {year}
  {2013})}\BibitemShut {NoStop}%
\bibitem [{\citenamefont {Tanigaki}\ \emph {et~al.}(2015)\citenamefont
  {Tanigaki}, \citenamefont {Shibata}, \citenamefont {Kanazawa}, \citenamefont
  {Yu}, \citenamefont {Aizawa}, \citenamefont {Onose}, \citenamefont {Park},
  \citenamefont {Shindo},\ and\ \citenamefont {Tokura}}]{2015:Tanigaki:arXiv}%
  \BibitemOpen
  \bibfield  {author} {\bibinfo {author} {\bibfnamefont {T.}~\bibnamefont
  {Tanigaki}}, \bibinfo {author} {\bibfnamefont {K.}~\bibnamefont {Shibata}},
  \bibinfo {author} {\bibfnamefont {N.}~\bibnamefont {Kanazawa}}, \bibinfo
  {author} {\bibfnamefont {X.~Z.}\ \bibnamefont {Yu}}, \bibinfo {author}
  {\bibfnamefont {S.}~\bibnamefont {Aizawa}}, \bibinfo {author} {\bibfnamefont
  {Y.}~\bibnamefont {Onose}}, \bibinfo {author} {\bibfnamefont {H.~S.}\
  \bibnamefont {Park}}, \bibinfo {author} {\bibfnamefont {D.}~\bibnamefont
  {Shindo}}, \ and\ \bibinfo {author} {\bibfnamefont {Y.}~\bibnamefont
  {Tokura}},\ }\bibfield  {title} {\enquote {\bibinfo {title} {{Real-space
  observation of short-period cubic lattice of skyrmions in MnGe}},}\ }\href
  {http://arxiv.org/abs/1503.03945} {\bibfield  {journal} {\bibinfo  {journal}
  {arXiv:1503.03945}\ } (\bibinfo {year} {2015})}\BibitemShut {NoStop}%
\bibitem [{\citenamefont {Janoschek}\ \emph {et~al.}(2010)\citenamefont
  {Janoschek}, \citenamefont {Bernlochner}, \citenamefont {Dunsiger},
  \citenamefont {Pfleiderer}, \citenamefont {B\"{o}ni}, \citenamefont
  {Roessli}, \citenamefont {Link},\ and\ \citenamefont
  {Rosch}}]{2010:Janoschek:PhysRevB}%
  \BibitemOpen
  \bibfield  {author} {\bibinfo {author} {\bibfnamefont {M.}~\bibnamefont
  {Janoschek}}, \bibinfo {author} {\bibfnamefont {F.}~\bibnamefont
  {Bernlochner}}, \bibinfo {author} {\bibfnamefont {S.}~\bibnamefont
  {Dunsiger}}, \bibinfo {author} {\bibfnamefont {C.}~\bibnamefont
  {Pfleiderer}}, \bibinfo {author} {\bibfnamefont {P.}~\bibnamefont
  {B\"{o}ni}}, \bibinfo {author} {\bibfnamefont {B.}~\bibnamefont {Roessli}},
  \bibinfo {author} {\bibfnamefont {P.}~\bibnamefont {Link}}, \ and\ \bibinfo
  {author} {\bibfnamefont {A.}~\bibnamefont {Rosch}},\ }\bibfield  {title}
  {\enquote {\bibinfo {title} {{Helimagnon bands as universal excitations of
  chiral magnets}},}\ }\href {\doibase 10.1103/PhysRevB.81.214436} {\bibfield
  {journal} {\bibinfo  {journal} {Phys. Rev. B}\ }\textbf {\bibinfo {volume}
  {81}},\ \bibinfo {pages} {214436} (\bibinfo {year} {2010})}\BibitemShut
  {NoStop}%
\bibitem [{\citenamefont {Koralek}\ \emph {et~al.}(2012)\citenamefont
  {Koralek}, \citenamefont {Meier}, \citenamefont {Hinton}, \citenamefont
  {Bauer}, \citenamefont {Parameswaran}, \citenamefont {Vishwanath},
  \citenamefont {Ramesh}, \citenamefont {Schoenlein}, \citenamefont
  {Pfleiderer},\ and\ \citenamefont {Orenstein}}]{2012:Koralek:PhysRevLett}%
  \BibitemOpen
  \bibfield  {author} {\bibinfo {author} {\bibfnamefont {J.~D.}\ \bibnamefont
  {Koralek}}, \bibinfo {author} {\bibfnamefont {D.}~\bibnamefont {Meier}},
  \bibinfo {author} {\bibfnamefont {J.~P.}\ \bibnamefont {Hinton}}, \bibinfo
  {author} {\bibfnamefont {A.}~\bibnamefont {Bauer}}, \bibinfo {author}
  {\bibfnamefont {S.~A.}\ \bibnamefont {Parameswaran}}, \bibinfo {author}
  {\bibfnamefont {A.}~\bibnamefont {Vishwanath}}, \bibinfo {author}
  {\bibfnamefont {R.}~\bibnamefont {Ramesh}}, \bibinfo {author} {\bibfnamefont
  {R.~W.}\ \bibnamefont {Schoenlein}}, \bibinfo {author} {\bibfnamefont
  {C.}~\bibnamefont {Pfleiderer}}, \ and\ \bibinfo {author} {\bibfnamefont
  {J.}~\bibnamefont {Orenstein}},\ }\bibfield  {title} {\enquote {\bibinfo
  {title} {{Observation of Coherent Helimagnons and Gilbert Damping in an
  Itinerant Magnet}},}\ }\href {\doibase 10.1103/PhysRevLett.109.247204}
  {\bibfield  {journal} {\bibinfo  {journal} {Phys. Rev. Lett.}\ }\textbf
  {\bibinfo {volume} {109}},\ \bibinfo {pages} {247204} (\bibinfo {year}
  {2012})}\BibitemShut {NoStop}%
\bibitem [{\citenamefont {Onose}\ \emph {et~al.}(2012)\citenamefont {Onose},
  \citenamefont {Okamura}, \citenamefont {Seki}, \citenamefont {Ishiwata},\
  and\ \citenamefont {Tokura}}]{2012:Onose:PhysRevLett}%
  \BibitemOpen
  \bibfield  {author} {\bibinfo {author} {\bibfnamefont {Y.}~\bibnamefont
  {Onose}}, \bibinfo {author} {\bibfnamefont {Y.}~\bibnamefont {Okamura}},
  \bibinfo {author} {\bibfnamefont {S.}~\bibnamefont {Seki}}, \bibinfo {author}
  {\bibfnamefont {S.}~\bibnamefont {Ishiwata}}, \ and\ \bibinfo {author}
  {\bibfnamefont {Y.}~\bibnamefont {Tokura}},\ }\bibfield  {title} {\enquote
  {\bibinfo {title} {{Observation of Magnetic Excitations of Skyrmion Crystal
  in a Helimagnetic Insulator Cu$_{2}$OSeO$_{3}$}},}\ }\href {\doibase
  10.1103/PhysRevLett.109.037603} {\bibfield  {journal} {\bibinfo  {journal}
  {Phys. Rev. Lett.}\ }\textbf {\bibinfo {volume} {109}},\ \bibinfo {pages}
  {037603} (\bibinfo {year} {2012})}\BibitemShut {NoStop}%
\bibitem [{\citenamefont {Heinze}\ \emph {et~al.}(2011)\citenamefont {Heinze},
  \citenamefont {v.~Bergmann}, \citenamefont {Menzel}, \citenamefont {Brede},
  \citenamefont {Kubetzka}, \citenamefont {Wiesendanger}, \citenamefont
  {Bihlmayer},\ and\ \citenamefont {Bl\"{u}gel}}]{2011:Heinze:NaturePhys}%
  \BibitemOpen
  \bibfield  {author} {\bibinfo {author} {\bibfnamefont {S.}~\bibnamefont
  {Heinze}}, \bibinfo {author} {\bibfnamefont {K.}~\bibnamefont {v.~Bergmann}},
  \bibinfo {author} {\bibfnamefont {M.}~\bibnamefont {Menzel}}, \bibinfo
  {author} {\bibfnamefont {J.}~\bibnamefont {Brede}}, \bibinfo {author}
  {\bibfnamefont {A.}~\bibnamefont {Kubetzka}}, \bibinfo {author}
  {\bibfnamefont {R.}~\bibnamefont {Wiesendanger}}, \bibinfo {author}
  {\bibfnamefont {G.}~\bibnamefont {Bihlmayer}}, \ and\ \bibinfo {author}
  {\bibfnamefont {S.}~\bibnamefont {Bl\"{u}gel}},\ }\bibfield  {title}
  {\enquote {\bibinfo {title} {{Spontaneous atomic-scale magnetic skyrmion
  lattice in two dimensions}},}\ }\href {\doibase 10.1038/nphys2045} {\bibfield
   {journal} {\bibinfo  {journal} {Nature Phys.}\ }\textbf {\bibinfo {volume}
  {7}},\ \bibinfo {pages} {713} (\bibinfo {year} {2011})}\BibitemShut {NoStop}%
\bibitem [{\citenamefont {Romming}\ \emph {et~al.}(2013)\citenamefont
  {Romming}, \citenamefont {Hanneken}, \citenamefont {Menzel}, \citenamefont
  {Bickel}, \citenamefont {Wolter}, \citenamefont {v.~Bergmann}, \citenamefont
  {Kubetzka},\ and\ \citenamefont {Wiesendanger}}]{2013:Romming:Science}%
  \BibitemOpen
  \bibfield  {author} {\bibinfo {author} {\bibfnamefont {N.}~\bibnamefont
  {Romming}}, \bibinfo {author} {\bibfnamefont {C.}~\bibnamefont {Hanneken}},
  \bibinfo {author} {\bibfnamefont {M.}~\bibnamefont {Menzel}}, \bibinfo
  {author} {\bibfnamefont {J.~E.}\ \bibnamefont {Bickel}}, \bibinfo {author}
  {\bibfnamefont {B.}~\bibnamefont {Wolter}}, \bibinfo {author} {\bibfnamefont
  {K.}~\bibnamefont {v.~Bergmann}}, \bibinfo {author} {\bibfnamefont
  {A.}~\bibnamefont {Kubetzka}}, \ and\ \bibinfo {author} {\bibfnamefont
  {R.}~\bibnamefont {Wiesendanger}},\ }\bibfield  {title} {\enquote {\bibinfo
  {title} {{Writing and Deleting Single Magnetic Skyrmions}},}\ }\href
  {\doibase 10.1126/science.1240573} {\bibfield  {journal} {\bibinfo  {journal}
  {Science}\ }\textbf {\bibinfo {volume} {341}},\ \bibinfo {pages} {636}
  (\bibinfo {year} {2013})}\BibitemShut {NoStop}%
\bibitem [{\citenamefont {Yu}\ \emph {et~al.}(2012{\natexlab{b}})\citenamefont
  {Yu}, \citenamefont {Mostovoy}, \citenamefont {Tokunaga}, \citenamefont
  {Zhang}, \citenamefont {Kimoto}, \citenamefont {Matsui}, \citenamefont
  {Kaneko}, \citenamefont {Nagaosa},\ and\ \citenamefont
  {Tokura}}]{2012:Yu:PNatlAcadSciUSA}%
  \BibitemOpen
  \bibfield  {author} {\bibinfo {author} {\bibfnamefont {X.}~\bibnamefont
  {Yu}}, \bibinfo {author} {\bibfnamefont {M.}~\bibnamefont {Mostovoy}},
  \bibinfo {author} {\bibfnamefont {Y.}~\bibnamefont {Tokunaga}}, \bibinfo
  {author} {\bibfnamefont {W.}~\bibnamefont {Zhang}}, \bibinfo {author}
  {\bibfnamefont {K.}~\bibnamefont {Kimoto}}, \bibinfo {author} {\bibfnamefont
  {Y.}~\bibnamefont {Matsui}}, \bibinfo {author} {\bibfnamefont
  {Y.}~\bibnamefont {Kaneko}}, \bibinfo {author} {\bibfnamefont
  {N.}~\bibnamefont {Nagaosa}}, \ and\ \bibinfo {author} {\bibfnamefont
  {Y.}~\bibnamefont {Tokura}},\ }\bibfield  {title} {\enquote {\bibinfo {title}
  {{Magnetic stripes and skyrmions with helicity reversals}},}\ }\href
  {\doibase 10.1073/pnas.1118496109} {\bibfield  {journal} {\bibinfo  {journal}
  {P. Natl. Acad. Sci. USA}\ }\textbf {\bibinfo {volume} {109}},\ \bibinfo
  {pages} {8856} (\bibinfo {year} {2012}{\natexlab{b}})}\BibitemShut {NoStop}%
\bibitem [{\citenamefont {Finazzi}\ \emph {et~al.}(2013)\citenamefont
  {Finazzi}, \citenamefont {Savoini}, \citenamefont {Khorsand}, \citenamefont
  {Tsukamoto}, \citenamefont {Itoh}, \citenamefont {Du{\`o}}, \citenamefont
  {Kirilyuk}, \citenamefont {Rasing},\ and\ \citenamefont
  {Ezawa}}]{2013:Finazzi:PhysRevLett}%
  \BibitemOpen
  \bibfield  {author} {\bibinfo {author} {\bibfnamefont {M.}~\bibnamefont
  {Finazzi}}, \bibinfo {author} {\bibfnamefont {M.}~\bibnamefont {Savoini}},
  \bibinfo {author} {\bibfnamefont {A.~R.}\ \bibnamefont {Khorsand}}, \bibinfo
  {author} {\bibfnamefont {A.}~\bibnamefont {Tsukamoto}}, \bibinfo {author}
  {\bibfnamefont {A.}~\bibnamefont {Itoh}}, \bibinfo {author} {\bibfnamefont
  {L.}~\bibnamefont {Du{\`o}}}, \bibinfo {author} {\bibfnamefont
  {A.}~\bibnamefont {Kirilyuk}}, \bibinfo {author} {\bibfnamefont {Th.}\
  \bibnamefont {Rasing}}, \ and\ \bibinfo {author} {\bibfnamefont
  {M.}~\bibnamefont {Ezawa}},\ }\bibfield  {title} {\enquote {\bibinfo {title}
  {{Laser-Induced Magnetic Nanostructures with Tunable Topological
  Properties}},}\ }\href {\doibase 10.1103/PhysRevLett.110.177205} {\bibfield
  {journal} {\bibinfo  {journal} {Phys. Rev. Lett.}\ }\textbf {\bibinfo
  {volume} {110}},\ \bibinfo {pages} {177205} (\bibinfo {year}
  {2013})}\BibitemShut {NoStop}%
\bibitem [{\citenamefont {Sampaio}\ \emph {et~al.}(2013)\citenamefont
  {Sampaio}, \citenamefont {Cros}, \citenamefont {Rohart}, \citenamefont
  {Thiaville},\ and\ \citenamefont {Fert}}]{2013:Sampaio:NatureNano}%
  \BibitemOpen
  \bibfield  {author} {\bibinfo {author} {\bibfnamefont {J.}~\bibnamefont
  {Sampaio}}, \bibinfo {author} {\bibfnamefont {V.}~\bibnamefont {Cros}},
  \bibinfo {author} {\bibfnamefont {S.}~\bibnamefont {Rohart}}, \bibinfo
  {author} {\bibfnamefont {A.}~\bibnamefont {Thiaville}}, \ and\ \bibinfo
  {author} {\bibfnamefont {A.}~\bibnamefont {Fert}},\ }\bibfield  {title}
  {\enquote {\bibinfo {title} {{Nucleation, stability and current-induced
  motion of isolated magnetic skyrmions in nanostructures}},}\ }\href {\doibase
  10.1038/nnano.2013.210} {\bibfield  {journal} {\bibinfo  {journal} {Nature
  Nano.}\ }\textbf {\bibinfo {volume} {8}},\ \bibinfo {pages} {839} (\bibinfo
  {year} {2013})}\BibitemShut {NoStop}%
\bibitem [{\citenamefont {Yu}\ \emph {et~al.}(2013)\citenamefont {Yu},
  \citenamefont {DeGrave}, \citenamefont {Hara}, \citenamefont {Hara},
  \citenamefont {Jin},\ and\ \citenamefont {Tokura}}]{2013:Yu:NanoLett}%
  \BibitemOpen
  \bibfield  {author} {\bibinfo {author} {\bibfnamefont {X.}~\bibnamefont
  {Yu}}, \bibinfo {author} {\bibfnamefont {J.~P.}\ \bibnamefont {DeGrave}},
  \bibinfo {author} {\bibfnamefont {Y.}~\bibnamefont {Hara}}, \bibinfo {author}
  {\bibfnamefont {T.}~\bibnamefont {Hara}}, \bibinfo {author} {\bibfnamefont
  {S.}~\bibnamefont {Jin}}, \ and\ \bibinfo {author} {\bibfnamefont
  {Y.}~\bibnamefont {Tokura}},\ }\bibfield  {title} {\enquote {\bibinfo {title}
  {{Observation of the Magnetic Skyrmion Lattice in a MnSi Nanowire by Lorentz
  TEM}},}\ }\href {\doibase 10.1021/nl401687d} {\bibfield  {journal} {\bibinfo
  {journal} {Nano Lett.}\ }\textbf {\bibinfo {volume} {13}},\ \bibinfo {pages}
  {3755} (\bibinfo {year} {2013})}\BibitemShut {NoStop}%
\bibitem [{\citenamefont {Iwasaki}\ \emph
  {et~al.}(2013{\natexlab{b}})\citenamefont {Iwasaki}, \citenamefont
  {Mochizuki},\ and\ \citenamefont {Nagaosa}}]{2013:Iwasaki:NatureNano}%
  \BibitemOpen
  \bibfield  {author} {\bibinfo {author} {\bibfnamefont {J.}~\bibnamefont
  {Iwasaki}}, \bibinfo {author} {\bibfnamefont {M.}~\bibnamefont {Mochizuki}},
  \ and\ \bibinfo {author} {\bibfnamefont {N.}~\bibnamefont {Nagaosa}},\
  }\bibfield  {title} {\enquote {\bibinfo {title} {{Current-induced skyrmion
  dynamics in constricted geometries}},}\ }\href {\doibase
  10.1038/nnano.2013.176} {\bibfield  {journal} {\bibinfo  {journal} {Nature
  Nano.}\ }\textbf {\bibinfo {volume} {8}},\ \bibinfo {pages} {742} (\bibinfo
  {year} {2013}{\natexlab{b}})}\BibitemShut {NoStop}%
\bibitem [{\citenamefont {Lin}\ \emph {et~al.}(2013)\citenamefont {Lin},
  \citenamefont {Reichhardt}, \citenamefont {Batista},\ and\ \citenamefont
  {Saxena}}]{2013:Lin:PhysRevLett}%
  \BibitemOpen
  \bibfield  {author} {\bibinfo {author} {\bibfnamefont {S.-Z.}\ \bibnamefont
  {Lin}}, \bibinfo {author} {\bibfnamefont {C.}~\bibnamefont {Reichhardt}},
  \bibinfo {author} {\bibfnamefont {C.~D.}\ \bibnamefont {Batista}}, \ and\
  \bibinfo {author} {\bibfnamefont {A.}~\bibnamefont {Saxena}},\ }\bibfield
  {title} {\enquote {\bibinfo {title} {{Driven Skyrmions and Dynamical
  Transitions in Chiral Magnets}},}\ }\href {\doibase
  10.1103/PhysRevLett.110.207202} {\bibfield  {journal} {\bibinfo  {journal}
  {Phys. Rev. Lett.}\ }\textbf {\bibinfo {volume} {110}},\ \bibinfo {pages}
  {207202} (\bibinfo {year} {2013})}\BibitemShut {NoStop}%
\bibitem [{\citenamefont {Iwasaki}\ \emph {et~al.}(2014)\citenamefont
  {Iwasaki}, \citenamefont {Koshibae},\ and\ \citenamefont
  {Nagaosa}}]{2014:Iwasaki:NanoLett}%
  \BibitemOpen
  \bibfield  {author} {\bibinfo {author} {\bibfnamefont {J.}~\bibnamefont
  {Iwasaki}}, \bibinfo {author} {\bibfnamefont {W.}~\bibnamefont {Koshibae}}, \
  and\ \bibinfo {author} {\bibfnamefont {N.}~\bibnamefont {Nagaosa}},\
  }\bibfield  {title} {\enquote {\bibinfo {title} {{Colossal Spin Transfer
  Torque Effect on Skyrmion along the Edge}},}\ }\href {\doibase
  10.1021/nl501379k} {\bibfield  {journal} {\bibinfo  {journal} {Nano Lett.}\
  }\textbf {\bibinfo {volume} {14}},\ \bibinfo {pages} {4432} (\bibinfo {year}
  {2014})}\BibitemShut {NoStop}%
\bibitem [{\citenamefont {Lin}\ \emph {et~al.}(2014)\citenamefont {Lin},
  \citenamefont {Batista}, \citenamefont {Reichhardt},\ and\ \citenamefont
  {Saxena}}]{2014:Lin:PhysRevLett}%
  \BibitemOpen
  \bibfield  {author} {\bibinfo {author} {\bibfnamefont {S.-Z.}\ \bibnamefont
  {Lin}}, \bibinfo {author} {\bibfnamefont {C.~D.}\ \bibnamefont {Batista}},
  \bibinfo {author} {\bibfnamefont {C.}~\bibnamefont {Reichhardt}}, \ and\
  \bibinfo {author} {\bibfnamefont {A.}~\bibnamefont {Saxena}},\ }\bibfield
  {title} {\enquote {\bibinfo {title} {{ac Current Generation in Chiral
  Magnetic Insulators and Skyrmion Motion induced by the Spin Seebeck
  Effect}},}\ }\href {\doibase 10.1103/PhysRevLett.112.187203} {\bibfield
  {journal} {\bibinfo  {journal} {Phys. Rev. Lett.}\ }\textbf {\bibinfo
  {volume} {112}},\ \bibinfo {pages} {187203} (\bibinfo {year}
  {2014})}\BibitemShut {NoStop}%
\bibitem [{\citenamefont {M\"{u}ller}\ and\ \citenamefont
  {Rosch}(2015)}]{2015:Muller:PhysRevB}%
  \BibitemOpen
  \bibfield  {author} {\bibinfo {author} {\bibfnamefont {J.}~\bibnamefont
  {M\"{u}ller}}\ and\ \bibinfo {author} {\bibfnamefont {A.}~\bibnamefont
  {Rosch}},\ }\bibfield  {title} {\enquote {\bibinfo {title} {{Capturing of a
  magnetic skyrmion with a hole}},}\ }\href {\doibase
  10.1103/PhysRevB.91.054410} {\bibfield  {journal} {\bibinfo  {journal} {Phys.
  Rev. B}\ }\textbf {\bibinfo {volume} {91}},\ \bibinfo {pages} {054410}
  (\bibinfo {year} {2015})}\BibitemShut {NoStop}%
\bibitem [{\citenamefont {Zhang}\ \emph {et~al.}(2015)\citenamefont {Zhang},
  \citenamefont {Zhao}, \citenamefont {Fangohr}, \citenamefont {Liu},
  \citenamefont {Xia}, \citenamefont {Xia},\ and\ \citenamefont
  {Morvan}}]{2015:Zhang:SciRep}%
  \BibitemOpen
  \bibfield  {author} {\bibinfo {author} {\bibfnamefont {X.}~\bibnamefont
  {Zhang}}, \bibinfo {author} {\bibfnamefont {G.~P.}\ \bibnamefont {Zhao}},
  \bibinfo {author} {\bibfnamefont {H.}~\bibnamefont {Fangohr}}, \bibinfo
  {author} {\bibfnamefont {J.~P.}\ \bibnamefont {Liu}}, \bibinfo {author}
  {\bibfnamefont {W.~X.}\ \bibnamefont {Xia}}, \bibinfo {author} {\bibfnamefont
  {J.}~\bibnamefont {Xia}}, \ and\ \bibinfo {author} {\bibfnamefont {F.~J.}\
  \bibnamefont {Morvan}},\ }\bibfield  {title} {\enquote {\bibinfo {title}
  {{Skyrmion-skyrmion and skyrmion-edge repulsions in skyrmion-based racetrack
  memory}},}\ }\href {\doibase 10.1038/srep07643} {\bibfield  {journal}
  {\bibinfo  {journal} {Sci. Rep.}\ }\textbf {\bibinfo {volume} {5}},\ \bibinfo
  {pages} {7643} (\bibinfo {year} {2015})}\BibitemShut {NoStop}%
\bibitem [{\citenamefont {Ogawa}\ \emph {et~al.}(2015)\citenamefont {Ogawa},
  \citenamefont {Seki},\ and\ \citenamefont {Tokura}}]{2015:Ogawa:SciRep}%
  \BibitemOpen
  \bibfield  {author} {\bibinfo {author} {\bibfnamefont {N.}~\bibnamefont
  {Ogawa}}, \bibinfo {author} {\bibfnamefont {S.}~\bibnamefont {Seki}}, \ and\
  \bibinfo {author} {\bibfnamefont {Y.}~\bibnamefont {Tokura}},\ }\bibfield
  {title} {\enquote {\bibinfo {title} {{Ultrafast optical excitation of
  magnetic skyrmions}},}\ }\href {\doibase 10.1038/srep09552} {\bibfield
  {journal} {\bibinfo  {journal} {Sci. Rep.}\ }\textbf {\bibinfo {volume}
  {5}},\ \bibinfo {pages} {9552} (\bibinfo {year} {2015})}\BibitemShut
  {NoStop}%
\bibitem [{\citenamefont {Xie}\ \emph {et~al.}(2013)\citenamefont {Xie},
  \citenamefont {Thimmaiah}, \citenamefont {Lamsal}, \citenamefont {Liu},
  \citenamefont {Heitmann}, \citenamefont {Quirinale}, \citenamefont {Goldman},
  \citenamefont {Pecharsky},\ and\ \citenamefont
  {Miller}}]{2013:Xie:InorgChem}%
  \BibitemOpen
  \bibfield  {author} {\bibinfo {author} {\bibfnamefont {W.}~\bibnamefont
  {Xie}}, \bibinfo {author} {\bibfnamefont {S.}~\bibnamefont {Thimmaiah}},
  \bibinfo {author} {\bibfnamefont {J.}~\bibnamefont {Lamsal}}, \bibinfo
  {author} {\bibfnamefont {J.}~\bibnamefont {Liu}}, \bibinfo {author}
  {\bibfnamefont {T.~W.}\ \bibnamefont {Heitmann}}, \bibinfo {author}
  {\bibfnamefont {D.}~\bibnamefont {Quirinale}}, \bibinfo {author}
  {\bibfnamefont {A.~I.}\ \bibnamefont {Goldman}}, \bibinfo {author}
  {\bibfnamefont {V.}~\bibnamefont {Pecharsky}}, \ and\ \bibinfo {author}
  {\bibfnamefont {G.~J.}\ \bibnamefont {Miller}},\ }\bibfield  {title}
  {\enquote {\bibinfo {title} {{$\beta$-Mn-Type Co$_{8+x}$Zn$_{12–x}$ as a
  Defect Cubic Laves Phase: Site Preferences, Magnetism, and Electronic
  Structure}},}\ }\href {\doibase 10.1021/ic4009653} {\bibfield  {journal}
  {\bibinfo  {journal} {Inorg. Chem.}\ }\textbf {\bibinfo {volume} {52}},\
  \bibinfo {pages} {9399} (\bibinfo {year} {2013})}\BibitemShut {NoStop}%
\bibitem [{\citenamefont {Pocha}\ \emph {et~al.}(2010)\citenamefont {Pocha},
  \citenamefont {Johrendt},\ and\ \citenamefont
  {P\"{o}ttgen}}]{2010:Pocha:ChemMater}%
  \BibitemOpen
  \bibfield  {author} {\bibinfo {author} {\bibfnamefont {R.}~\bibnamefont
  {Pocha}}, \bibinfo {author} {\bibfnamefont {D.}~\bibnamefont {Johrendt}}, \
  and\ \bibinfo {author} {\bibfnamefont {R.}~\bibnamefont {P\"{o}ttgen}},\
  }\bibfield  {title} {\enquote {\bibinfo {title} {{Electronic and Structural
  Instabilities in GaV$_{4}$S$_{8}$ and GaMo$_{4}$S$_{8}$}},}\ }\href {\doibase
  10.1021/cm001099b} {\bibfield  {journal} {\bibinfo  {journal} {Chem. Mater.}\
  }\textbf {\bibinfo {volume} {12}},\ \bibinfo {pages} {2882} (\bibinfo {year}
  {2010})}\BibitemShut {NoStop}%
\bibitem [{\citenamefont {Yadav}\ \emph {et~al.}(2008)\citenamefont {Yadav},
  \citenamefont {Nigam},\ and\ \citenamefont {Rastogi}}]{2008:Yadav:PhysicaB}%
  \BibitemOpen
  \bibfield  {author} {\bibinfo {author} {\bibfnamefont {C.~S.}\ \bibnamefont
  {Yadav}}, \bibinfo {author} {\bibfnamefont {A.~K.}\ \bibnamefont {Nigam}}, \
  and\ \bibinfo {author} {\bibfnamefont {A.~K.}\ \bibnamefont {Rastogi}},\
  }\bibfield  {title} {\enquote {\bibinfo {title} {{Thermodynamic properties of
  ferromagnetic Mott-insulator GaV$_{4}$S$_{8}$}},}\ }\href {\doibase
  10.1016/j.physb.2007.10.172} {\bibfield  {journal} {\bibinfo  {journal}
  {Physica B}\ }\textbf {\bibinfo {volume} {403}},\ \bibinfo {pages} {1474}
  (\bibinfo {year} {2008})}\BibitemShut {NoStop}%
\end{thebibliography}%

\end{document}